\documentclass[english]{article}
\usepackage[LGR,T1]{fontenc}
\usepackage[latin9]{inputenc}
\usepackage{color}
\usepackage{units}
\usepackage{textcomp}
\usepackage{mathrsfs}
\usepackage{amsmath}
\usepackage{amssymb}
\usepackage{cancel}
\usepackage{setspace}
\usepackage{esint}
\makeatletter

\DeclareRobustCommand{\greektext}{%
  \fontencoding{LGR}\selectfont\def\encodingdefault{LGR}}
\DeclareRobustCommand{\textgreek}[1]{\leavevmode{\greektext #1}}
\ProvideTextCommand{\~}{LGR}[1]{\char126#1}

\usepackage{fullpage}

\makeatother

\usepackage{babel}
\begin{document}
\title{Yang-Mills sources in biconformal gravity}
\author{Davis W. Muhwezi\thanks{Utah State University Department of Physics, dvsmuhwezi@gmail.com}
$\;$and James T. Wheeler\thanks{Utah State University Department of Physics, jim.wheeler@usu.edu}}
\maketitle
\begin{abstract}
Biconformal gravity, based on gauging of the conformal group to $2n$
dimensions, reproduces $n$-dim scale-covariant general relativity
on the co-tangent bundle in any dimension. We generalize this result
to include Yang-Mills matter sources formulated as $SU\left(N\right)$
gauge theories with a twisted action on the full $2n$-dimensional
biconformal space. We show that the coupling of the sources to gravity
does not stop the reduction of effective dimension $2n\rightarrow n$
of the gravity theory, and instead forces the Yang-Mills source to
reduce to ordinary $n$-dimensional Yang-Mills theory on the gravitating
cotangent bundle, with the usual Yang-Mills energy tensor as gravitational
source. The results apply as well to gravity with sources on Kähler
manifolds and in double field theories.
\end{abstract}
\newpage{}

\section{Introduction}

In general relativity the coupling of matter sources to gravity is
accomplished by making the matter action invariant under general coordinate
transformations, then adding it to the Einstein-Hilbert action. In
formulations of general relativity based on the conformal group, there
may be additional conditions. Here we examine $SU\left(N\right)$
gauge theories as sources for gravity in a large class of spaces of
doubled dimension.

Doubled dimension of spacetime arises in various contexts, frequently
related to the idea of a relativistic phase space. Born reciprocity
(M. Born, \cite{Born1938,Born1949}) was one early suggestion aimed
at unifying relativity and quantum theory. The reciprocity involves
the scaled symplectic exchange $x^{\alpha}\rightarrow ap^{\alpha},p^{\beta}\rightarrow-bx^{\beta}$,
thereby preserving Hamilton's equations. Further developments include
the study of Kähler manifolds, with mutually compatible metric, symplectic,
and complex structures such that any two of the structures yield the
third.

In 1982, using a gauge theory approach to gravity, Ivanov and Niederle
\cite{IvanovI,Ivanov} showed that general relativity can arise in
a space of doubled dimension called biconformal space. Generalizing
the 8-dimensional quotient of the conformal group of spacetime by
its homogeneous Weyl subgroup led the authors to a class of curved
geometries. From an action quadratic in the curvatures they found
that suitable constraints reduced the field equations to the Einstein
equation in 4-dimensions.

The group quotient approach used in \cite{IvanovI,Ivanov} generalizes
to arbitrary dimension $n$ and signature $\left(p,q\right)$. The
quotient of conformal group of an $n$ dimensional space by its Weyl
subgroup yields a $2n$ dimensional biconformal manifold. In \cite{WW}
it was shown that biconformal spaces of any dimension $2n$ admit
an action \emph{linear} in the curvature with field equations reducing
to the vacuum Einstein equation in $n$ dimensions. Unlike the other
double field theories cited below, this reduction in both field count
and the number of independent variables occurs \emph{by virtue of
the field equations}. Starting from the most general action linear
in the curvatures--which takes essentially the same form in any dimension--it
has now been shown that the field equations generically lead to scale
invariant general relativity on the co-tangent bundle \cite{WW,WheelerAugust}.
As in Riemannian geometry, these biconformal spaces are taken to be
torsion free.

Because biconformal spaces have natural symplectic structure \cite{Hazboun Wheeler}
they give an arena appropriate to quantum problems \cite{AWQM}, and
when they are required to fully reproduce the properties of phase
space, the $\left(3,1\right)$ signature of spacetime emerges necessarliy
from an originally Euclidean space \cite{Spencer Wheeler,Hazboun Wheeler,Hazboun dissertation}.
It is further shown in \cite{WheelerAugust,Hazboun Wheeler,Hazboun dissertation,Hazboun}
that these spaces are generically Kähler, and that they share the
properties of double field theories.

Some years after the first biconformal spaces, another form of doubled
dimension called double field theories arose as a means of making
the $O(d,d)$ symmetry of $T$-duality manifest. By introducing scalars
to produce an additional $d$ dimensions, Duff \cite{Duff}{]} doubled
the X(\textgreek{sv},\textgreek{t}) string variables to make this
$O(d,d)$ symmetry manifest. Siegel brought the idea to full fruition
by deriving results from superstring theory \cite{Siegel1,Siegel2,Siegel3}.
Allowing fields to depend on all $2d$ coordinates, Siegel introduced
generalized Lie brackets, gauge transformations, covariant derivatives,
and a section condition on the full doubled space, thereby introducing
torsions and curvatures in addition to the manifest $T$-duality.
By restricting half the coordinates--called imposing a \emph{section
condition}--one recovers the $d$-dimensional theory.

In any of these doubled dimension gravity theories it is desirable
to understand matter couplings, and preferable to introduce the matter
fields \emph{ab initio} in the doubled space. Carrying out such an
investigation in biconformal spaces simultaneously gives results applicable
to other doubled spaces. Because biconformal gravity is a gauge theory,
it allows direct extension of the symmetry to include gravitational
sources from $SU\left(N\right)$ gauge theories. However, such Yang-Mills
type sources must be written in the space of doubled dimension. This
means the introduction of far more potential source fields, $\frac{2n\left(2n-1\right)}{2}\times\left(N^{2}-1\right)$
instead of only $\frac{n\left(n-1\right)}{2}\times\left(N^{2}-1\right)$,
each depending on $2n$ independent variables instead of only $n$.
There are two principal questions we address here. First is the question
of the correct form of the Yang-Mills action in the doubled dimension.
We find that the usual $\int tr\left(\boldsymbol{\mathcal{F}}\,^{*}\boldsymbol{\mathcal{F}}\right)$
form must be augmented with a ``twist'' to reproduce familiar results,
similar to the twist found in studies of double field theories \cite{SUSY Twist,SUSY double gauge theory,Twisted DFT,Gauged double field theory}.
The second question is whether the increase in fields and independent
variables spoils the gravitational reduction, or at the other extreme,
shares the reduction to $n$ dimensions with gravity. We find that
both gravity and sources reduce to $n$-dimensions. The reduction
again occurs by virtue of the field equations, with no need for section
conditions.

A further result arises from this investigation. With the exceptions
of a study of biconformal supersymmetry \cite{AW}, which necessarily
includes matter fields, and of a brief study \cite{WWMatter} with
scalar field sources, previous solutions and further properties of
biconformal gravity \cite{Ivanov,WW,WheelerAugust,Hazboun Wheeler,Hazboun dissertation,Lovelady,NCG}
have been based on pure gravity solutions. The field equations for
pure biconformal gravity arise by variation of the gauge fields, and
it has not been necessary to introduce a metric. However, the actions
for Yang-Mills theories involve the Hodge dual and therefore a metric.
But biconformal spaces possess both conformal and Kähler structures,
it is not clear which of these should take precedence in defining
the orthonormality of the basis gauge fields. The ambiguity in identifying
the metric is confounded with the specification of the biconformal
matter action. The outcome of the current investigation is that only
the Killing form reproduces the expected coupling to general relativity.

Concretely, biconformal spaces are spanned by two sets of frame fields,
$\left(\mathbf{e}^{a},\mathbf{f}_{b}\right)$, called the solder form
and the co-solder form. In solutions, the solder form, $\mathbf{e}^{a}$,
reduces via the field equations to the usual solder form on spacetime.
In order to write any Yang-Mills action we are compelled to place
an orthonormality condition on these forms,
\begin{eqnarray*}
\left\langle \mathbf{e}^{a},\mathbf{e}^{b}\right\rangle  & = & M^{ab}\\
\left\langle \mathbf{e}^{a},\mathbf{f}_{b}\right\rangle  & = & M_{\;\;\;b}^{a}\\
\left\langle \mathbf{f}_{a},\mathbf{f}_{b}\right\rangle  & = & M_{ab}
\end{eqnarray*}
where some invariant matrix,
\[
M^{AB}=\left(\begin{array}{cc}
M^{ab} & M_{\;\;\;b}^{a}\\
M_{a}^{\;\;\;b} & M_{ab}
\end{array}\right)
\]
must be specified. A central hurdle in the course of our study was
that there are two natural candidates for $M^{AB}$: the Killing form
of the conformal group restricted to the base manifold, and the Kähler
metric of the Kähler structure. We carry out the search for a suitable
Yang-Mills action for each candidate symmetric form. Ultimately, we
find that only the Killing form can give the expected coupling to
gravitation. It is therefore the Killing form that provides the orthonormality
of the solder and co-solder forms throughout the remainder of the
paper.

Before proceeding with our investigation of $SU\left(N\right)$ sources
in biconformal/double-field-theory/Kahler gravity, we look briefly
at sources in other gravity theories. We first exhibit general relativity
with Yang-Mills sources as a gauge theory. This displays our general
approach to gauging, as well as the reduced result we hope to achieve.
In the remainder of this Section, we look at the modifications of
other conformally based gravity theories required in order to include
sources. In general relativity, we must extend the symmetry of source
actions to general coordinate from global Lorentz. By analogy, we
expect that including matter in conformally based theories may require
some additional conditions. We briefly discuss sources in scale-invariant
gravity and Weyl gravity, before concluding the Section with a discussion
of sources in doubled dimension theories.

\subsection{Yang-Mills matter in general relativity}

Defining a projection on the quotient of the $\frac{n\left(n+1\right)}{2}$-dimensional
Poincarè group $\mathcal{P}$ by its $\frac{n\left(n-1\right)}{2}$-dimensional
Lorentz subgroup $\mathcal{L}$ gives a principal fiber bundle with
Lorentz fibers over an $n$-dimensional Minkowski spacetime. Generalizing
the base space by changing the Maurer-Cartan connection forms of the
Poincarè group and perhaps changing the manifold, the fiber structure
is maintained by demanding horizontality of the curvature and torsion.
The result is a Riemann-Cartan geometry characterized by curvature
and torsion with local Lorentz symmetry.

Concretely, the generalization of the connection $\left(\tilde{\mathbf{e}}^{b},\tilde{\boldsymbol{\omega}}_{\;\;\;b}^{a}\right)\Rightarrow\left(\mathbf{e}^{b},\boldsymbol{\omega}_{\;\;\;b}^{a}\right)$
takes the Maurer-Cartan equations of the Poincarè group,
\begin{eqnarray*}
\mathbf{d}\tilde{\boldsymbol{\omega}}_{\;\;\;b}^{a} & = & \tilde{\boldsymbol{\omega}}_{\;\;\;b}^{c}\land\tilde{\boldsymbol{\omega}}_{\;\;\;c}^{a}\\
\mathbf{d}\tilde{\mathbf{e}}^{a} & = & \tilde{\mathbf{e}}^{b}\land\tilde{\boldsymbol{\omega}}_{\;\;\;b}^{a}
\end{eqnarray*}
to the Cartan equations,
\begin{eqnarray}
\mathbf{d}\boldsymbol{\omega}_{\;\;\;b}^{a} & = & \boldsymbol{\omega}_{\;\;\;b}^{c}\land\boldsymbol{\omega}_{\;\;\;c}^{a}+\mathbf{R}_{\;\;\;b}^{a}\label{Curvature Cartan equation}\\
\mathbf{d}\mathbf{e}^{a} & = & \mathbf{e}^{b}\land\boldsymbol{\omega}_{\;\;\;b}^{a}+\mathbf{T}^{a}\label{Torsion Cartan equation}
\end{eqnarray}
where horizontality of the curvature $\mathbf{R}^{ab}$ and the torsion
$\mathbf{T}^{a}$ is captured by omitting any occurrence of the spin
connection when writing them expressly as 2-forms
\begin{eqnarray*}
\mathbf{R}_{\;\;\;b}^{a} & = & \frac{1}{2}R_{\;\;\;bcd}^{a}\mathbf{e}^{c}\land\mathbf{e}^{d}\\
\mathbf{T}^{a} & = & \frac{1}{2}T_{\;\;\;bc}^{a}\mathbf{e}^{b}\land\mathbf{e}^{c}
\end{eqnarray*}
Horizontality insures the survival of the principal fiber bundle by
guaranteeing that integrals of the curvatures over an area, or equivalenty
of the connection forms over closed curves, are independent of lifting.

Completing the description of the Riemann-Cartan geometry is the demand
for integrability of the Cartan equations, which follows by exterior
differentiation of Eqs.(\ref{Curvature Cartan equation}) and (\ref{Torsion Cartan equation}):
\begin{eqnarray*}
\mathbf{D}\mathbf{R}_{\;\;\;b}^{a} & = & 0\\
\mathbf{D}\mathbf{T}^{a} & = & \mathbf{e}^{b}\land\mathbf{R}_{\;\;\;b}^{a}
\end{eqnarray*}
When torsion vanishes, this construction describes a general $n$-dimensional
Riemannian spacetime with local Lorentz symmetry.

To include an additional $SU\left(N\right)$ Yang-Mills symmetry in
the fiber bundle we extend the $\mathcal{P}/\mathcal{L}$ quotient
to the quotient of the product $\mathcal{P}\times SU\left(N\right)$
by the product $\mathcal{L}\times SU\left(N\right)$:
\[
\mathcal{P}\times SU\left(N\right)/\mathcal{L}\times SU\left(N\right)
\]
and carry out the same procedure. This still results in an $n$-dimensional
spacetime but now the fibers of the principal bundle are isomorphic
to $\mathcal{L}\times SU\left(N\right)$. The Cartan Eqs.(\ref{Curvature Cartan equation})
and (\ref{Torsion Cartan equation}) are augmented by a third equation,
\begin{eqnarray*}
\mathbf{d}\mathbf{A}^{i} & = & -\frac{1}{2}c_{\;\;\;jk}^{i}\mathbf{A}^{j}\land\mathbf{A}^{k}+\mathbf{F}^{i}
\end{eqnarray*}
where indices beginning with $i$ have range $i,j,k,\ldots=1,2,\ldots,N^{2}-1$,
and $\text{\ensuremath{\mathbf{F}^{i}} }$ is horizontal
\begin{eqnarray*}
\mathbf{F}^{i} & = & \frac{1}{2}F_{\;\;\;ab}^{i}\mathbf{e}^{a}\land\mathbf{e}^{b}
\end{eqnarray*}
The integrability condition is $\boldsymbol{\mathcal{D}}\mathbf{F}^{i}=0$,
where $\boldsymbol{\mathcal{D}}$ is the $SU\left(N\right)$ covariant
derivative. Here, indices from the first part of the alphabet have
range $a,b,\ldots=1,\ldots,n$.

To build a physical theory, we write an action functional using any
of the tensor fields arising from the construction, $\mathbf{R}_{\;\;\;b}^{a},\mathbf{T}^{a},\mathbf{F}^{i},\mathbf{e}^{a},\eta_{ab},e_{ab\cdots c}$,
together with any representations of the original group.

In any dimension of spacetime, the action coupling the $SU\left(N\right)$
Yang-Mills field to general relativity is written as
\begin{eqnarray*}
S & = & S_{GR}+S_{\unit{YM}}\\
 & = & \intop\mathbf{R}^{ab}\wedge\mathbf{e}^{c}\wedge\ldots\wedge\mathbf{e}^{d}e_{abc\ldots d}-\frac{\kappa}{2}\intop\mathbf{F}^{i}\wedge^{*}\mathbf{F}_{i}
\end{eqnarray*}
where $^{*}\mathbf{F}_{i}$ is the Hodge dual of the 2-form $\mathbf{F}_{i}$.
We vary the action with respect to the solder form, $\mathbf{e}^{a}$,
the spin connection, $\boldsymbol{\omega}_{\;\;\;b}^{a}$, and the
Yang-Mills connection $\mathbf{A}^{i}$. Making the usual assumptions
for the gravity theory to reduce to general relativity, this results
in
\begin{eqnarray}
R_{ab}-\frac{1}{2}\eta_{ab}R & = & \kappa\left(\eta^{cd}F_{\;\;\;ac}^{i}F_{i\;bd}-\frac{1}{4}\eta_{ab}F^{i\,cd}F_{i\,cd}\right)\nonumber \\
\tilde{D}^{c}F_{\;\;\;ac}^{i} & = & 0\label{Standard YM equations}
\end{eqnarray}
where $\tilde{D}^{c}$ is covariant with respect to both local Lorentz
and local $SU\left(N\right)$ transformations. These methods generalize
immediately to additional internal symmetries, such as the $SU\left(3\right)\times SU\left(2\right)\times U\left(1\right)$
of the standard model.

Our main result is to show that Eqs.(\ref{Standard YM equations})
in $n$-dimensions follow from the field equations of biconformal
gravity coupled to a twisted Yang-Mills matter action, formulated
in $2n$-dimensions.

\subsection{Sources for scale-invariant gravity}

In addition to general covariance, scale-invariant gravity typically
requires tracelessness of the energy tensor. This condition arises
when local scale invariance is a symmetry of the action. Let $S_{g}$
be any locally scale invariant gravity action, with sources supplied
by adding a matter action,
\[
S_{matter}=\int\mathcal{L}_{matter}\sqrt{-g}d^{4}x
\]
Then the source for gravity is the energy-momentum tensor given by
the metric variation,
\begin{eqnarray*}
\delta_{g}S_{matter} & = & \int\frac{\delta\left(\mathcal{L}_{matter}\sqrt{-g}\right)}{\delta g^{\alpha\beta}}\delta g^{\alpha\beta}d^{4}x\\
 & = & \int T_{\alpha\beta}\delta g^{\alpha\beta}\sqrt{-g}d^{4}x
\end{eqnarray*}
If the energy-momentum tensor is scale \emph{invariant}, we may apply
Noether's theorem to the scaling symmetry transformation, $\delta g^{\alpha\beta}=-2\phi g^{\alpha\beta}$
immediately. This gives
\[
0\equiv\delta_{symmetry}S_{matter}=-2\int T_{\alpha\beta}\phi g^{\alpha\beta}\sqrt{-g}d^{4}x=-2\int T_{\;\;\;\alpha}^{\alpha}\phi\sqrt{-g}d^{4}x
\]
and since this must hold for any $\phi$, the energy tensor must be
traceless, $T_{\;\;\;\alpha}^{\alpha}=0$.

The tracelessness follows from the assumption that the only conformal
dependence resides in the single metric factor. However, it is more
often the case that the energy tensor $T_{\alpha\beta}$ is \emph{covariant}
under conformal transformations. For example, consider a simple massless
scalar field $\phi$ with action
\[
S_{scalar}=\frac{1}{2}\int g^{\mu\nu}\partial_{\mu}\phi\partial_{\nu}\phi\sqrt{-g}d^{4}x
\]
This action is scale invariant under the simultaneous transformations
\begin{eqnarray*}
\widetilde{g}_{\alpha\beta} & = & e^{2\alpha}g_{\alpha\beta}\\
\widetilde{\phi} & = & e^{-\alpha}\phi
\end{eqnarray*}
since then
\begin{eqnarray*}
\tilde{S}_{scalar} & = & \frac{1}{2}\int\tilde{g}^{\mu\nu}\partial_{\mu}\tilde{\phi}\partial_{\nu}\tilde{\phi}\sqrt{-\tilde{g}}d^{4}x\\
 & = & \frac{1}{2}\int e^{-2\alpha}g^{\mu\nu}\partial_{\mu}\left(e^{-\alpha}\phi\right)\partial_{\nu}\left(e^{-\alpha}\phi\right)e^{4\alpha}\sqrt{-g}d^{4}x\\
 & = & S_{scalar}
\end{eqnarray*}

The energy-momentum tensor for $\phi$ is given by metric variation
\begin{eqnarray*}
\delta_{g}S_{scalar} & = & \frac{1}{2}\int\left(\delta g^{\mu\nu}\partial_{\mu}\phi\partial_{\nu}\phi\sqrt{-g}-\frac{1}{2}\left(g^{\alpha\beta}\partial_{\alpha}\phi\partial_{\beta}\phi\right)g_{\mu\nu}\delta g^{\mu\nu}\sqrt{-g}\right)d^{4}x\\
 & = & \frac{1}{2}\int\delta g^{\mu\nu}\left(\partial_{\mu}\phi\partial_{\nu}\phi-\frac{1}{2}g_{\mu\nu}\left(g^{\alpha\beta}\partial_{\alpha}\phi\partial_{\beta}\phi\right)\right)\sqrt{-g}d^{4}x
\end{eqnarray*}
as
\[
T_{\mu\nu}=\partial_{\mu}\phi\partial_{\nu}\phi-\frac{1}{2}g_{\mu\nu}\left(g^{\alpha\beta}\partial_{\alpha}\phi\partial_{\beta}\phi\right)
\]
This is clearly not traceless even though the action is scale invariant.
Indeed, under conformal transformation $T_{\mu\nu}$ has conformal
weight $-2$,
\[
\tilde{T}_{\mu\nu}=e^{-2\alpha}T_{\mu\nu}
\]
and this covariance is all that is required to make the action invariant.

In cases such as this, an energy-momentum tensor with definite, nonzero
conformal weight may serve as the source for the curvature of a Weyl
geometry. By including the Weyl vector in these geometries, even the
Weyl-Ricci tensor takes on definite conformal weight:
\begin{eqnarray*}
\tilde{C}_{\;\;\;bcd}^{a} & = & e^{-2\alpha}C_{\;\;\;bcd}^{a}\\
\tilde{\mathcal{R}}_{ab} & = & e^{-2\alpha}\left(\mathcal{R}_{ab}+W_{\left(a;b\right)}-W_{a}W_{b}+\frac{1}{2}W^{2}\eta_{ab}\right)\\
\tilde{\mathcal{R}} & = & e^{-2\alpha}\left(\tilde{\mathcal{R}}+W_{\;\;\;;a}^{a}+W^{2}\right)
\end{eqnarray*}
where $\mathcal{R}_{ab}$ is the Schouten tensor and $W_{a}$ the
Weyl vector. A scale-invariant action functional in a Weyl geometry
differs from the Einstein-Hilbert action. A full discussion of such
theories is given in \cite{WeylGeom}.

\subsection{Sources for Weyl gravity}

We present two possibilities for the inclusion of sources in Weyl
gravity.

Weyl, or conformal, gravity begins with the free action,
\begin{eqnarray*}
S_{Weyl} & = & \intop C^{\alpha\beta\mu\nu}C_{\alpha\beta\mu\nu}\sqrt{-g}d^{4}x
\end{eqnarray*}
which is quadratic in the conformal curvature, with metric variation
leading to the Bach equation \cite{Bach}, 
\begin{equation}
W^{\mu\nu}\;\;=\;\;D_{\alpha}D_{\beta}C^{\mu\alpha\nu\beta}-\frac{1}{2}C^{\mu\alpha\nu\beta}R_{\alpha\beta}=0\label{Bach equation}
\end{equation}
The Bach tensor $W^{\mu\nu}$ is automatically traceless by the tracelessness
of the conformal curvature, $g_{\mu\nu}C^{\mu\alpha\nu\beta}=0$.
Because Eq.(\ref{Bach equation}) arises by metric variation it is
necessarily symmetric. Since the metric variation has the form
\[
\delta_{g}S_{Weyl}=\intop W^{\mu\nu}\delta g_{\mu\nu}\sqrt{-g}d^{4}x
\]
the change of the metric, $\delta g_{\mu\nu}=h_{\mu;\nu}+h_{\nu;\mu}$
under a change of coordinates implies
\begin{eqnarray*}
0 & = & \intop W^{\mu\nu}\left(h_{\mu;\nu}+h_{\nu;\mu}\right)\sqrt{-g}d^{4}x\\
 & = & -2\intop W_{\;\;\quad;\nu}^{\mu\nu}h_{\mu}\sqrt{-g}d^{4}x
\end{eqnarray*}
and since $h_{\mu}$ is arbitrary, the Bach tensor is divergence free,
\begin{eqnarray*}
W_{\;\;\quad;\nu}^{\mu\nu} & = & 0
\end{eqnarray*}

By rewriting the conformal curvature $C^{\alpha\beta\mu\nu}$ in terms
of the Riemann tensor minus its traces, and using the Gauss-Bonnet
density for the Euler character to remove the Riemann squared term,
$S_{Weyl}$ becomes proportional to $S'=\int\left(R_{\alpha\beta}R^{\alpha\beta}-\frac{1}{3}R^{2}\right)\sqrt{-g}d^{4}x$.
Variation of the metric now leads to a field equation written entirely
in terms of the Ricci tensor,
\begin{eqnarray}
0 & = & R_{\quad\;\;\;;\beta}^{\mu\nu;\beta}\text{\textminus}R_{\quad\;\;\;;\beta}^{\mu\beta;\nu}-R_{\quad\;\;\;;\beta}^{\nu\beta;\mu}\text{\textminus}2R^{\mu\beta}R_{\;\;\;\beta}^{\nu}\text{\textminus}\frac{1}{6}g^{\mu\nu}R{}_{\;\;\;;\beta}^{;\beta}\nonumber \\
 &  & +\frac{1}{2}g^{\mu\nu}R^{\alpha\beta}R_{\alpha\beta}+\frac{2}{3}R{}^{;\mu\nu}+2RR^{\mu\nu}\text{\textminus}\frac{1}{6}g^{\mu\nu}R{}^{2}\label{Bach in terms of Ricci}
\end{eqnarray}
Clearly, vacuum solutions to the Einstein equation are also solutions
to Eq.(\ref{Bach in terms of Ricci}). Thus, it is fair to argue that
evidence for vacuum general relativity is also evidence for vacuum
Weyl gravity.

It is well known that the converse is false. There are explicit solutions
to Eq.(\ref{Bach in terms of Ricci}) that are not solutions to vacuum
general relativity.

We cannot make a similar claim that solutions to general relativity
with sources gives evidence for Weyl gravity with sources. As we show
below, there are alternative candidates for sources for the Bach equation.
Without a clear specification of the source, the question of the relationship
between solutions in general relativity and solutions in Weyl gravity
is ill-posed.

Nonetheless, many argue for or assume \cite{MannheimGeneral,Mannheim}
direct addition of a matter action with traceless energy tensor,
\begin{equation}
S=\intop C^{\alpha\beta\mu\nu}C_{\alpha\beta\mu\nu}\sqrt{-g}d^{4}x+S_{matter}\label{Weyl grav with additive matter}
\end{equation}
The resulting metric variation leads to
\begin{equation}
W^{\mu\nu}=\kappa T^{\mu\nu}\label{Weyl gravity with T source}
\end{equation}
which is consistent with the properties of the Bach tensor as long
as the energy-momentum tensor is traceless.

However, the resulting solutions may differ substantially from the
verified predictions of general relativity. Specifically, Flanagan
has argued that the fourth order theory fails to agree with solar
system values, and Yoon has criticized the Newtonian limit and the
application to galactic rotation curves {[}for the critique and response
see \cite{Flanagan,Yoon,MannheimtoYoon}{]}. On the other hand, Palatini-style
variation has been shown for the vacuum case to reduce to scale covariant
general relativity \cite{Weyl grav as GR}, and a novel coupling to
matter might give better agreement with known results.

Because of the fourth order character of Eq.(\ref{Weyl gravity with T source}),
the coupling constant must have units that differ from the Einstein
coupling by some inverse length scale squared, $L^{-2}$. This is
expressible as a (very small) mass, $m\sim\frac{\hbar}{Lc}$. Therefore,
the nonvacuum solutions to Eq.(\ref{Weyl gravity with T source})
and solutions to general relativity with the same energy-momentum
tensor will differ by terms proportional to $m^{2}$. For sufficiently
close agreement with general relativity, this mass must be extremely
small. The length scale $L$ might represent a radius of inversion
symmetry. Since inversion is a conformal symmetry, this is a possible
interpretation. Consistency would require this length scale to be
cosmologically large, with the mass $m$ correspondingly small. The
mass-dependent corrections might then be consistent with experiment.

As a second possible coupling of matter to Weyl gravity, we note that
the introduction of a scale between general relativity and Weyl gravity
is eliminated if the source field equations are also modified to fourth
order equations. For example, we may modify the Klein-Gordon action
to
\begin{eqnarray*}
S_{MKG} & = & \frac{1}{2}\int\left(D_{\alpha}D_{\mu}\phi D^{\alpha}D^{\mu}\phi+m^{2}D_{\mu}\phi D^{\mu}\phi\right)\sqrt{-g}d^{4}x
\end{eqnarray*}
so that the field equation becomes
\[
D_{\mu}\square D^{\mu}\phi-m^{2}\square\phi=0
\]
and commuting $\square D^{\mu}\phi=D^{\mu}\square\phi+R^{\mu\alpha}\phi_{;\alpha}$
this becomes
\[
\square\left(\square\phi-m^{2}\phi\right)+\frac{1}{2}R^{;\alpha}\phi_{;\alpha}+R^{\mu\alpha}\phi_{;\alpha\mu}=0
\]
Because we have modified the mass term as well as the original kinetic
term, solutions to the original Klein-Gordon equation satisfy this
modified field equation up to terms proportional to the curvature.
In regions of weak gravity, at least some solutions will be close
to the corresponding solution fields in general relativity. Notice
that if we identify $B_{\alpha}=D_{\alpha}\phi$, the action $S_{MKG}$
describes a restriction of a Proca field.

Varying the metric, the energy tensor for this modified Klein-Gordon
field is found to be
\[
\mathcal{T}_{\alpha\beta}=g^{\mu\nu}D_{\alpha}D_{\mu}\phi D_{\beta}D_{\nu}\phi+\frac{1}{2}m^{2}D_{\alpha}\phi D_{\beta}\phi-\frac{1}{4}\left(D_{\nu}D_{\mu}\phi D^{\nu}D^{\mu}\phi+m^{2}D_{\mu}\phi D^{\mu}\phi\right)g_{\alpha\beta}
\]
This is more transparent if we let $B_{\alpha}=D_{\alpha}\phi$,
\[
\mathcal{T}_{\alpha\beta}=D_{\alpha}B^{\mu}D_{\beta}B_{\mu}+\frac{1}{2}m^{2}B_{\alpha}B_{\beta}-\frac{1}{4}g_{\alpha\beta}\left(D_{\nu}B_{\mu}D^{\nu}B^{\mu}+m^{2}B_{\mu}B^{\mu}\right)
\]
Notice that $\mathcal{T}_{\alpha\beta}$ is traceless if we set $m^{2}=0$.
To provide a conformal source, we therefore consider a massless modified
Klein-Gordon field. The resulting conserved gravitational equation
becomes
\begin{eqnarray*}
W_{\alpha\beta} & = & \kappa\mathcal{T}_{\alpha\beta}\\
 & = & \kappa\left(g^{\mu\nu}D_{\alpha}D_{\mu}\phi D_{\beta}D_{\nu}\phi-\frac{1}{4}g_{\alpha\beta}\left(D_{\nu}D_{\mu}\phi D^{\nu}D^{\mu}\phi\right)\right)
\end{eqnarray*}

Additional curvature terms will occur for higher spin fields, but
the basic conclusion remains the same. Solutions to the usual field
equation continue to solve the field equation, up to terms proportional
to the curvature. As a second example, the action for the general
fourth order modification of the Maxwell field may be written as
\[
S=\frac{1}{4}\int D_{\alpha}F_{\mu\nu}D^{\alpha}F^{\mu\nu}\sqrt{-g}d^{4}x
\]
where $F_{\left[\alpha\beta;\mu\right]}=0$ guarantees the existence
of a potential and makes this form unique. This results in the higher
order field equation
\[
D^{\nu}\square F_{\mu\nu}=0
\]
which we may once again commute to express in terms of the usual lower
order expression $D_{\nu}F^{\alpha\nu}$ plus curvature terms:
\[
\square D_{\nu}F^{\alpha\nu}+terms\:linear\:in\:FR=0
\]
As we found for the scalar field, symmetry and vanishing divergence
of the energy tensor,
\[
\mathcal{T}_{\rho\sigma}=D_{\rho}F^{\alpha\beta}D_{\sigma}F_{\alpha\beta}+2D^{\beta}F_{\rho}^{\;\;\;\mu}D_{\beta}F_{\sigma\mu}-\frac{1}{2}\left(D_{\alpha}F_{\mu\nu}D^{\alpha}F^{\mu\nu}\right)g_{\rho\sigma}
\]
are guaranteed by the metric variation and diffeomorphism invariance.

Tracelessness of $\mathcal{T}_{\alpha\beta}$, however, requires an
additional constraint on the derivatives of the field, $\mathcal{T}_{\;\;\;\alpha}^{\alpha}=D^{\sigma}F^{\alpha\beta}D_{\sigma}F_{\alpha\beta}$.
Using the field equation, we may express this as a vanishing divergence,
\begin{eqnarray*}
\mathcal{T}_{\;\;\;\alpha}^{\alpha} & = & D_{\sigma}\left(F_{\alpha\beta}D^{\sigma}F^{\alpha\beta}-2\square F^{\alpha\sigma}A_{\alpha}\right)\overset{!}{=}0
\end{eqnarray*}
indicating the presence of an additional conserved current.

Other field equations might be modified in a similar way. Of course,
though these modifications do not necessarily introduce a new mass
scale, they still require additional initial conditions.

\subsection{Sources for biconformal gravity, Kähler gravity, and double field
theory}

The situation may not be as complicated for biconformal gravity \cite{Ivanov,NCG,WW,WheelerAugust},
double field theory \cite{Duff,Siegel1,Siegel2,Siegel3,Brandenburger,Pedagogical double field theory},
or gravity on a Kähler manifold \cite{Hazboun Wheeler,WheelerAugust}.
Each of these cases starts as a fully $2n$-dimensional theory but
ultimately describes gravity on an $n$-dimensional submanifold. It
is desirable to have a fully $2n$-dimensional form of the matter
action which nonetheless also reduces to the expected $n$-dimensional
source as a consequence of the field equations. It is this condition
we address. We discuss the issue in biconformal space, since biconformal
gravity is already a gauge theory and it generically includes the
structures of both double field theory and Kähler manifolds \cite{Hazboun Wheeler,WheelerAugust}.

For matter fields we restrict our attention to Yang-Mills type sources.
We find that although the usual form of $2n$-dimensional Yang-Mills
action gives nonstandard coupling to gravity, including a ``twist''
matrix in the action corrects the problem.

Biconformal gravity arises as follows. The quotient of the conformal
group $\mathcal{C}_{p,q}=SO\left(p+1,q+1\right)$ of an $SO\left(p,q\right)$-symmetric
space $(p+q=n$, metric $\eta_{ab})$ by its homogeneous Weyl subgroup,
$\mathcal{W}_{p,q}\equiv SO\left(p,q\right)\times SO\left(1,1\right)$,
leads to a $2n$-dimensional homogeneous space with local $\mathcal{W}_{p,q}$
symmetry. This homogeneous space, discussed in \cite{Ivanov} and
\cite{NCG} and studied extensively in \cite{WheelerAugust,Hazboun Wheeler,Hazboun dissertation},
is found to have compatible symplectic, metric and complex structures,
making it Kähler \cite{Hazboun Wheeler}. In addition, the restriction
of the Killing form to the base manifold is nondegenerate and scale
invariant, and the volume form of the base manifold is scale invariant.
The homogeneous space and its curved generalizations are called biconformal
spaces.

Ivanov and Niederle \cite{Ivanov}, wrote a gravity theory on an 8-dimensional
biconformal space, using the curvature-quadratic action of Weyl gravity.
By a suitable restriction of the coordinate transformations of the
extra $4$-dimensions, they showed that $4$-dimensional general relativity
describes the remaining subspace.\emph{ }Subsequently, Wehner and
Wheeler \cite{WW} introduced a class of $\mathcal{W}$-invariant
actions \emph{linear} in the curvatures, defining biconformal gravity.
Curvature-linear actions are possible because the $2n$-dimensional
volume element is scale invariant. Unlike the 4-dimensional theories
above with actions quadratic in the curvature, the linear action functionals
take the same form in any dimension. The doubled dimension is understood
in terms of the symplectic structure, leading to a phase space interpretation
for generic solutions. Lagrangian submanifolds represent the physical
spacetime and have the original dimension. The class of torsion-free
biconformal spaces has been shown to reduce to general relativity
on the cotangent bundle of spacetime \cite{WheelerAugust}. These
reductions of the model work for any signature $\left(p,q\right)$.

The most general action linear in the biconformal curvatures is given
by
\begin{equation}
S=\lambda\int e_{ac\cdots d}^{\quad\quad be\cdots f}\left(\alpha\boldsymbol{\Omega}_{\;\;b}^{a}+\beta\delta_{\;b}^{a}\boldsymbol{\Omega}+\gamma\mathbf{e}^{a}\wedge\mathbf{f}_{b}\right)\land\mathbf{e}^{c}\land\cdots\land\mathbf{e}^{d}\land\mathbf{f}_{e}\land\cdots\land\mathbf{f}_{f}\label{Action}
\end{equation}
where $\boldsymbol{\Omega}_{\;\;b}^{a}$ is the curvature of the spin
connection and $\boldsymbol{\Omega}$ is the dilatational curvature.
Here $\lambda=\frac{\left(-1\right)^{n}}{\left(n-1\right)!\left(n-1\right)!}$
is a convenient constant, chosen to eliminate a combinatoric factor
and to make our sign conventions agree with \cite{WheelerAugust}.
The cotangent bundle is spanned by the pair, $\left(\mathbf{e}^{a},\mathbf{f}_{b}\right)$,
called the solder form and the co-solder form, respectively. The variation
is taken with respect to all $\frac{\left(n+1\right)\left(n+2\right)}{2}$
gauge fields.

The reduction of a fully $2n$-dimensional gravity theory to dependence
only on the fields of $n$-dimensional gravity is a remarkable feature
of biconformal gravity. While it has been shown to be a double field
theory \cite{Duff,Siegel1,Siegel2,Siegel3,Berman Blair and Perry,Pedagogical double field theory,WheelerAugust},
double field theories require the assumption that fields depend on
only half the coordinates. This artificial constraint is called a
section condition. In sharp contrast, biconformal solutions do \emph{not}
require a section condition, \emph{reducing as a consequence of the
field equations} of torsion-free biconformal spaces. Thus, using the
torsion-free field equations, the components of the $\frac{\left(n+1\right)\left(n+2\right)}{2}$
curvatures--initially dependent on $2n$ independent coordinates--reduce
to the usual Riemannian curvature tensor in $n$ dimensions. Correspondingly,
the $n$-dim solder form determines all fields, up to coordinate and
gauge transformations. Generic, torsion-free, vacuum solutions describe
$n$-dimensional scale-covariant general relativity on the co-tangent
bundle.

As noted in the introduction, with the exceptions biconformal supergravity
\cite{AW} and a scalar field example \cite{BCSMatterWW}, studies
of biconformal spaces \cite{NCG,WW,WheelerAugust,Spencer Wheeler,Hazboun Wheeler,Hazboun dissertation,AWQM,Lovelady,Hazboun}
have considered pure gravity biconformal spaces, leading to vacuum
general relativity.

Here we consider $SU\left(N\right)$ Yang-Mills fields as gravitational
sources. The central issue is to show that even with a completely
general $SU\left(N\right)$ gauge theory over a $2n$-dimensional
biconformal space, a full $2n\rightarrow n$ reduction occurs, both
for gravity and the Yang-Mills field.

As with the Riemann-Cartan construction of general relativity above,
the development of biconformal spaces from group symmetry makes it
straightforward to include the additional symmetry of sources. By
extending the quotient to
\begin{eqnarray*}
\mathcal{M}^{2n} & = & \mathcal{C}_{p,q}\times SU\left(N\right)/\mathcal{W}_{p,q}\times SU\left(N\right)
\end{eqnarray*}
the local symmetry is enlarged by $SU\left(N\right)$ and we may add
a Yang-Mills or similar action to Eq.(\ref{Action}). This construction
gives the form of the Yang-Mills field in terms of the potentials,
but not the form of the action.

There are then two basic parts to our investigation.

First, we must determine a suitable $2n$-dimensional action functional
for the sources. To accomplish this requires two interdependent specifications:
\begin{itemize}
\item Find a form of the Yang-Mills action which gives the usual $n$-dimensional
Yang-Mills source to the Einstein tensor.
\item Fix the orthornormality condition for the solder and co-solder form,
$M^{AB}=\left\langle \tilde{\mathbf{e}}^{A},\tilde{\mathbf{e}}^{B}\right\rangle $
where $\tilde{\mathbf{e}}^{A}=\left(\mathbf{e}^{a},\mathbf{f}_{b}\right)$.
This determines for form of the Hodge dual and ultimately the metric
variation of the matter action.
\end{itemize}
In accomplishing these steps we show that the standard Yang-Mills
action
\[
S_{YM}^{0}=\int tr\left(\mathbf{F}\land^{*}\mathbf{F}\right)
\]
cannot give the right couplings, and find a satisfactory modification
\[
S_{YM}=\int tr\left(\bar{\mathbf{F}}\land^{*}\mathbf{F}\right)
\]
where the twisted field, $\bar{\mathbf{F}}$ is defined below. The
twisted action, together with the restricted Killing form as orthonormal
metric, give the desired reduction.

The second challenge is then to use the field equations to show:
\begin{itemize}
\item The number of field components in $2n$ dimensions reduces to the
expected number on $n$ dimensional spacetime.
\item The functional dependence of the fields reduces from $2n$ to $n$
independent variables.
\item The gravitational source is the usual Yang-Mills stress-energy tensor.
\item The $SU\left(N\right)$ field equation is the usual $n$-dim Yang-Mills
field equation.
\end{itemize}
\medskip{}
The remainder of our presentation proceeds as follows. In the next
Section, we introduce our notation and other conventions. In Section
3, we show that the usual form of Yang-Mills action, $S_{YM}^{0}$,
cannot produce the usual coupling to gravity. The twisted form $\bar{\mathbf{F}}$
is developed in Section 4, and the variation of $S_{YM}$ carried
out. The Yang-Mills potentials are identified and varied in Section
5. The next Section contains the reduction of the gravitational field
equations, as far as possible with the presence of sources. This reduction
closely follows reference \cite{WheelerAugust}. We find that the
field equations imply certain restrictions that must be applied to
the matter sources.

The reduction of the number of fields and the number of independent
variables is shown in Section 7 to follow from the full gravitational
equations. We find that the reduction of fields that is necessary
in the purely gravitational sector also forces reduction of the source
fields. The Section concludes with the emergence of both the usual
Yang-Mills gravitational source in $n$-dimensions, and the usual
$n$-dimensional Yang-Mills equation. The final Section includes a
brief review of the main results.

\section{Notation and conventions}

\subsection{Conventions with biconformal tensors}

\subsubsection{Differential forms}

The co-tangent space of biconformal manifolds are spanned by two sets
of opposite conformal weight orthonormal frame fields, $\tilde{\mathbf{e}}^{A}=\left(\mathbf{e}^{a},\mathbf{f}_{b}\right)$,
with lowercase Latin indices $a,b,\ldots=1,2,\ldots,n$ indicating
the use of these frames and upper case Latin $A,B,\ldots=1,2,\ldots,2n$
to denote the pair. Coordinate indices are lower case Greek, $\mu,\nu,\ldots=1,2,\ldots,n$
so that we have, for example,
\[
\mathbf{e}^{a}=e_{\mu}^{\;\;\;a}\mathbf{d}x^{\mu}+e^{\mu a}\mathbf{d}y_{\mu}
\]
A general 2-form may be written in the orthonormal basis as
\begin{eqnarray*}
\boldsymbol{\mathcal{F}} & = & \frac{1}{2}F_{ab}\mathbf{e}^{a}\wedge\mathbf{e}^{b}+F_{\;\;\;b}^{a}\mathbf{f}_{a}\wedge\mathbf{e}^{b}+\frac{1}{2}F^{ab}\mathbf{f}_{a}\wedge\mathbf{f}_{b}
\end{eqnarray*}
It is important to realize that $F_{ab},F_{\;\;\;b}^{a}$ and $F^{ab}$
are distinct fields. Therefore, we cannot raise and lower indices
in the usual way unless we choose different names for the separate
independent components. As compensation for this, the raised or lowered
position of an index reflects its conformal weight. Thus, $F_{ab}$
has weight $-2$ while $F^{ab}$ has weight $+2$. When practical
these distinct fields will be given different names,
\begin{eqnarray*}
\boldsymbol{\mathcal{F}} & = & \frac{1}{2}F_{ab}\mathbf{e}^{a}\wedge\mathbf{e}^{b}+G_{\;\;\;b}^{a}\mathbf{f}_{a}\wedge\mathbf{e}^{b}+H^{ab}\mathbf{f}_{a}\wedge\mathbf{f}_{b}
\end{eqnarray*}
but this can lead to an unnecessary profusion of field names.

When we need to explicitly refer to internal $SU\left(N\right)$ indices,
the field and its components will be given an additional index from
the lower case Latin set $\left\{ i,j,k\right\} $. Other lower case
Latin indices refer to the null orthonormal frame field, $\left(\mathbf{e}^{b},\mathbf{f}_{a}\right)$.
Thus $F_{\;\;\;ab}^{i}$ represents the components of $\frac{1}{2}F_{\;\;\;ab}^{i}G_{i}\mathbf{e}^{a}\wedge\mathbf{e}^{b}$
where $G_{i}$ is a generator of $SU\left(N\right)$ as $i$ runs
from 1 to $N^{2}-1$. In most cases the internal index is suppressed.

Differential forms are written in boldface and always multiplied with
the wedge product. For brevity in some longer expressions we omit
the explicit wedge between forms. Thus, for example, 
\[
\mathbf{f}_{a}\wedge\mathbf{f}_{b}\wedge\mathbf{f}_{c}\wedge\mathbf{e}^{d}\wedge\mathbf{e}^{e}\Longleftrightarrow\mathbf{f}_{abc}\mathbf{e}^{de}
\]
The bold font shows that these are differential forms, and therefore
are to be wedged together.

As a compromise between keeping track of conformal weights, while
being able to assess the symmetry or antisymmetry of components, we
introduce a weight +1 basis $\mathbf{e}^{A}\equiv\left(\mathbf{e}^{a},\eta^{ab}\mathbf{f}_{b}\right)$,
where $\eta_{ab}$ is the $n$-dimensional metric of the original
$SO\left(p,q\right)$-symmetric space, \emph{not} the metric of the
biconformal space. Thus, we may write
\begin{eqnarray*}
\boldsymbol{\mathcal{F}} & = & \frac{1}{2}F_{AB}\mathbf{e}^{A}\wedge\mathbf{e}^{B}\\
 & = & \frac{1}{2}F_{ab}\mathbf{e}^{a}\wedge\mathbf{e}^{b}+\mathcal{G}_{ab}\eta^{ac}\mathbf{f}_{c}\wedge\mathbf{e}^{b}+\frac{1}{2}\mathcal{H}_{ab}\eta^{ac}\eta^{bd}\mathbf{f}_{c}\wedge\mathbf{f}_{d}
\end{eqnarray*}
where we have defined
\begin{eqnarray*}
\mathcal{G}_{ab} & \equiv & \eta_{ac}G_{\;\;\;b}^{c}\\
\mathcal{H}_{ab} & \equiv & \eta_{ac}\eta_{bd}H^{cd}
\end{eqnarray*}
The use of a different font is important because we are not using
the biconformal metric, $K_{AB}$, to change index positions. It is
the original index positions and font, $G_{\quad b}^{a}$, $H^{ab}$,
that are the proper field components.

The matrix components $F_{AB}$ are then written as
\begin{eqnarray}
F_{AB} & = & \left(\begin{array}{cc}
F_{ab} & F_{a}^{\;\;\;b}\\
F_{\;\;\;b}^{a} & F^{ab}
\end{array}\right)\nonumber \\
 & = & \left(\begin{array}{cc}
F_{ab} & -\mathcal{G}_{ba}\\
\mathcal{G}_{ab} & \mathcal{H}_{ab}
\end{array}\right)\label{Matrix form}
\end{eqnarray}
where the form of the upper right corner follows because
\begin{eqnarray*}
-\mathcal{G}_{ba}\mathbf{e}^{a}\land\eta^{bc}\mathbf{f}_{c} & = & \mathcal{G}_{ba}\eta^{bc}\mathbf{f}_{c}\land\mathbf{e}^{a}\\
 & = & G_{\;\;\;a}^{c}\mathbf{f}_{c}\land\mathbf{e}^{a}
\end{eqnarray*}
With this convention, we can meaningfully define the transpose of
matrices. Specifically, while the transpose of
\[
\left(\begin{array}{cc}
F_{ab} & F_{a}^{\;\;\;b}\\
F_{\;\;\;b}^{a} & F^{ab}
\end{array}\right)
\]
is ill-defined because of the mixed indices on the off-diagonal terms,
the transpose of $F_{AB}$ in the weight +1 basis is
\begin{equation}
\left(\begin{array}{cc}
F_{ab} & -\mathcal{G}_{ba}\\
\mathcal{G}_{ab} & \mathcal{H}_{ab}
\end{array}\right)^{t}=\left(\begin{array}{cc}
F_{ba} & \mathcal{G}_{ba}\\
-\mathcal{G}_{ab} & \mathcal{H}_{ba}
\end{array}\right)\label{Transpose}
\end{equation}
For $\mathcal{H}_{ba}=-\mathcal{H}_{ab}$ and $F_{ba}=-F_{ab}$ this
is manifestly antisymmetric, as befits a 2-form. Notice that the effect
of two transposes is the identity, so this operation provides an involutive
automorphism even though $\eta_{ab}$ is not the biconformal metric.

From $\left[\mathcal{G}_{ab}\right]^{t}=\mathcal{G}_{ba}$ we have
\[
\left[\eta_{ac}G_{\;\;\;b}^{c}\right]^{t}=\left[\eta_{bc}G_{\;\;\;a}^{c}\right]
\]
and therefore, $\left[G_{\;\;\;b}^{d}\right]^{t}=\eta^{ad}\eta_{bc}G_{\;\;\;a}^{c}$.

\subsubsection{Metric}

Because all quadrants of the metric are used in the variation, it
is desirable to retain both the name and index positions throughout.
Since
\[
K_{AB}=\left(\begin{array}{cc}
K_{ab} & K_{a}^{\;\;\;b}\\
K_{\;\;\;b}^{a} & K^{ab}
\end{array}\right)
\]
contains all index positions, the inverse metric and its components
are written with an overbar,
\[
\bar{K}^{AB}=\left(\begin{array}{cc}
\bar{K}^{ab} & \bar{K}_{\;\;\;b}^{a}\\
\bar{K}_{a}^{\;\;\;b} & \bar{K}_{ab}
\end{array}\right)
\]
Thus $K_{ab}$ is the first quadrant of the metric, while $\bar{K}_{ab}$
is the final quadrant of the inverse metric. Here any changes of index
position must be indicated with an explicit factor of $\eta_{ab}$
or $\eta^{ab}$

\subsubsection{Volume form}

It is convenient to define a volume form as $\boldsymbol{\Phi}\equiv{}^{*}1$
but in defining the Hodge dual operation a number of ambiguities need
to be clarified. Because up and down indices have distinct conformal
weight, we may partially order the indices on the Levi-Civita tensor.
We establish the following conventions:
\begin{enumerate}
\item All factors of $\mathbf{f}_{a}$ are written first, followed by all
of the $\mathbf{e}^{b}$.
\item The Levi-Civita \emph{tensor} is written as $e_{\quad\quad de\cdots f}^{ab\cdots c}$,
with all $n$ up indices first. The partially ordered antisymmetric
\emph{symbol} is written as $\varepsilon_{\quad\quad de\cdots f}^{ab\cdots c}$.
\item When taking the dual of a $p$-form, we sum the components on the
first $p$ indices of the Levi-Civita tensor, then introduce factors
of $-1$ to move indices to the default positions. For example, the
dual of $\mathcal{\boldsymbol{H}}=H_{\;\;\;b}^{a}\mathbf{f}_{a}\wedge\mathbf{e}^{b}$
is
\begin{eqnarray*}
^{*}\mathcal{\boldsymbol{H}} & = & \frac{1}{\left(n-1\right)!\left(n-1\right)!}H_{\;\;\;b}^{a}e_{a\quad\quad\;\;\;e\cdots f}^{\;\;\;bc\cdots d}\mathbf{f}_{c}\wedge\cdots\wedge\mathbf{f}_{d}\wedge\mathbf{e}^{e}\wedge\cdots\wedge\mathbf{e}^{f}\\
 & = & \frac{\left(-1\right)^{n}}{\left(n-1\right)!\left(n-1\right)!}H_{\;\;\;b}^{a}e_{\quad\quad ae\cdots f}^{bc\cdots d}\mathbf{f}_{c}\wedge\cdots\wedge\mathbf{f}_{d}\wedge\mathbf{e}^{e}\wedge\cdots\wedge\mathbf{e}^{f}\\
 & = & \frac{\left(-1\right)^{n}}{\left(n-1\right)!\left(n-1\right)!}H_{\;\;\;b}^{a}e_{\quad\quad ae\cdots f}^{bc\cdots d}\mathbf{f}_{c\cdots d}\mathbf{e}^{e\cdots f}
\end{eqnarray*}
Notice that in the final step, the number of wedged forms in $\mathbf{f}_{c\cdots d}$
may be inferred from the Levi-Civita tensor. Since the Levi-Civita
tensor always has $n$ up and $n$ down indices, the number of basis
forms of each type is unambiguous. For example,
\[
e_{\qquad e\cdots f}^{abc\cdots d}\mathbf{f}_{c\cdots d}\wedge\mathbf{e}^{e\cdots f}=e_{\qquad e\cdots f}^{abc\cdots d}\underbrace{\mathbf{f}_{c}\wedge\ldots\wedge\mathbf{f}_{d}}_{n-2}\wedge\underbrace{\mathbf{e}^{e}\wedge\ldots\wedge\mathbf{e}^{f}}_{n}
\]
is a $2n-2$ form that includes the wedge product of $n-2$ factors
of $\mathbf{f}_{a}$ and $n$ factors of $\mathbf{e}^{a}$.
\item An $m$-form is a polynomial with each term having different numbers
of $\mathbf{e}$'s and $\mathbf{f}$'s, we write the terms in order
of increasing number of $\mathbf{f}$'s.
\item The partial ordering of indices on the Levi-Civita tensor reduces
the normalization of the volume element from $\frac{1}{\left(2n\right)!}$
to $\frac{1}{n!n!}$.
\end{enumerate}
It has been noted elsewhere \cite{AWBCYMGrav} that there are alternative
duals in biconformal space. For instance, we may use the symplectic
form instead of the metric to connect indices. The difference resides
in the relative signs between the $\mathbf{e}^{a}$, $\mathbf{f}_{a}$,
and mixed terms. Here we use only the Hodge dual, taking care to keep
the correct signs.

With these conventions in mind, we define
\begin{eqnarray}
\boldsymbol{\Phi} & \equiv & ^{*}1\nonumber \\
 & = & \frac{1}{n!n!}e_{\qquad e\cdots f}^{c\cdots d}\mathbf{f}_{c\cdots d}\wedge\mathbf{e}^{e\cdots f}\nonumber \\
 & = & \frac{1}{n!n!}\sqrt{K}\varepsilon_{\qquad e\cdots f}^{c\cdots d}\mathbf{f}_{c\cdots d}\wedge\mathbf{e}^{e\cdots f}\label{Volume form}
\end{eqnarray}
and consequently,
\begin{eqnarray}
\mathbf{f}_{c\cdots d}\wedge\mathbf{e}^{e\cdots f} & = & \frac{1}{\sqrt{K}}\varepsilon_{\qquad c\cdots d}^{e\cdots f}\boldsymbol{\Phi}\nonumber \\
 & = & \bar{e}_{c\cdots d}^{\qquad e\cdots f}\boldsymbol{\Phi}\label{Volume form from basis forms}
\end{eqnarray}
where the overbar denotes the contravariant form of the Levi-Civita
tensor. The contravariant form satisfies
\begin{eqnarray*}
e_{\qquad c\cdots d}^{a\cdots b}\bar{e}_{a\cdots b}^{\qquad c\cdots d} & = & n!n!
\end{eqnarray*}
We also need the reduction formulas,
\begin{eqnarray}
e_{\qquad mnc\cdots d}^{e\cdots f}\bar{e}_{\qquad e\cdots f}^{ghc\cdots d} & = & n!\left(n-2\right)!\left(\delta_{m}^{g}\delta_{n}^{h}-\delta_{m}^{h}\delta_{n}^{g}\right)\nonumber \\
e_{\qquad ne\cdots f}^{mc\cdots d}\bar{e}_{\qquad gc\cdots d}^{he\cdots f} & = & \left(n-1\right)!\left(n-1\right)!\delta_{n}^{h}\delta_{g}^{m}\label{Reduction formulas}
\end{eqnarray}

\subsubsection{Further notation}

The antisymmetric projection operator mapping for $\left(\begin{array}{c}
1\\
1
\end{array}\right)$ tensors,
\begin{eqnarray*}
\Delta_{db}^{ac} & \equiv & \frac{1}{2}\left(\delta_{d}^{a}\delta_{b}^{c}-\eta^{ac}\eta_{db}\right)\\
 & = & \frac{1}{2}\eta^{ce}\eta_{bf}\left(\delta_{d}^{a}\delta_{e}^{f}-\delta_{e}^{a}\delta_{d}^{f}\right)
\end{eqnarray*}
arises frequently.

\subsection{Conventions for invariant matrices}

Possible actions can be constructed using curvatures naturally arising
in a theory, together with any invariant tensors consistent with the
gauging. The biconformal gauging of the conformal group has a surprising
number of invariant objects. These invariant structures arise from
internal symmetries of the conformal group are induced into generic
biconformal spaces \cite{Hazboun Wheeler}.

The conformally invariant Killing form, restricted to the base manifold:
\begin{equation}
K_{AB}=\left(\begin{array}{cc}
0 & \delta_{a}^{\;\;\;b}\\
\delta_{\;\;\;b}^{a} & 0
\end{array}\right)\label{Killing form}
\end{equation}
The symplectic form, underlying dilatations,
\begin{equation}
\Omega_{AB}=\left(\begin{array}{cc}
0 & \delta_{a}^{\;\;\;b}\\
-\delta_{\;\;\;b}^{a} & 0
\end{array}\right)\label{Symplectic form}
\end{equation}
Interestingly, this form manifests Born reciprocity \cite{Born1938,Born1949}.
The complex structure, arising from the symmetry between translations
of the origin and translations of the point at infinity (i.e., special
conformal transformations),
\begin{eqnarray}
J_{\;\;\;B}^{A} & := & \left(\begin{array}{cc}
0 & -\eta^{ab}\\
\eta_{ab} & 0
\end{array}\right)\label{Almost complex structure}
\end{eqnarray}
The symmetric Kähler form, arising from the compatibility, $g\left(u,v\right)=\Omega\left(u,Jv\right)$,
of the symplectic and complex structures,
\begin{eqnarray}
g_{AB} & = & \Omega_{AC}J_{\;\;\;B}^{C}\nonumber \\
 & = & \left(\begin{array}{cc}
0 & \delta_{a}^{\;\;\;c}\\
-\delta_{\;\;\;c}^{a} & 0
\end{array}\right)\left(\begin{array}{cc}
0 & -\eta^{cb}\\
\eta_{cb} & 0
\end{array}\right)\nonumber \\
 & = & \left(\begin{array}{cc}
\eta_{ab} & 0\\
0 & \eta^{ab}
\end{array}\right)\label{K=0000E4hler metric}
\end{eqnarray}

These three Kähler structures satisfy
\begin{equation}
g\left(u,v\right)=\Omega\left(u,Jv\right)\label{Compatibility condition}
\end{equation}
Notice that the symmetric Kähler form is not invariant under the conformal
structure.

\subsection{Kähler or Killing?}

As noted in the introduction, vacuum biconformal gravity depends only
on the variation of the gauge fields and does not require introduction
of a metric while Yang-Mills actions make use of the Hodge dual and
a metric is required. Given the presence of two natural symmetric
forms in biconformal spaces, i.e., the scale invariant Killing form,
Eq.(\ref{Killing form}), and the Kähler metric, Eq.(\ref{K=0000E4hler metric}),
we must make a choice of how to specify the orthonormality relation
of the solder and co-solder forms. With $\tilde{\mathbf{e}}^{A}=\left(\mathbf{e}^{a},\mathbf{f}_{b}\right)$,
and using an overbar to denote the inverse, we may specify orthonormality
by either
\begin{equation}
\left\langle \tilde{\mathbf{e}}^{A},\tilde{\mathbf{e}}^{B}\right\rangle \equiv\bar{K}^{AB}\label{Killing inner product}
\end{equation}
or
\begin{equation}
\left\langle \tilde{\mathbf{e}}^{A},\tilde{\mathbf{e}}^{B}\right\rangle \equiv\bar{g}^{AB}\label{Kahler inner product}
\end{equation}
but clearly not both. Once we choose either Eq.(\ref{Killing inner product})
or Eq.(\ref{Kahler inner product}), the corresponding symmetric form
becomes the metric, with its variation following from the variation
of the solder and co-solder forms.

This choice between the symmetric Killing and Kähler forms is not
arbitrary. We therefore studied both cases, ultimately showing that
the correct matter couplings arise only if we use the \emph{Killing
form}, $K_{AB}$, to determine orthonormality of the basis forms.
By inverting the components of the basis forms, we then have the metric
in arbitrary coordinates (indicated by \emph{coordinate }indices $M,N$,
distinct from \emph{orthonormal} indices $A,B$)
\begin{equation}
K_{MN}=\tilde{e}_{M}^{\;\;\;\;A}\tilde{e}_{N}^{\;\;\;\;B}K_{AB}\label{Killing metric}
\end{equation}
This is our final choice of metric, and is the only choice which yields
the expected Yang-Mills source for general relativity.

The gravitational field equations follow by variation of the connection
forms, including the solder and co-solder forms $\tilde{e}_{M}^{\;\;\;\;A}$.
The variation of the metric then follows from Eq.(\ref{Killing metric}).
Until the variation is complete, we need the general form of the inverse
metric,
\[
\bar{K}^{AB}=\left(\begin{array}{cc}
\bar{K}^{ab} & \bar{K}_{\;\;\;b}^{a}\\
\bar{K}_{a}^{\;\;\;b} & \bar{K}_{ab}
\end{array}\right)
\]
Once the variations are expressed in terms of $\delta\bar{K}^{AB}$
or $\delta\bar{g}^{AB}$ any remaining components may be returned
to the orthonormal form, Eq.(\ref{Killing form}) or Eq.(\ref{K=0000E4hler metric}).

Let the solder and co-solder variations be given by
\begin{eqnarray}
\delta\mathbf{e}^{a} & = & A_{\;\;\;c}^{a}\mathbf{e}^{c}+B^{ac}\mathbf{f}_{c}\label{Solder form variation}\\
\delta\mathbf{f}_{c} & = & C_{cd}\mathbf{e}^{d}+D_{c}^{\;\;\;d}\mathbf{f}_{d}\label{Co-solder form variation}
\end{eqnarray}
Then for variation of the components of the Killing metric we expand
Eq.(\ref{Killing inner product})
\begin{eqnarray*}
\delta\bar{K}^{ab} & = & \delta\left\langle \mathbf{e}^{a},\mathbf{e}^{b}\right\rangle \\
 & = & \left\langle A_{\;\;\;c}^{a}\mathbf{e}^{c}+B^{ac}\mathbf{f}_{c},\mathbf{e}^{b}\right\rangle +\left\langle \mathbf{e}^{a},A_{\;\;\;c}^{b}\mathbf{e}^{c}+B^{bc}\mathbf{f}_{c}\right\rangle \\
 & = & A_{\;\;\;c}^{a}\bar{K}^{cb}+B^{ac}\bar{K}_{c}^{\;\;\;b}+A_{\;\;\;c}^{b}\bar{K}^{ac}+B^{bc}\bar{K}_{\;\;\;c}^{a}
\end{eqnarray*}
Since the coefficients $A_{\;\;\;c}^{a},B^{ac},C_{cd},D_{c}^{\;\;\;d}$
now represent the variation, we may return the remaining components
of $\bar{K}^{AB}$ to the null orthonormal form of Eq.(\ref{Killing form}),
\[
\delta\bar{K}^{ab}=B^{ab}+B^{ba}
\]
Computing the remaining components in the same way, we arrive at the
full set,
\begin{eqnarray}
\delta\bar{K}^{ab} & = & B^{ab}+B^{ba}\nonumber \\
\delta\bar{K}_{\;\;\;b}^{a} & = & A_{\;\;\;b}^{a}+D_{b}^{\;\;\;a}\nonumber \\
\delta\bar{K}_{a}^{\;\;\;b} & = & D_{a}^{\;\;\;b}+A_{\;\;\;a}^{b}\nonumber \\
\delta\bar{K}_{ab} & = & C_{ab}+C_{ba}\label{Variation of the Killing form}
\end{eqnarray}

We also considered the analogous calculation if we were to choose
the Kähler inner product, Eq.(\ref{Kahler inner product}). This variation
gives
\begin{eqnarray}
\delta\bar{g}^{ab} & = & A_{\;\;\;c}^{a}\eta^{cb}+A_{\;\;\;c}^{b}\eta^{ac}\nonumber \\
\delta\bar{g}_{\;\;\;b}^{a} & = & B^{ac}\eta_{bc}+C_{bc}\eta^{ac}\nonumber \\
\delta\bar{g}_{a}^{\;\;\;b} & = & C_{ac}\eta^{cb}+B^{bc}\eta_{ac}\nonumber \\
\delta\bar{g}_{ab} & = & D_{a}^{\;\;\;c}\eta_{cb}+D_{b}^{\;\;\;c}\eta_{ac}\label{Variation of the Kahler metric}
\end{eqnarray}

As we indicate when we perform the gravitational variation, Eq.(\ref{Gravitational variation})
in Section 6 below, it is the $B^{ab}$ part of the variation which
ultimately reduces to the Einstein equation. From Eqs.(\ref{Variation of the Killing form})
and (\ref{Variation of the Kahler metric}) above we see that the
different possible choices for the metric lead to completely different
sources for gravity. Using the Killing form to define orthonormality
of the basis, it is the coefficient of the $\delta\bar{K}^{ab}$ variation
that provides the gravitational source, while the opposite choice
of the Kähler form couples the coefficients of the cross-terms $\delta\bar{g}_{\;\;\;b}^{a}$
and $\delta\bar{g}_{a}^{\;\;\;b}$ to gravity. Only one of these can
give the usual Yang-Mills energy tensor. To determine which, we checked
each of the two inner products, for each proposed action functional
below until it became clear that \emph{we must use the Killing form}.

Rather than presenting these distinct variations here, we continue
with the Killing metric and its variation, as given by Eqs.(\ref{Killing form}),
(\ref{Killing inner product}), and (\ref{Variation of the Killing form}).
Details of the failure of the Kähler choice are given in Appendix
A.

\section{Constructing the Yang-Mills action}

In spacetime, the action for a Yang-Mills field may be written as
\begin{eqnarray}
S_{\unit{YM}} & = & -\frac{\kappa}{2}\intop tr\boldsymbol{\mathcal{F}}\wedge^{*}\boldsymbol{\mathcal{F}}\label{Yang-Mills action}
\end{eqnarray}
and it is natural to consider the same form in biconformal space.
However, as we show in this Section, this usual form leads to nonstandard
coupling to gravity. We show in the next Section that a twisted action
is required to give the usual coupling.

Expanded into independent components in the $\tilde{\mathbf{e}}^{A}$
basis,

\begin{eqnarray*}
\boldsymbol{\mathcal{F}}^{i} & = & \frac{1}{2}\mathcal{F}_{\quad ab}^{i}\mathbf{e}^{a}\wedge\mathbf{e}^{b}+\mathcal{F}_{\quad b}^{ia}\mathbf{f}_{a}\wedge\mathbf{e}^{b}+\frac{1}{2}\mathcal{F}^{iab}\mathbf{f}_{a}\wedge\mathbf{f}_{b}
\end{eqnarray*}
where $i$ is an index of the internal Lie algebra. This index can
be suppressed without loss of generality in the action and gravitational
field equations. Also, for the Yang-Mills field it proves more transparent
to give the three coefficients distinct names. Finally, since we must
insure antisymmetry of the twisted field, it is most transparent to
use the uniform weight basis, $\mathbf{e}^{A}=\left(\mathbf{e}^{a},\eta^{bc}\mathbf{f}_{c}\right)=\left(\mathbf{e}^{a},\mathbf{f}^{b}\right)$.
We therefore write
\begin{eqnarray}
\boldsymbol{\mathcal{F}}\;\;=\;\;\frac{1}{2}\mathcal{F}_{AB}\mathbf{e}^{A}\wedge\mathbf{e}^{b} & = & \frac{1}{2}F_{ab}\mathbf{e}^{a}\wedge\mathbf{e}^{b}+\mathcal{G}_{ab}\mathbf{f}^{a}\wedge\mathbf{e}^{b}+\frac{1}{2}\mathcal{H}_{ab}\mathbf{f}^{a}\wedge\mathbf{f}^{b}\label{YM field in raised basis}
\end{eqnarray}
In discussions of the Yang-Mills field equations (as opposed to the
gravitational equations), the internal index becomes important and
will be shown where necessary.

Note that $F_{ab}$ and $\mathcal{H}_{ab}$ are antisymmetric. The
cross-term may be written as
\begin{eqnarray*}
G_{\quad b}^{a}\mathbf{f}_{a}\wedge\mathbf{e}^{b} & = & \frac{1}{2}\left(G_{\quad b}^{a}\right)\mathbf{f}_{a}\wedge\mathbf{e}^{b}+\frac{1}{2}\left(-G_{\quad a}^{b}\right)\mathbf{e}^{a}\wedge\mathbf{f}_{b}\\
 & = & \mathcal{G}_{ab}\mathbf{f}^{a}\wedge\mathbf{e}^{b}
\end{eqnarray*}
where $\mathcal{G}_{ab}$ may be asymmetric. As a matrix, $\mathcal{F}_{AB}$
takes the form given in Eq.(\ref{Matrix form}).

\subsection{The Hodge dual and standard Yang-Mills action}

With these factors in mind, we find the Hodge dual of the Yang-Mills
field. In general terms the Hodge dual of a 2-form is given by
\begin{eqnarray*}
^{*}\boldsymbol{\mathcal{F}} & = & ^{*}\left(\frac{1}{2}\mathcal{F}_{AB}\tilde{\mathbf{e}}^{A}\wedge\tilde{\mathbf{e}}^{B}\right)\\
 & = & \frac{1}{\left(2n-2\right)!}\frac{1}{2}\mathcal{F}_{AB}\bar{K}^{AC}\bar{K}^{BD}\varepsilon_{CDE\ldots F}\tilde{\mathbf{e}}^{E}\wedge\ldots\wedge\tilde{\mathbf{e}}^{F}
\end{eqnarray*}
However, we need to separate the distinct quadrants of each inverse
metric. Expanding each upper case index $A,B,\ldots$, as a raised,
weight +1 index and a lowered weight -1 index in turn, then collecting
like terms leads to
\begin{eqnarray}
^{*}\boldsymbol{\mathcal{F}} & = & \frac{1}{n!\left(n-2\right)!}\left(\frac{1}{2}F_{ab}\bar{K}^{am}\bar{K}^{bn}+G_{\;\;\;b}^{a}\bar{K}_{a}^{\;\;\;m}\bar{K}^{bn}+\frac{1}{2}H^{ab}\bar{K}_{a}^{\;\;\;m}\bar{K}_{b}^{\;\;\;n}\right)\varepsilon_{\;\;\;\qquad mne\cdots f}^{c\cdots d}\mathbf{f}_{c\cdots d}\mathbf{e}^{e\cdots f}\nonumber \\
 &  & +\frac{\left(-1\right)^{n-1}}{\left(n-1\right)!\left(n-1\right)!}\left(\frac{1}{2}F_{ab}\bar{K}_{\;\;\;m}^{a}\bar{K}^{bn}+G_{\;\;\;b}^{a}\bar{K}_{am}\bar{K}^{bn}+\frac{1}{2}H^{ab}\bar{K}_{am}\bar{K}_{b}^{\;\;\;n}\right)\varepsilon_{\;\;\;\qquad ne\cdots f}^{mc\cdots d}\mathbf{f}_{c\cdots d}\mathbf{e}^{e\cdots f}\nonumber \\
 &  & +\frac{\left(-1\right)^{n}}{\left(n-1\right)!\left(n-1\right)!}\left(\frac{1}{2}F_{ab}\bar{K}^{am}\bar{K}_{\;\;\;n}^{b}+G_{\;\;\;b}^{a}\bar{K}_{a}^{\;\;\;m}\bar{K}_{\;\;\;n}^{b}+\frac{1}{2}H^{ab}\bar{K}_{a}^{\;\;\;m}\bar{K}_{bn}\right)\varepsilon_{\;\;\;\qquad me\cdots f}^{nc\cdots d}\mathbf{f}_{c\cdots d}\mathbf{e}^{e\cdots f}\nonumber \\
 &  & +\frac{1}{n!\left(n-2\right)!}\left(\frac{1}{2}F_{ab}\bar{K}_{\;\;\;m}^{a}\bar{K}_{\;\;\;n}^{b}+G_{\;\;\;b}^{a}\bar{K}_{am}\bar{K}_{\;\;\;n}^{b}+\frac{1}{2}H^{ab}\bar{K}_{am}\bar{K}_{bn}\right)\varepsilon_{\;\;\;\qquad e\cdots f}^{mnc\cdots d}\mathbf{f}_{c\cdots d}\mathbf{e}^{e\cdots f}\label{Hodge dual of a 2 form}
\end{eqnarray}
where one readily sees the advantage of omitting the wedge between
forms, $\mathbf{e}^{e}\wedge\ldots\wedge\mathbf{e}^{f}\longleftrightarrow\mathbf{e}^{e\ldots f}$.

We form the usual Yang-Mills Lagrangian density as the wedge product,
$\boldsymbol{\mathcal{F}}\wedge{}^{*}\boldsymbol{\mathcal{F}}$, eliminating
the basis forms in favor of the volume form $\boldsymbol{\Phi}$.
After a bit of algebra, we find

\begin{eqnarray}
\boldsymbol{\mathcal{F}}\wedge{}^{*}\boldsymbol{\mathcal{F}} & = & \left(\frac{1}{2}F_{mn}\bar{K}^{am}\bar{K}^{bn}+G_{\;\;\;n}^{m}\bar{K}_{\;\;\;m}^{a}\bar{K}^{bn}+\frac{1}{2}H^{mn}\bar{K}_{\;\;\;m}^{a}\bar{K}_{\;\;\;n}^{b}\right)F_{ab}\boldsymbol{\Phi}\nonumber \\
 &  & +\left(F_{mn}\bar{K}_{a}^{\;\;\;m}\bar{K}^{bn}+G_{\;\;\;n}^{m}\left(\bar{K}_{am}\bar{K}^{bn}-\bar{K}_{a}^{\;\;\;n}\bar{K}_{\;\;\;m}^{b}\right)+H^{mn}\bar{K}_{am}\bar{K}_{\;\;\;n}^{b}\right)G_{\;\;\;b}^{a}\boldsymbol{\Phi}\nonumber \\
 &  & +\left(\frac{1}{2}F_{mn}\bar{K}_{a}^{\;\;\;m}\bar{K}_{b}^{\;\;\;n}+G_{\;\;\;n}^{m}\bar{K}_{am}\bar{K}_{b}^{\;\;\;n}+\frac{1}{2}H^{mn}\bar{K}_{am}\bar{K}_{bn}\right)H^{ab}\boldsymbol{\Phi}\label{YM Lagrange density}
\end{eqnarray}
and the matter action is given by Eq.(\ref{Yang-Mills action}). The
full action is the combination of Eq.(\ref{Action}) and Eq.(\ref{Yang-Mills action}),

\[
S=S_{G}+S_{YM}
\]
Substituting the null orthonormal form of the Killing metric, Eq.(\ref{Killing form})
into Eq.(\ref{YM Lagrange density}) the lagrange density reduces
to
\begin{equation}
\boldsymbol{\mathcal{F}}\:{}^{*}\boldsymbol{\mathcal{F}}=\left(H^{ab}F_{ab}-G_{\;\;\;a}^{b}G_{\;\;\;b}^{a}\right)\boldsymbol{\Phi}\label{F F dual lagrange density Killing}
\end{equation}
We find that this Hodge dual form of the action is identical to the
form of the action given in \cite{AW}, despite the claim in \cite{AW}
that the action is independent of the metric. The presence of the
metric is concealed in \cite{AW} because the Killing metric in this
basis is comprised of Kronecker deltas.

Although varying the Yang-Mills potentials in Eq.(\ref{YM Lagrange density}),
or equivalently in Eq.(\ref{F F dual lagrange density Killing}),
yields the usual Yang-Mills field equation, we show below that varying
the metric gives a nonstandard coupling to gravity. In the next Section
we define a twisted action that gives the usual coupling to gravity,
while still leading to the correct Yang-Mills field equations.

\subsection{The failure of the $\mathcal{F}\wedge^{*}\mathcal{F}$ action}

Variation of the usual Lagrangian Eq.(\ref{YM Lagrange density})
gives
\begin{eqnarray*}
\delta_{g}\int\boldsymbol{\mathcal{F}}\wedge{}^{*}\boldsymbol{\mathcal{F}} & = & \int\left(F_{mn}F_{ab}\delta\bar{K}^{am}\eta^{bn}\right)\boldsymbol{\Phi}\\
 &  & +\left(G_{\;\;\;n}^{m}F_{ab}\eta^{bn}\delta\bar{K}_{\;\;\;m}^{a}+F_{mn}G_{\;\;\;b}^{a}\eta^{bn}\delta\bar{K}_{a}^{\;\;\;m}\right)\boldsymbol{\Phi}\\
 &  & +\left(G_{\;\;\;n}^{m}G_{\;\;\;b}^{a}\eta^{bn}\delta\bar{K}_{am}+G_{\;\;\;n}^{m}G_{\;\;\;b}^{a}\eta_{am}\delta\bar{K}^{bn}\right)\boldsymbol{\Phi}\\
 &  & +\left(H^{mn}G_{\;\;\;b}^{a}\eta_{am}\delta\bar{K}_{\;\;\;n}^{b}+G_{\;\;\;n}^{m}H^{ab}\eta_{am}\delta\bar{K}_{b}^{\;\;\;n}\right)\boldsymbol{\Phi}\\
 &  & +\left(H^{mn}H^{ab}\eta_{bn}\delta\bar{K}_{am}\right)\boldsymbol{\Phi}
\end{eqnarray*}
where after variation we returned the remaining inverse metric components
to the orthonormal form. The metric variations are now given by Eq.(\ref{Variation of the Killing form}).
Substituting, the variation yields the gravitational field equations,
\begin{eqnarray}
\alpha\Omega_{\;\;\;b\quad m}^{n\quad b}-\alpha\Omega_{\;\;\;b\quad a}^{a\quad b}\delta_{m}^{n}+\beta\Omega_{\;\;\;m}^{n}-\beta\Omega_{\;\;\;a}^{a}\delta_{m}^{n}+\Lambda\delta_{m}^{n} & = & -\kappa\left(H^{bn}F_{bm}-G_{\;\;\;m}^{b}G_{\;\;\;b}^{n}\right)\nonumber \\
 &  & -\kappa\left(\frac{1}{2}\delta_{m}^{n}\left(F_{bc}H^{bc}-G_{\;\;\;c}^{b}G_{\;\;\;b}^{c}\right)\right)\label{FEQ1}\\
\alpha\Omega_{\;\;\;n\quad a}^{a\quad m}-\alpha\Omega_{\;\;\;b\quad a}^{a\quad b}\delta_{n}^{m}+\beta\Omega_{\;\;\;n}^{m}-\beta\Omega_{\;\;\;a}^{a}\delta_{n}^{m}+\Lambda\delta_{n}^{m} & = & -\kappa\left(H^{bm}F_{bn}-G_{\;\;\;n}^{b}G_{\;\;\;b}^{m}\right)\nonumber \\
 &  & -\kappa\left(\frac{1}{2}\delta_{n}^{m}\left(F_{bc}H^{bc}-G_{\;\;\;c}^{b}G_{\;\;\;b}^{c}\right)\right)\label{FEQ2}\\
\alpha\Omega_{\;\;\;nam}^{a}+\beta\Omega_{nm} & = & \kappa\left(F_{am}G_{\;\;\;n}^{a}+F_{an}G_{\;\;\;m}^{a}\right)\label{FEQ3}\\
\alpha\Omega_{\;\;\;b}^{n\quad bm}+\beta\Omega^{nm} & = & -\kappa\left(H^{nb}G_{\;\;\;b}^{m}+H^{mb}G_{\;\;\;b}^{n}\right)\label{FEQ4}
\end{eqnarray}

The problem with this coupling to gravity becomes evident when the
gravitational equations require vanishing momentum curvatures on the
left side of Eq.(\ref{FEQ4}). With this, Eq.(\ref{FEQ4}) implies
\begin{eqnarray*}
H^{nb}G_{\;\;\;b}^{m}+H^{mb}G_{\;\;\;b}^{n} & = & 0
\end{eqnarray*}
The failure of this result is not immediate, but the reduction of
Eqs.(\ref{FEQ1}) and (\ref{FEQ2}) gives a second constraint on $H^{ab}$
and $G_{\;\;\;b}^{a}$. The two constraints lead, at the most general,
to vanishing $H^{ab}$ and symmetric $G_{\;\;\;b}^{a}$, leaving the
source for the Einstein equation, Eq.(\ref{FEQ3}) linear in $F_{ab}$
and therefore incompatible with the usual energy source for general
relativity.

The problem cannot be altered by a different choice of the inner product
(see Appendix A), but must lie in the use of the usual action. In
the context of other double field theories, a twist allows dimensional
reduction to preserve gauging of supersymmetries \cite{SUSY Twist,Twisted DFT,Gauged double field theory}.
Here we find that including a twist insures consistency under dimensional
reduction not only for supersymmetry, but also gives the correct coupling
of Yang-Mills sources to gravity.

We now turn to the definition of the twisted Yang-Mills action, and
its metric variation. Then, in Section 5, we vary the Yang-Mills potentials.
The remainder of our presentation details the reduction of the full
set of field equations to reproduce scale covariant general relativity
with Yang-Mills sources in the usual $n$-dimensional form.

\section{Metric variation of the twisted Yang-Mills Action}

Instead of the usual spacetime action, Eq.(\ref{Yang-Mills action}),
we consider a biconformal Yang-Mills theory with an action functional
of the form 
\begin{equation}
S_{\unit{TYM}}=-\frac{\kappa}{2}\intop tr\bar{\mathcal{F}}\wedge^{*}\mathcal{F}\label{Twisted Yang-Mills action}
\end{equation}
where $*$ is the usual Hodge dual, $\mathcal{F}$ is a curvature
$2$-form, $\bar{\mathcal{F}}$ is a twisted conjugate curvature,
and the trace is over the $SU\left(N\right)$ generators. The twist
matrix is formed using both the Killing metric and the Kähler form,
$K_{\;\;\;B}^{A}\equiv\bar{K}^{AC}g_{CB}$. While the twist matrix
is similar to that used to preserve supersymmetry in other double
field theories, \cite{SUSY Twist,Twisted DFT,Gauged double field theory},
we define the twisted Yang-Mills field by
\[
\bar{\mathcal{F}}_{AB}=\frac{1}{2}\left(K_{A}^{\;\;\;C}\mathcal{F}_{CB}+\mathcal{F}_{AC}K_{\;\;\;B}^{C}\right)
\]
We find that this form is necessary to preserve the antisymmetry of
the field while giving the required interchange of source fields.
Variation of the inverse Killing form $\bar{K}^{AB}$ then gives the
source for the gravitational field equations.

\subsection{Details of the twist}

The twist is accomplished using $K_{\;\;\;B}^{A}\equiv\bar{K}^{AC}g_{CB}$
where $K_{\;\;\;B}^{A}=K_{B}^{\;\;\;A}$, since both $g_{AB}$ and
$K_{AB}$ are symmetric. In the null-orthonormal form and the $\mathbf{e}^{A}$
basis, this matrix is simply
\begin{eqnarray*}
K_{\;\;\;B}^{A} & = & K_{B}^{\;\;\;A}=\left(\begin{array}{cc}
0 & \delta_{b}^{a}\\
\delta_{b}^{a} & 0
\end{array}\right)
\end{eqnarray*}
and the required form of the field is
\begin{eqnarray}
\bar{\mathcal{F}}_{AB} & = & \frac{1}{2}\left(K_{A}^{\;\;\;C}\mathcal{F}_{CB}+\mathcal{F}_{AC}K_{\;\;\;B}^{C}\right)\nonumber \\
 & = & \left(\begin{array}{cc}
\mathcal{G}_{\left[ab\right]} & \frac{1}{2}\left(F_{ab}+\mathcal{H}_{ab}\right)\\
\frac{1}{2}\left(F_{ab}+\mathcal{H}_{ab}\right) & \mathcal{G}_{\left[ab\right]}
\end{array}\right)\label{Twist in the NON frame}
\end{eqnarray}
where $\mathcal{F}_{AB}$ is given by Eq.(\ref{Matrix form}). Note
that this transformation maintains the antisymmetry while interchanging
the diagonal and anti-diagonal elements.

However, the reduced form of the twisted field in the orthonormal
frame given in Eq.(\ref{Twist in the NON frame}) is insufficient.
Until we complete the metric variation we must use the generic form
of the metric in computing the twist matrix,
\begin{eqnarray*}
K_{\;\;\;B}^{A} & = & \left(\begin{array}{cc}
\bar{K}^{ac} & \bar{K}_{\;\;\;e}^{a}\eta^{ec}\\
\eta^{ae}\bar{K}_{e}^{\;\;\;c} & \eta^{ae}\eta^{cf}\bar{K}_{ef}
\end{array}\right)\left(\begin{array}{cc}
\eta_{cb} & 0\\
0 & \eta_{bc}
\end{array}\right)\\
 & = & \left(\begin{array}{cc}
\bar{K}^{ac}\eta_{cb} & \bar{K}_{\;\;\;b}^{a}\\
\bar{K}_{\;\;\;b}^{a} & \eta^{ae}\bar{K}_{eb}
\end{array}\right)\\
K_{A}^{\;\;\;B} & = & \left(\begin{array}{cc}
\eta_{ad}\bar{K}^{db} & \bar{K}_{a}^{\;\;\;b}\\
\bar{K}_{a}^{\;\;\;b} & \bar{K}_{ae}\eta^{eb}
\end{array}\right)
\end{eqnarray*}
with symmetry given by $\bar{K}_{\;\;\;b}^{a}=\eta^{ae}\bar{K}_{e}^{\;\;\;c}\eta_{cb}$
(see Appendix B for details of the symmetry). Then $\bar{\mathcal{F}}_{AB}$
becomes

\begin{eqnarray}
\bar{\mathcal{F}}_{AB} & = & \frac{1}{2}\left(K_{A}^{\;\;\;C}\mathcal{F}_{CB}+\mathcal{F}_{AC}K_{\;\;\;B}^{C}\right)\nonumber \\
 & = & \frac{1}{2}\left(\begin{array}{cc}
\eta_{ad}\bar{K}^{dc}F_{cb}+\bar{K}_{a}^{\;\;\;c}\mathcal{G}_{cb} & -\eta_{ad}\bar{K}^{dc}\mathcal{G}_{bc}+\bar{K}_{a}^{\;\;\;c}\mathcal{H}_{cb}\\
\eta_{ad}\bar{K}_{\;\;\;e}^{d}\eta^{ec}F_{cb}+\bar{K}_{ae}\eta^{ec}\mathcal{G}_{cb}\quad & -\eta_{ad}\bar{K}_{\;\;\;e}^{d}\eta^{ec}\mathcal{G}_{bc}+\bar{K}_{ae}\eta^{ec}\mathcal{H}_{cb}
\end{array}\right)\nonumber \\
 &  & +\frac{1}{2}\left(\begin{array}{cc}
F_{ac}\bar{K}^{ce}\eta_{eb}-\mathcal{G}_{ca}\eta^{cd}\bar{K}_{d}^{\;\;\;e}\eta_{eb} & F_{ac}\bar{K}_{\;\;\;b}^{c}-\mathcal{G}_{ca}\eta^{ce}\bar{K}_{eb}\\
\mathcal{G}_{ac}\bar{K}^{ce}\eta_{eb}+\mathcal{H}_{ac}\eta^{cd}\bar{K}_{d}^{\;\;\;e}\eta_{eb}\quad & \mathcal{G}_{ac}\bar{K}_{\;\;\;b}^{c}+\mathcal{H}_{ac}\eta^{ce}\bar{K}_{eb}
\end{array}\right)\label{Twisted F}
\end{eqnarray}
and we check that \ref{Twisted F} agrees with Eq.(\ref{Twist in the NON frame})
when we restore the null orthonormal frame for the metric. The twisted
field is simpler when written as a 2-form,
\begin{eqnarray}
\bar{\boldsymbol{\mathcal{F}}} & = & \frac{1}{2}\bar{\mathcal{F}}_{AB}\mathbf{e}^{A}\wedge\mathbf{e}^{B}\nonumber \\
 & = & \frac{1}{2}\left(F_{ac}\bar{K}^{ce}\eta_{eb}+\bar{K}_{a}^{\;\;\;c}\mathcal{G}_{cb}\right)\mathbf{e}^{a}\wedge\mathbf{e}^{b}\nonumber \\
 &  & +\frac{1}{2}\left(F_{ac}\bar{K}_{\;\;\;b}^{c}\eta^{bf}+\bar{K}_{a}^{\;\;\;c}\mathcal{H}_{cb}\eta^{bf}-\mathcal{G}_{ca}\eta^{ce}\bar{K}_{eb}\eta^{bf}-\mathcal{G}_{bc}\bar{K}^{ce}\eta_{ea}\eta^{bf}\right)\mathbf{e}^{a}\wedge\mathbf{f}_{f}\nonumber \\
 &  & +\frac{1}{2}\left(\mathcal{H}_{ac}\eta^{ce}\bar{K}_{eb}+\mathcal{G}_{ac}\bar{K}_{\;\;\;b}^{c}\right)\eta^{af}\mathbf{f}_{f}\wedge\eta^{bg}\mathbf{f}_{g}\label{F bar as a 2 form}
\end{eqnarray}
We may interchange $\bar{K}_{a}^{\;\;\;b}$ and $\bar{K}_{\;\;\;a}^{b}$
when convenient since these have identical variations and both restrict
to $\delta_{b}^{a}$ in the null orthonormal basis.

\subsection{The action}

The dual field is given by Eq.(\ref{Hodge dual of a 2 form}). To
avoid duplication of indices when we wedge the dual field together
with the twisted field, we rename the indices in the twisted field
\begin{eqnarray*}
\bar{\boldsymbol{\mathcal{F}}} & = & \frac{1}{2}\left(F_{rq}\bar{K}^{qt}\eta_{ts}+\bar{K}_{r}^{\;\;\;q}\mathcal{G}_{qs}\right)\mathbf{e}^{r}\wedge\mathbf{e}^{s}\\
 &  & +\frac{1}{2}\left(F_{rc}\bar{K}_{\;\;\;s}^{c}+\bar{K}_{r}^{\;\;\;c}\mathcal{H}_{cs}\right)\eta^{sw}\mathbf{e}^{r}\wedge\mathbf{f}_{w}\\
 &  & +\frac{1}{2}\left(-\mathcal{G}_{sq}\bar{K}^{qt}\eta_{tr}-\mathcal{G}_{qr}\eta^{qt}\bar{K}_{ts}\right)\eta^{sw}\mathbf{e}^{r}\wedge\mathbf{f}_{w}\\
 &  & +\frac{1}{2}\left(\mathcal{G}_{rq}\bar{K}_{\;\;\;s}^{q}+\mathcal{H}_{rq}\eta^{qt}\bar{K}_{ts}\right)\eta^{rw}\eta^{sx}\mathbf{f}_{w}\wedge\mathbf{f}_{x}
\end{eqnarray*}
Each term of the wedge product is proportional to the volume form,
using $\mathbf{f}_{c\cdots d}\wedge\mathbf{e}^{e\cdots f}=\bar{e}_{c\cdots d}^{\qquad e\cdots f}\boldsymbol{\Phi}$.
Then, replacing the double Levi-Civita tensors with Kronecker deltas
according to Eqs.(\ref{Reduction formulas}), we fully distribute
the lengthy expression. We summarize the essential features here,
but details including the full wedge product $\bar{\boldsymbol{\mathcal{F}}}\wedge{}^{*}\boldsymbol{\mathcal{F}}$
and its reduction to $\left.\bar{\boldsymbol{\mathcal{F}}}\wedge{}^{*}\boldsymbol{\mathcal{F}}\right|_{contributing}$
below are given in Appendix C.

Since we are interested only in the variation, and will return the
metric to the null orthonormal form of Eq.(\ref{Killing form}) after
variation, certain terms clearly do not contribute to the field equations.
For example, in the product 
\begin{eqnarray*}
 &  & \left(-\mathcal{G}_{sq}\bar{K}^{qt}\eta_{tr}-\mathcal{G}_{qr}\eta^{qt}\bar{K}_{ts}\right)\eta^{sw}\left(\frac{1}{2}F_{ab}\bar{K}_{\;\;\;m}^{a}\bar{K}^{bn}+\frac{1}{2}\mathcal{H}_{gh}\eta^{ga}\eta^{hb}\bar{K}_{am}\bar{K}_{b}^{\;\;\;n}\right)
\end{eqnarray*}
none of the four terms after distribution will contribute to the field
equation because the variation of any one factor of metric components
always leaves an unvaried $\bar{K}_{ts}$ or $\bar{K}^{bn}$,
\[
\delta\left(\bar{K}_{\;\;\;m}^{a}\bar{K}_{ts}\bar{K}^{bn}\right)=\delta\bar{K}_{\;\;\;m}^{a}\,\cancelto{0}{\bar{K}_{ts}\bar{K}^{bn}}+\bar{K}_{\;\;\;m}^{a}\delta\bar{K}_{ts}\,\cancelto{0}{\bar{K}^{bn}}+\bar{K}_{\;\;\;m}^{a}\,\cancelto{0}{\bar{K}_{ts}}\delta\bar{K}^{bn}
\]
Dropping such terms, we are only required to vary terms linear in
$\bar{K}_{ab}$ or $\bar{K}^{ab}$, or cubic in the off diagonal components
$\bar{K}_{\;\;\;b}^{a}$. Collecting these and using the symmetries
of the fields, finally yields,
\begin{eqnarray}
\left.\bar{\boldsymbol{\mathcal{F}}}\wedge{}^{*}\boldsymbol{\mathcal{F}}\right|_{contributing} & = & \left(\frac{1}{2}F_{bc}F_{da}\eta^{ac}+F_{bc}\mathcal{H}^{ac}\eta_{ad}+\frac{1}{2}\left(\mathcal{G}_{ad}-2\mathcal{G}_{da}\right)\eta_{bc}\mathcal{G}^{ac}\right)\bar{K}^{bd}\boldsymbol{\Phi}\nonumber \\
 &  & +\left(\mathcal{G}_{ba}H^{dc}+F_{ab}\mathcal{G}^{cd}\right)\bar{K}_{\;\;\;e}^{b}\bar{K}_{\;\;\;d}^{e}\bar{K}_{\;\;\;c}^{a}\boldsymbol{\Phi}\nonumber \\
 &  & +\left(\frac{1}{2}\mathcal{H}_{ac}\mathcal{H}^{ab}\eta^{cd}+F_{ac}\mathcal{H}^{ab}\eta^{cd}+\frac{1}{2}\left(\mathcal{G}^{ba}-2\mathcal{G}^{ab}\right)\eta^{cd}\mathcal{G}_{ca}\right)\bar{K}_{bd}\boldsymbol{\Phi}\label{YM Lagrangian}
\end{eqnarray}
We may now vary the metric.

To carry out the variation of the Yang-Mills potentials we may write
the action in the null orthonormal frame. This form is still contained
in the expression above, given by the purely off-diagonal terms, cubic
in $\bar{K}_{\;\;\;b}^{a}$. This simpler form of the action follows
immediately as
\begin{equation}
S_{TYM}=\kappa\int tr\left(\bar{\boldsymbol{\mathcal{F}}}\wedge{}^{*}\boldsymbol{\mathcal{F}}\right)=\kappa\int tr\left(\mathcal{G}^{ab}\left(\mathcal{H}_{ab}+F_{ab}\right)\right)\boldsymbol{\Phi}\label{YM action NON frame}
\end{equation}
The variation of the potentials is carried out in Section \ref{sec:Yang-Mills-potentials-and}.

\subsection{Metric variation}

Using the variation of the Killing metric given in Eq.(\ref{Variation of the Killing form})
and the variation of the volume form given by
\begin{eqnarray*}
\delta\boldsymbol{\Phi} & = & -\frac{1}{2}K_{AB}\delta\bar{K}^{AB}\boldsymbol{\Phi}\\
 & = & -\frac{1}{2}\left[K_{ab}\left(B^{ab}+B^{ba}\right)+K_{\;\;\;b}^{a}\left(D_{a}^{\;\;\;b}+A_{\;\;\;a}^{b}\right)+K_{b}^{\;\;\;a}\left(A_{\;\;\;a}^{b}+D_{a}^{\;\;\;b}\right)+K^{ab}\left(C_{ab}+C_{ba}\right)\right]\boldsymbol{\Phi}\\
 & = & -\delta_{\;\;\;b}^{a}\left(A_{\;\;\;a}^{b}+D_{a}^{\;\;\;b}\right)\boldsymbol{\Phi}
\end{eqnarray*}
the variation of Eq.(\ref{YM Lagrangian}) yields
\begin{eqnarray}
\delta\left(\bar{\boldsymbol{\mathcal{F}}}\wedge{}^{*}\boldsymbol{\mathcal{F}}\right) & = & \left(\frac{1}{2}F_{bc}F_{da}\eta^{ac}+F_{bc}\mathcal{H}^{ac}\eta_{ad}+\frac{1}{2}\left(\mathcal{G}_{ad}-2\mathcal{G}_{da}\right)\eta_{bc}\mathcal{G}^{ac}\right)2B^{\left(bd\right)}\boldsymbol{\Phi}\nonumber \\
 &  & +\left(\left(2\mathcal{G}^{ad}-\mathcal{G}^{da}\right)F_{ab}-\left(2\mathcal{G}^{da}-\mathcal{G}^{ad}\right)\mathcal{H}_{ab}-\left(F_{ac}+\mathcal{H}_{ac}\right)\mathcal{G}^{ac}\delta_{\;\;\;b}^{d}\right)\left(A_{\;\;\;d}^{b}+D_{d}^{\;\;\;b}\right)\boldsymbol{\Phi}\nonumber \\
 &  & +\left(\frac{1}{2}\mathcal{H}_{ac}\mathcal{H}^{ab}\eta^{cd}+F_{ac}\mathcal{H}^{ab}\eta^{cd}+\frac{1}{2}\left(\mathcal{G}^{ba}-2\mathcal{G}^{ab}\right)\eta^{cd}\mathcal{G}_{ca}\right)2C_{\left(bd\right)}\boldsymbol{\Phi}\label{YM sources for gravity}
\end{eqnarray}
This variation couples to the $\left(\mathbf{e}^{a},\mathbf{f}_{a}\right)$
variation of the gravity action, Eq.(\ref{Action}). The resulting
field equations and their reduction to the gravity theory are given
in Section (\ref{sec:Reduction-of-the}) below. Before considering
reduction of the field equations, we turn to variation of the Yang-Mills
potentials to find the Yang-Mills equations.

\section{Yang-Mills potentials and field equations\label{sec:Yang-Mills-potentials-and}}

We also need to vary the potential to find the Yang-Mills field equations.
We start with the action of Eq.(\ref{YM action NON frame}) in the
null orthonormal basis,
\begin{eqnarray}
S_{TYM} & = & \kappa\int tr\bar{\boldsymbol{\mathcal{F}}^{i}}\wedge{}^{*}\boldsymbol{\mathcal{F}}^{i}=\kappa\int\mathcal{G}^{iab}\left(\mathcal{H}_{i\;ab}+F_{i\;ab}\right)\boldsymbol{\Phi}\label{YM action in NON metric}
\end{eqnarray}
In this Section, we make the internal symmetry explicit, varying $S_{YM}$
with respect to the $SU\left(N\right)$ potentials. It is most convenient
to work in the $\bar{\mathbf{e}}^{A}=\left(\mathbf{e}^{a},\mathbf{f}_{b}\right)$
basis. Internal indices are labeled with letters $i,j,k$, while frame
indices are chosen from the beginning of the alphabet, $a,b,c,\ldots$.

\subsection{The field components and potentials}

The $SU\left(N\right)$ field is given in the $\bar{\mathbf{e}}^{A}$
basis by Eq.(\ref{SU(N) field in components}),
\begin{eqnarray}
\boldsymbol{\mathcal{F}}^{i} & = & \frac{1}{2}F_{\;\;\;ab}^{i}\mathbf{e}^{a}\wedge\mathbf{e}^{b}+\mathcal{G}_{\;\;\;cb}^{i}\eta^{ac}\mathbf{f}_{a}\wedge\mathbf{e}^{b}+\frac{1}{2}\mathcal{H}_{\;\;\;ab}^{i}\eta^{ac}\eta^{bd}\mathbf{f}_{c}\wedge\mathbf{f}_{c}\label{SU(N) field in components}
\end{eqnarray}
where as a matrix,
\begin{eqnarray*}
\mathcal{F}_{\;\;\;AB}^{i} & = & \left(\begin{array}{cc}
F_{\;\;\;ab}^{i} & G_{\;\;a}^{i\quad\;\;b}\\
G_{\;\;\quad b}^{i\;a} & \mathcal{H}_{\;\;\;ab}^{i}
\end{array}\right)
\end{eqnarray*}
with $G_{\;\;a}^{i\quad\;\;b}=-G_{\;\;\quad b}^{i\;a}$.

The field is given in terms of its $U\left(1\right)$ or $SU\left(N\right)$
potential by the Cartan equation,
\begin{eqnarray*}
\boldsymbol{\mathcal{F}}^{i} & = & \mathbf{d}\boldsymbol{\mathcal{A}}^{i}-\frac{1}{2}c_{\;\;\;jk}^{i}\boldsymbol{\mathcal{A}}^{j}\land\boldsymbol{\mathcal{A}}^{k}
\end{eqnarray*}
where the potentials are biconformal 1-forms,
\[
\boldsymbol{\mathcal{A}}^{i}=A_{\;\;\;a}^{i}\mathbf{e}^{a}+B^{i\;a}\mathbf{f}_{a}
\]
In terms of $A_{\;\;\;a}^{i},B^{i\;a}$, the field becomes

\begin{eqnarray}
\boldsymbol{\mathcal{F}}^{i} & = & \mathbf{D}A_{\;\;\;a}^{i}\land\mathbf{e}^{a}-\frac{1}{2}\boldsymbol{\mathcal{A}}_{\;\;\;k}^{i}\land A_{\;\;\;b}^{k}\mathbf{e}^{b}\nonumber \\
 &  & +\mathbf{D}B^{i\;a}\land\mathbf{f}_{a}+B^{i\;a}\mathbf{S}{}_{a}-\frac{1}{2}\boldsymbol{\mathcal{A}}_{\;\;\;k}^{i}\land B^{k\;b}\mathbf{f}_{b}\label{SU(N) field in terms of potentials}
\end{eqnarray}
where $\boldsymbol{\mathcal{A}}_{\;\;\;k}^{i}$ is the $SU\left(N\right)$
connection in the adjoint representation,
\begin{eqnarray*}
\boldsymbol{\mathcal{A}}_{\;\;\;k}^{i} & \equiv & c_{\;\;\;jk}^{i}\boldsymbol{\mathcal{A}}^{j}
\end{eqnarray*}
and the Weyl-covariant derivatives of the potentials are given by
\begin{eqnarray*}
\mathbf{D}A_{\;\;\;a}^{i} & = & \mathbf{d}A_{\;\;\;a}^{i}-\boldsymbol{\omega}_{\;\;\;c}^{a}A_{\;\;\;a}^{i}+A_{\;\;\;a}^{i}\boldsymbol{\omega}\\
\mathbf{D}B^{i\;a} & = & \mathbf{d}B^{i\;a}-B^{i\;a}\boldsymbol{\omega}_{\;\;\;a}^{c}-B^{i\;a}\boldsymbol{\omega}
\end{eqnarray*}
Note that the covariant derivative of $\eta^{ab}$ does not necessarily
vanish, $\mathbf{D}\eta^{ab}=\mathbf{d}\eta^{ab}-2\boldsymbol{\omega}\eta^{ab}$
where $\mathbf{d}\eta^{ab}$ takes into account the conformal equivalence
class, $\eta_{ab}\in\left\{ \left.e^{2\varphi}\eta_{ab}^{0}\right|\varphi=\varphi\left(x,y\right)\right\} $.

Now we separate Eq.(\ref{SU(N) field in terms of potentials}) into
its independent $\mathbf{e}^{a}\land\mathbf{e}^{b}$, $\mathbf{f}_{a}\land\mathbf{e}^{b}$
and $\mathbf{f}_{a}\land\mathbf{f}_{b}$ parts. We begin by expanding
the exterior derivatives in the orthonormal basis 
\begin{eqnarray*}
\mathbf{d}A_{\;\;\;a}^{i} & = & A_{\;\;\;a,b}^{i}\mathbf{e}^{b}+A_{\;\;\;a}^{i\quad\;\;,b}\mathbf{f}_{b}\\
\mathbf{d}B^{i\;a} & = & B_{\;\;\;\;\;,b}^{i\;a}\mathbf{e}^{b}+B^{i\;\;a,b}\mathbf{f}_{b}
\end{eqnarray*}
and similarly the covariant exterior derivatives 
\begin{eqnarray*}
\mathbf{D}A_{\;\;\;a}^{i} & = & A_{\;\;\;a;b}^{i}\mathbf{e}^{b}+A_{\;\;\;a}^{i\quad\;\;;b}\mathbf{f}_{b}\\
\mathbf{D}B^{i\;a} & = & B_{\;\;\;\;\;;b}^{i\;a}\mathbf{e}^{b}+B^{i\;a;b}\mathbf{f}_{b}
\end{eqnarray*}
We also expand the $SU\left(N\right)$ connection, $\boldsymbol{\mathcal{A}}^{j}G_{j}$,
in the adjoint representation:
\begin{eqnarray*}
\boldsymbol{\mathcal{A}}_{\;\;\;k}^{i} & = & c_{\;\;\;jk}^{i}A_{\;\;\;a}^{j}\mathbf{e}^{a}+c_{\;\;\;jk}^{i}B^{j\;a}\mathbf{f}_{a}\\
 & = & \boldsymbol{\alpha}_{\;\;\;k}^{i}+\boldsymbol{\beta}_{\;\;\;k}^{i}
\end{eqnarray*}
Finally, writing the general form of the co-torsion
\begin{eqnarray*}
\mathbf{S}{}_{a} & = & \frac{1}{2}S_{abc}\mathbf{e}^{b}\wedge\mathbf{e}^{c}+S_{a\quad\;\;c}^{\;\;\;b}\mathbf{f}_{b}\wedge\mathbf{e}^{c}+\frac{1}{2}S_{a}^{\;\;\;bc}\mathbf{f}_{b}\land\mathbf{f}_{c}
\end{eqnarray*}
we arrive the component fields in terms of the potentials:
\begin{eqnarray}
F_{\;\;\;ab}^{i} & = & A_{\;\;\;b;a}^{i}-A_{\;\;\;a;b}^{i}-c_{\;\;\;jk}^{i}A_{\;\;\;a}^{j}A_{\;\;\;b}^{k}+B^{i\;c}S_{cab}\label{F field in terms of potential}\\
G_{\;\;\quad b}^{i\;a} & = & A_{\;\;\;b}^{i\quad\;\;;a}-B_{\;\;\;;b}^{i\;a}-c_{\;\;\;jk}^{i}B^{j\;a}A_{\;\;\;b}^{k}+B^{i\;c}S_{c\quad\;\;b}^{\;\;\;a}\label{G field in terms of potentials}\\
H^{i\;\;ab} & = & B^{i\;\;b;a}-B^{i\;\;a;b}-c_{\;\;\;jk}^{i}B^{j\;a}B^{k\;b}+B^{i\;c}S{}_{c}^{\;\;\;ab}\label{H field in terms of potentials}
\end{eqnarray}
These expressions are covariant with respect to both $SU\left(N\right)$
and Weyl group transformations.

\subsection{The field equations for the potentials}

Using Eqs.(\ref{F field in terms of potential})-(\ref{H field in terms of potentials})
in the action, Eq.(\ref{YM action in NON metric}), the variation
of $A_{\;\;\;a}^{i}$ gives
\begin{eqnarray*}
\delta_{A}S_{TYM} & = & \kappa\int\left(\delta_{A}\mathcal{G}^{iab}\left(\mathcal{H}_{i\;ab}+F_{i\;ab}\right)\right)\boldsymbol{\Phi}\\
 &  & +\kappa\int\mathcal{G}^{iab}\left(\cancel{\delta_{A}\mathcal{H}_{i\;ab}}+\delta_{A}F_{i\;ab}\right)\boldsymbol{\Phi}\\
 & = & \kappa\int\left(\delta A_{\;\;\;b}^{k}\eta^{bc}\left(-\left(\mathcal{H}_{k\;ac}+F_{k\;ac}\right)^{;a}-c_{\;\;\;jk}^{i}B^{j\;a}\left(\mathcal{H}_{i\;ac}+F_{i\;ac}\right)\right)\right)\boldsymbol{\Phi}\\
 &  & +\kappa\int\delta A_{\;\;\;b}^{k}\left(-\left(\mathcal{G}_{k}^{\;\;\;ab}-\mathcal{G}_{k}^{\;\;\;ba}\right)_{;a}-c_{\;\;\;kj}^{i}A_{\;\;\;a}^{j}\left(\mathcal{G}_{i}^{\;\;\;ba}-\mathcal{G}_{i}^{\;\;\;ab}\right)\right)\boldsymbol{\Phi}
\end{eqnarray*}
Therefore the field equation becomes
\begin{eqnarray*}
0 & = & \eta^{bc}\left(\mathcal{H}_{k\;ac}+F_{k\;ac}\right)^{;a}+\eta^{bc}\left(\mathcal{H}_{i\;ac}+F_{i\;ac}\right)\beta_{\;\;\;k}^{i\;\;\quad a}\\
 &  & +\left(\mathcal{G}_{k}^{\;\;\;ab}-\mathcal{G}_{k}^{\;\;\;ba}\right)_{;a}+\left(\mathcal{G}_{i}^{\;\;\;ab}-\mathcal{G}_{i}^{\;\;\;ba}\right)\alpha_{\;\;\;ka}^{i}
\end{eqnarray*}
For the $B^{i\;a}$ variation, recalling that $i,j,$ are internal
indices and $a,b,c,e$ orthonormal indices, we find
\begin{eqnarray*}
0 & = & \eta^{bc}\left(\left(\mathcal{H}_{j\;ac}+F_{j\;ac}\right)_{;b}+\alpha_{\;\;\;jb}^{i}\left(\mathcal{H}_{i\;ac}+F_{i\;ac}\right)+S_{a\quad\;\;b}^{\;\;\;e}\left(\mathcal{H}_{j\;ec}+F_{j\;ec}\right)\right)\\
 &  & +\left(\mathcal{G}_{j\;ab}-\mathcal{G}_{j\;ba}\right)^{;b}+\left(\mathcal{G}_{i\;ab}-\mathcal{G}_{i\;ba}\right)\beta_{\;\;\;j}^{i\;\;\quad b}+\frac{1}{2}\left(\mathcal{G}_{j\;bc}-\mathcal{G}_{j\;cb}\right)S{}_{a}^{\;\;\;bc}\\
 &  & +\frac{1}{2}\left(\mathcal{G}_{j}^{\;\;\;bc}-\mathcal{G}_{j}^{\;\;\;cb}\right)S_{abc}
\end{eqnarray*}

Notice that there is no dynamical equation for the symmetric part
of $\mathcal{G}_{\;\;\;cd}^{k}$. Moreover, the action depends only
on the antisymmetric part of $\mathcal{G}_{\;\;\;cd}^{k}$. Therefore,
from here on we assume that like $F_{\;\;\;ab}^{i}$ and $H^{i\,ab}$,
$\mathcal{G}_{\;\;\;cd}^{k}$ is antisymmetric, $\mathcal{G}_{\;\;\;ab}^{k}\equiv\mathcal{G}_{\;\;\;\left[ab\right]}^{k}$.
Also notice that there is no separate Yang-Mills field equation for
$F_{ab}$ and $H_{ab}$. Both field equations contain only their sum,
$\mathcal{H}_{\;\;\;ab}^{k}+F_{\;\;\;ab}^{k}$, although $F_{ab}$
and $\mathcal{H}_{ab}$ enter the gravitational equations separately.
We therefore define a new field,
\begin{eqnarray*}
K_{\;\;\;ab}^{k} & \equiv & \frac{1}{2}\left(\mathcal{H}_{\;\;\;ab}^{k}+F_{\;\;\;ab}^{k}\right)
\end{eqnarray*}
In terms of these, the field equations become
\begin{eqnarray}
0 & = & \eta^{bc}K_{k\;ac}^{\;\;\quad;a}+\eta^{bc}K_{i\;ac}\beta_{\;\;\;k}^{i\;\;\quad a}+\mathcal{G}_{k\;\;\quad;a}^{\;\;\;ab}+\mathcal{G}_{i}^{\;\;\;ab}\alpha_{\;\;\;ka}^{i}\label{Nonabelian YM field equation 1}\\
0 & = & \eta^{bc}\left(K_{j\;ac;b}+K_{i\;ac}\alpha_{\;\;\;jb}^{i}+K_{j\;ec}S_{a\quad\;\;b}^{\;\;\;e}\right)\nonumber \\
 &  & +\mathcal{G}_{j\;ab}^{\;\;\quad;b}+\mathcal{G}_{i\;ab}\beta_{\;\;\;j}^{i\;\;\quad b}+\frac{1}{2}\mathcal{G}_{j\;bc}S{}_{a}^{\;\;\;bc}+\frac{1}{2}\mathcal{G}_{j}^{\;\;\;bc}S_{abc}\label{Nonabelian YM field equation 2}
\end{eqnarray}
These have the expected form of a divergence of the field strength.

The field equations of an $SU\left(N\right)$ gauge theory on a $2n$-dimensional
space given by Eqs.(\ref{Nonabelian YM field equation 1}) and (\ref{Nonabelian YM field equation 2})
together with the gravitational sources of Eqs.(\ref{YM sources for gravity})
complete the first stage of our investigation. These expressions give
a satisfactory formulation of biconformal Yang-Mills theory.

The second stage of this study is to understand how the $SU\left(N\right)$
gravitational sources together with their field equations affect the
gravitational and Yang-Mills solutions. Specifically, we want to know
if the reduction to the co-tangent bundle of $n$-dimensional spacetime
still occurs, and if so, whether the usual Yang-Mills field equations
and gravitational sources result.

This is indeed what happens. Once we have reduced the gravitational
field equations to more simply describe the underlying geometry, we
will return to Eqs.(\ref{Nonabelian YM field equation 1}) and (\ref{Nonabelian YM field equation 2}).
Our final result is to show that the reduction of the underlying geometry
to the co-tangent bundle of general relativity simultaneously forces
the reduction of the source equations to the usual Yang-Mills sources
for general relativity.

In the next Section we carry out the gravitational reduction. Since
the pure gravitational field case is presented in detail elsewhere
\cite{WheelerAugust}, we are able to simply state some of the conclusions,
focussing on those features which are different in the presence of
matter. Subsequently, in Sec.(\ref{sec:The-source-for}), we give
particular attention to the resulting twofold reduction of the $SU\left(N\right)$
fields: (1) from $\left(N^{2}-1\right)\times\frac{2n\left(2n-1\right)}{2}$
field components to $\left(N^{2}-1\right)\times\frac{n\left(n-1\right)}{2}$
components, and (2) the restriction of the number of effective independent
variables, $\mathcal{\boldsymbol{F}}\left(x,y\right)\rightarrow\mathcal{\boldsymbol{F}}\left(x\right)$.

\section{Reduction of the gravitational field equations\label{sec:Reduction-of-the}}

Although the reduction of the gravitational field equations has been
presented in detail elsewhere \cite{WheelerAugust} and we only highlight
the features which differ in the presence of matter, the presentation
is still somewhat lengthy. To aid in following the discussion, we
begin with a short description of the steps before providing details.
The basic steps are the following:
\begin{itemize}
\item Present the structure equations and Bianchi identities and their immediate
consequences for the form of the torsion-free biconformal curvatures
(Subsections \ref{subsec:Field-equations-for} - \ref{subsec:Vanishing-torsion}).
By manipulating the field equations, we arrive at a reduced form for
the components of the curvature tensors (Subsections \ref{subsec:Vanishing-torsion}
- \ref{subsec:Dilatation}). (These calculations involve the sources
in essential ways that require more detailed presentation here.)
\item The Cartan structure equations show that the solder form $\mathbf{e}^{a}$
is in involution. The Frobenius theorem therefore allows us to set
$\mathbf{e}^{a}=e_{\mu}^{\;\;\;a}\mathbf{d}x^{\mu}$ and study the
restricted solution on the $x^{\mu}=constant$ submanifolds spanned
by $n$ additional coordinates $y_{\mu}$. Because of the reduced
form of the curvature components, this restricted solution completely
determines the $\mathbf{e}^{a}=0$ pieces of the connection forms.
We then extend the connection forms back to the full biconformal space.
At this point, the connection forms have far fewer than their original
number of components (Summarized in Subsection \ref{subsec:The-Frobenius-theorem}).
\item Substitute the reduced connection forms into the structure equations
and impose the field equations. This is done one structure equation
at a time, each time reducing the degrees of freedom of the full system.
Ultimately, all connection and curvature components are determined
by the solder form, $\mathbf{e}^{a}$, and the components of the solder
form itself only depend on half the original coordinates, $e_{\mu}^{\;\;\;a}\left(x,y\right)\Rightarrow e_{\mu}^{\;\;\;a}\left(x\right)$
(Subsections \ref{subsec:The-curvature}, \ref{subsec:The-co-torsion}).
\end{itemize}

\subsection{Field equations for the twisted action \label{subsec:Field-equations-for}}

The full action is
\begin{eqnarray}
S & = & S_{G}+S_{YM}\nonumber \\
 & = & \lambda\int e_{ac\cdots d}^{\quad\quad be\cdots f}\left(\alpha\boldsymbol{\Omega}_{\;\;\;b}^{a}+\beta\delta_{\;\;\;b}^{a}\boldsymbol{\Omega}+\gamma\mathbf{e}^{a}\wedge\mathbf{f}_{b}\right)\land\mathbf{e}^{c}\land\cdots\land\mathbf{e}^{d}\land\mathbf{f}_{e}\land\cdots\land\mathbf{f}_{f}\nonumber \\
 &  & -\frac{\kappa}{2}\intop tr\overline{\mathcal{F}}\wedge^{*}\mathcal{F}\label{Full action with twist}
\end{eqnarray}
We choose the combinatoric factor $\lambda$ so the final coupling
is $\kappa$.

\subsubsection{Curvatures, Bianchi identities, and gravity variation}

The gravitational field equations are given by varying $S_{G}$ in
Eq.(\ref{Full action with twist}) with respect to all connection
1-forms, then combining with the Yang-Mills sources found in Eq.(\ref{YM sources for gravity}).
The curvature components are given in terms of the connection by the
Cartan structure equations,
\begin{eqnarray}
\mathbf{d}\boldsymbol{\omega}_{\;\;\;b}^{a} & = & \boldsymbol{\omega}_{\;\;\;b}^{c}\boldsymbol{\omega}_{\;\;\;c}^{a}+2\Delta_{db}^{ac}\mathbf{f}_{c}\mathbf{e}^{d}+\boldsymbol{\Omega}_{\;\;\;b}^{a}\label{Curvature}\\
\mathbf{d}\mathbf{e}^{a} & = & \mathbf{e}^{c}\boldsymbol{\omega}_{\;\;\;c}^{a}+\boldsymbol{\omega}\mathbf{e}^{a}+\mathbf{T}{}^{a}\label{Torsion}\\
\mathbf{d}\mathbf{f}_{a} & = & \boldsymbol{\omega}_{\;\;\;a}^{c}\mathbf{f}_{c}+\mathbf{f}_{a}\boldsymbol{\omega}+\mathbf{S}{}_{a}\label{Co-torsion}\\
\mathbf{d}\boldsymbol{\omega} & = & \mathbf{e}^{c}\mathbf{f}_{c}+\boldsymbol{\Omega}\label{Dilatation}
\end{eqnarray}
We also have the integrability conditions--generalized Bianchi identities--which
follow from the Poincarè lemma, $\mathbf{d}^{2}\equiv0$. Exterior
differentiation of the Cartan equations, Eqs.(\ref{Curvature})-(\ref{Dilatation})
yields conditions which must be satisfied in order for solutions to
exist. These take the form
\begin{eqnarray}
\mathbf{D}\boldsymbol{\Omega}_{\;\;\;b}^{a} & = & 2\Delta_{db}^{ac}\mathbf{f}_{c}\mathbf{T}^{d}-2\Delta_{db}^{ac}\mathbf{S}_{c}\mathbf{e}^{d}\label{Curvature Bianchi}\\
\mathbf{D}\mathbf{T}{}^{a} & = & \mathbf{e}^{c}\boldsymbol{\Omega}_{\;\;\;c}^{a}-\boldsymbol{\Omega}\mathbf{e}^{a}\label{Torsion Bianchi}\\
\mathbf{D}\mathbf{S}{}_{a} & = & -\boldsymbol{\Omega}_{\;\;\;a}^{c}\mathbf{f}_{c}+\mathbf{f}_{a}\boldsymbol{\Omega}\label{Co-torsion Bianchi}\\
\mathbf{D}\boldsymbol{\Omega} & = & -\mathbf{T}^{c}\mathbf{f}_{c}+\mathbf{e}^{c}\mathbf{S}_{c}\label{Dilatation Bianchi}
\end{eqnarray}
Conversely, the converse to the Poincarè lemma tells us that in a
star-shaped region the integrability conditions Eqs.(\ref{Curvature Bianchi})-(\ref{Dilatation Bianchi})
imply the original Cartan equations, up to boundary terms.

For example, consider a Newtonian force written as minus the gradient
of a potential, $\mathbf{F}=-\mathbf{d}V$. The Poincarè lemma shows
that this equation has no solution unless
\begin{eqnarray*}
0 & \equiv & \mathbf{d}^{2}V=-\mathbf{d}\mathbf{F}
\end{eqnarray*}
so the force must be curl-free. Conversely, we may solve the this
``Bianchi identity'' first. In a star-shaped region any curl-free
force must be a gradient, $\mathbf{F}=\mathbf{d}U$. This reproduces
the potential up to a ``boundary'' term satisfying
\begin{eqnarray*}
\mathbf{d}\left(U+V\right) & = & 0\\
U+V & = & constant
\end{eqnarray*}
This in turn provides the usual additive constant to the potential.
These principles hold for any set of equations where we may apply
the Poincarè lemma and its converse, allowing us use both the original
equations and their integrability conditions throughout the development
of a solution. In the end, one set of equations cannot be satisfied
without simultaneously satisfying the other up to boundary terms.

The variation of the gravitational part of the action, Eq.(\ref{Full action with twist}),
is discussed in detail in \cite{WheelerAugust}, and involves only
variation of the connection forms, so we simply state the result.
The variation of the spin connnection $\boldsymbol{\omega}_{\;\;\;b}^{a}$
and Weyl vector $\boldsymbol{\omega}$ give
\begin{eqnarray}
T_{\quad e}^{ae}-T_{\quad e}^{ea}-S_{e}^{\;\;ae} & = & 0\label{TFE1}\\
T_{\;\;ca}^{a}+S_{c\quad a}^{\;\;a}-S_{a\quad c}^{\;\;a} & = & 0\label{TFE2}\\
\alpha\Delta_{sb}^{ar}\left(T_{\quad a}^{mb}-\delta_{a}^{m}T_{\quad e}^{eb}-\delta_{a}^{m}S_{c}^{\;\;bc}\right) & = & 0\label{TFE3}\\
\alpha\Delta_{sb}^{ar}\left(\delta_{c}^{b}T_{\;\;ad}^{d}+S_{c\quad a}^{\;\;\;b}-\delta_{c}^{b}S_{d\quad a}^{\;\;\;d}\right) & = & 0\label{TFE4}
\end{eqnarray}
and these acquire no sources since the Yang-Mills action is independent
of the these connection forms, as noted in \cite{WW}. The variation
of the solder and co-solder forms lead to
\begin{eqnarray}
\left[\alpha\left(\Omega_{\;\;\;b\quad m}^{n\quad b}-\Omega_{\;\;\;b\quad a}^{a\quad b}\delta_{m}^{n}\right)+\beta\left(\Omega_{\;\;\;m}^{n}-\Omega_{\;\;\;a}^{a}\delta_{m}^{n}\right)+\Lambda\delta_{m}^{n}\right]A_{\;\;\;n}^{m}\nonumber \\
\left[\alpha\left(\Omega_{\;\;\;n\quad a}^{a\quad m}-\Omega_{\;\;\;b\quad a}^{a\quad b}\delta_{n}^{m}\right)+\beta\left(\Omega_{\;\;\;n}^{m}-\Omega_{\;\;\;a}^{a}\delta_{n}^{m}\right)+\Lambda\delta_{n}^{m}\right]D_{n}^{\;\;\;m}\nonumber \\
-\left[\alpha\Omega_{\;\;\;nam}^{a}+\beta\Omega_{nm}\right]B^{mn}\nonumber \\
\left[\alpha\Omega_{\;\;\;b}^{n\quad bm}+\beta\Omega^{nm}\right]C_{mn}\label{Gravitational variation}
\end{eqnarray}
with the arbitrary variations $A_{\;\;\;n}^{m},B^{mn},C_{mn}$ and
$D_{n}^{\;\;\;m}$ defined in Eqs.(\ref{Solder form variation}) and
(\ref{Co-solder form variation}). In \cite{WheelerAugust} the expressions
above are equated to zero, but they now acquire sources.

\subsubsection{Combined equations}

Equating corresponding parts of Eqs.(\ref{YM sources for gravity})
and Eqs.(\ref{Gravitational variation}) and symmetrizing appropriately,
\begin{eqnarray}
\left[\alpha\left(\Omega_{\;\;\;c\quad b}^{a\quad c}-\Omega_{\;\;\;d\quad c}^{c\quad d}\delta_{b}^{a}\right)+\beta\left(\Omega_{\;\;\;b}^{a}-\Omega_{\;\;\;c}^{c}\delta_{b}^{a}\right)+\Lambda\delta_{b}^{a}\right] & = & \kappa W_{\;\;\;b}^{a}\label{Curv eq 1}\\
\left[\alpha\left(\Omega_{\;\;\;b\quad c}^{c\quad a}-\Omega_{\;\;\;d\quad c}^{c\quad d}\delta_{b}^{a}\right)+\beta\left(\Omega_{\;\;\;b}^{a}-\Omega_{\;\;\;c}^{c}\delta_{b}^{a}\right)+\Lambda\delta_{b}^{a}\right] & = & \kappa W_{\;\;\;b}^{a}\label{Curv eq 2}\\
-\left[\alpha\Omega_{\;\;\;bca}^{c}+\beta\Omega_{ba}\right] & = & \kappa T_{ab}\label{Ricci eq 1}\\
\left[\alpha\Omega_{\;\;\;c}^{b\quad ca}+\beta\Omega^{ba}\right] & = & \kappa S^{ab}\label{Ricci eq 2}
\end{eqnarray}
where, recalling the antisymmetry of $\mathcal{G}^{ab}$,
\begin{eqnarray}
T_{ab} & \equiv & tr\left(F_{ac}F_{bd}\eta^{cd}+H_{ac}F_{bd}\eta^{cd}+H_{bc}F_{ad}\eta^{cd}+3\eta^{cd}\mathcal{G}_{ac}\mathcal{G}_{bd}\right)\label{Definition of Tab}\\
S^{ab} & \equiv & \eta^{bd}\eta^{ac}tr\left(\eta^{ef}H_{ec}H_{fd}+\eta^{ef}H_{ec}F_{fd}+\eta^{ef}H_{ed}F_{fc}+3\eta^{fe}\mathcal{G}_{ec}\mathcal{G}_{fd}\right)\label{Definition of Sab}\\
W_{\;\;\;b}^{a} & \equiv & tr\left(3\mathcal{G}^{ca}F_{cb}+3\mathcal{G}^{ca}\mathcal{H}_{cb}-\left(F_{cd}+\mathcal{H}_{cd}\right)\mathcal{G}^{cd}\delta_{\;\;\;b}^{a}\right)\label{Definition of Wab}
\end{eqnarray}
In addition, we have the field equations for the $U\left(1\right)$
field,
\begin{eqnarray}
0 & = & \eta^{bc}K_{k\;ac}^{\;\;\quad;a}+\eta^{bc}K_{i\;ac}\beta_{\;\;\;k}^{i\;\;\quad a}+\mathcal{G}_{k\;\;\quad;a}^{\;\;\;ab}+\mathcal{G}_{i}^{\;\;\;ab}\alpha_{\;\;\;ka}^{i}\label{YM1}\\
0 & = & \eta^{bc}\left(K_{j\;ac;b}+K_{i\;ac}\alpha_{\;\;\;jb}^{i}+K_{j\;ec}S_{a\quad\;\;b}^{\;\;\;e}\right)\nonumber \\
 &  & +\mathcal{G}_{j\;ab}^{\;\;\quad;b}+\mathcal{G}_{i\;ab}\beta_{\;\;\;j}^{i\;\;\quad b}+\frac{1}{2}\mathcal{G}_{j\;bc}S{}_{a}^{\;\;\;bc}+\frac{1}{2}\mathcal{G}_{j}^{\;\;\;bc}S_{abc}\label{YM2}
\end{eqnarray}
Our gravitational solution now follows many of the steps presented
in detail in \cite{WheelerAugust}.

\subsection{Solving the field equations for the twisted action}

For the remainder of the gravitational solution, the particular forms
of $T_{ab},S^{ab}$ and $W_{\;\;\;b}^{a}$ make little difference;
indeed, the solution of this Section holds for the metric variation
of any sources at all. While the form of these source tensors varies
with the fields and the matter action, the positions in which they
occur in the field equations and their symmetries follow knowing only
the variation of $\bar{K}^{AB}$. For the present, this is all we
need.

We first turn to the consequences of vanishing torsion, $\mathbf{T}^{a}=0$.

\subsubsection{Vanishing torsion \label{subsec:Vanishing-torsion}}

Similarly to general relativity, with vanishing torsion the torsion
Bianchi identity, Eq.(\ref{Torsion Bianchi}), simplifies to an algebraic
relation,
\begin{eqnarray*}
0 & = & \mathbf{e}^{c}\boldsymbol{\Omega}_{\;\;\;c}^{a}-\boldsymbol{\Omega}\mathbf{e}^{a}
\end{eqnarray*}
which must hold independently of any sources. The algebraic condition
$\mathbf{e}^{c}\boldsymbol{\Omega}_{\;\;\;c}^{a}=\boldsymbol{\Omega}\mathbf{e}^{c}$
expands to three independent component equations,
\begin{eqnarray}
\Omega_{\;\;\left[bcd\right]}^{a} & = & \delta_{[b}^{a}\Omega_{cd]}\label{Vanishing torsion effect on curvature 1}\\
\Omega_{\;\;b\quad d}^{a\quad c}-\Omega_{\;\;d\quad b}^{a\quad c} & = & \delta_{b}^{a}\Omega_{\quad d}^{c}-\delta_{d}^{a}\Omega_{\quad b}^{c}\label{Vanishing torsion effect on curvature 2}\\
\Omega_{\;\;\;b}^{a\quad cd} & = & \delta_{b}^{a}\Omega^{cd}\label{Vanishing torsion effect on curvature 3}
\end{eqnarray}
Since $\eta_{ea}\Omega_{\;\;b}^{a\quad cd}=-\eta_{ba}\Omega_{\;\;e}^{a\quad cd}$,
the $ab$ trace of Eq.(\ref{Vanishing torsion effect on curvature 3})
leaves $\Omega^{cd}=0$. Therefore each term vanishes separately,
\begin{eqnarray}
\Omega_{\;\;b}^{a\quad cd} & = & 0\label{Vanishing momentum curvature}\\
\Omega^{cd} & = & 0\label{Vanishing momentum dilatation}
\end{eqnarray}
The $ad$ contraction of Eq.(\ref{Vanishing torsion effect on curvature 2})
gives
\begin{equation}
\Omega_{\;\;b\quad a}^{a\quad c}=-\left(n-1\right)\Omega_{\quad b}^{c}\label{Cross-curvature trace}
\end{equation}

Combining Eqs.(\ref{Vanishing momentum curvature}) and (\ref{Vanishing momentum dilatation})
with Eq.(\ref{Ricci eq 2}) we immediately find that the gravitational
fields force a constraint on the source fields. This is our \emph{first
source constraint}:
\begin{equation}
S^{ab}=0\label{First source constraint}
\end{equation}
We next look at the field equations for the curvature.

\subsubsection{Curvature equations}

We now combine the vanishing torsion simplifications with the curvature
and dilatation field equations, Eqs.(\ref{Curv eq 1}) and (\ref{Curv eq 2}).
The reduction of these equations begins by noting that the difference
between Eq.(\ref{Curv eq 1}) and Eq.(\ref{Curv eq 2}) immediately
gives equality of the traces,
\begin{eqnarray}
\Omega_{\;\;\;b\quad c}^{c\quad a} & = & \Omega_{\;\;\;c\quad b}^{a\quad c}\label{trace symmetry}
\end{eqnarray}

Next, formally lowering an index in Eq.(\ref{Vanishing torsion effect on curvature 2})
\[
\eta_{ea}\Omega_{\;\;b\quad d}^{a\quad c}-\eta_{ea}\Omega_{\;\;d\quad b}^{a\quad c}=\eta_{eb}\Omega_{\quad d}^{c}-\eta_{ed}\Omega_{\quad b}^{c}
\]
we cycle $ebd$, then add the first two and subtract the third. Using
the the antisymmetry of the curvature on the first two indices, $\eta_{ea}\Omega_{\;\;d\quad b}^{a\quad c}=-\eta_{da}\Omega_{\;\;e\quad b}^{a\quad c}$
we find
\begin{eqnarray}
\Omega_{\;\;e\quad d}^{a\quad c} & = & -2\Delta_{de}^{ab}\Omega_{\quad b}^{c}\label{Cross-curvature in terms of dilatation}
\end{eqnarray}

Substituting Eq.(\ref{Cross-curvature in terms of dilatation}) into
the trace symmetry, Eq.(\ref{trace symmetry}) to the two contractions
of Eq.(\ref{Cross-curvature in terms of dilatation}) constrains the
cross-dilatation,
\begin{eqnarray*}
-\delta_{d}^{a}\Omega_{\quad c}^{c}+\eta^{ae}\eta_{cd}\Omega_{\quad e}^{c} & = & -\left(n-1\right)\Omega_{\quad d}^{a}
\end{eqnarray*}
Contracting with $\eta_{ba}$, we see that the antisymmetric part
vanishes,
\begin{eqnarray*}
\left(n-2\right)\left(\eta_{bc}\Omega_{\quad d}^{c}-\eta_{cd}\Omega_{\quad b}^{c}\right) & = & 0
\end{eqnarray*}
in dimensions greater than 2, while an explicit check confirms the
vanishing antisymmetry in 2-dimensions as well. Therefore, the symmetric
part, $\eta_{bc}\Omega_{\quad d}^{c}+\eta_{dc}\Omega_{\quad b}^{c}=\frac{2}{n}\eta_{bd}\Omega_{\quad c}^{c}$,
becomes a solution for the full cross-dilatation in terms of its trace,
\begin{equation}
\Omega_{\quad d}^{c}=\frac{1}{n}\delta_{b}^{a}\Omega_{\quad c}^{c}\label{Cross-dilatation in terms of its trace}
\end{equation}
This, in turn, combines with Eq.(\ref{Cross-curvature in terms of dilatation})
to give the cross-curvature in terms of the trace of the dilatation,
\begin{eqnarray}
\Omega_{\;\;b\quad d}^{a\quad c} & = & -\frac{2}{n}\Delta_{db}^{ac}\Omega_{\quad e}^{e}\label{Cross-curvature in terms of the dilatation trace}
\end{eqnarray}

We have one remaining cross-curvature field equation, Eq.(\ref{Curv eq 1}),
which couples the cross-dilatation trace, $\Omega_{\;\;\;a}^{a}$,
to the Yang-Mills source fields. Using Eqs.(\ref{Cross-dilatation in terms of its trace})
and (\ref{Cross-curvature in terms of the dilatation trace}) to replace
the cross-curvature and the cross-dilatation in Eq.(\ref{Curv eq 1}),
and simplifying,
\begin{eqnarray*}
\frac{1}{n}\left(n-1\right)\left[\left(\left(n-1\right)\alpha-\beta\right)\Omega_{\quad c}^{c}+n\Lambda\right]\delta_{b}^{a} & = & W_{\;\;\;b}^{a}
\end{eqnarray*}
so that $W_{\;\;\;b}^{a}=f\delta_{b}^{a}$ for some function $f$.
The constant $\Lambda$ is given by $\Lambda=\left(n-1\right)\alpha-\beta+n^{2}\gamma$.
Contracting then substituting back, the gravitational field equations
force a \emph{second source constraint}:
\begin{eqnarray}
W_{\;\;\;b}^{a} & = & \frac{1}{n}W_{\;\;\;c}^{c}\delta_{b}^{a}\label{Second source constraint}
\end{eqnarray}
where
\begin{eqnarray*}
W_{\;\;\;c}^{c} & = & \left(n-1\right)\left[\left(\left(n-1\right)\alpha-\beta\right)\Omega_{\quad c}^{c}+n\Lambda\right]
\end{eqnarray*}
The traced source tensor on the right, $W_{\;\;\;c}^{c}$, therefore
drives the entire cross-curvature and cross-dilatation. It is striking
that the only source dependence for these components is the Yang-Mills
Lagrangian density,
\begin{eqnarray*}
W_{\;\;\;a}^{a} & = & 3\mathcal{G}^{i\;ac}\left(\mathcal{H}_{i\;ac}+F_{i\;ac}\right)=3\mathcal{L}
\end{eqnarray*}

\subsubsection{Spacetime terms}

Finally, we combine the remaining field equation, Eq.(\ref{Ricci eq 1}),
\[
\alpha\Omega_{\;\;\;bca}^{c}+\beta\Omega_{ba}=-\kappa T_{ab}
\]
and the corresponding part of the vanishing torsion Bianchi, Eq.(\ref{Vanishing torsion effect on curvature 1}),
which expanded becomes
\[
\Omega_{\;\;bcd}^{a}+\Omega_{\;\;cdb}^{a}+\Omega_{\;\;dbc}^{a}=\delta_{b}^{a}\Omega_{cd}+\delta_{c}^{a}\Omega_{db}+\delta_{d}^{a}\Omega_{bc}
\]
The $ac$ trace reduces this to
\[
\Omega_{\;\;bcd}^{c}-\Omega_{\;\;dcb}^{c}=-\left(n-2\right)\Omega_{bd}
\]
Combining this with the antisymmetric part of the field equation,
$\alpha\left(\Omega_{\;\;\;nam}^{a}-\Omega_{\;\;\;man}^{a}\right)=-2\beta\Omega_{nm}$
shows that
\begin{eqnarray*}
\left(\left(n-2\right)\alpha-2\beta\right)\Omega_{ab} & = & 0
\end{eqnarray*}
so that generically (i.e., unless $\left(n-2\right)\alpha=2\beta$),
the spacetime dilatation vanishes. Note that this is true for any
symmetric source tensor, so spacetime dilatation is never driven by
ordinary matter. As a result,
\begin{eqnarray}
\Omega_{\;\;\;acb}^{c}=\Omega_{\;\;\;\left(a\left|c\right|b\right)}^{c} & = & -\frac{\kappa}{\alpha}T_{ab}\label{Spacetime curvature equation}\\
\Omega_{nm} & = & 0\label{Vanishing spacetime dilatation}
\end{eqnarray}

\subsubsection{Dilatation \label{subsec:Dilatation}}

Having reduced the dilatational curvature to a single function,
\begin{eqnarray}
\boldsymbol{\Omega} & = & \chi\mathbf{e}^{a}\wedge\mathbf{f}_{a}\label{Fully reduced dilatation}
\end{eqnarray}
where $\chi\equiv-\frac{1}{n}\Omega_{\;\;\;a}^{a}$, we can now use
the dilatational integrability condition, Eq.(\ref{Dilatation Bianchi}),
to press further. Substituting Eq.(\ref{Fully reduced dilatation})
into the Bianchi identity, Eq.(\ref{Dilatation Bianchi}),
\begin{eqnarray*}
0 & = & \mathbf{d}\boldsymbol{\Omega}-\mathbf{e}^{b}\wedge\mathbf{S}_{b}\\
 & = & \mathbf{d}\chi\wedge\mathbf{e}^{a}\wedge\mathbf{f}_{a}-\left(1+\chi\right)\mathbf{e}^{a}\wedge\mathbf{S}_{a}
\end{eqnarray*}
where we have used $\mathbf{d}\left(\mathbf{e}^{a}\wedge\mathbf{f}_{a}\right)=\mathbf{D}\left(\mathbf{e}^{a}\wedge\mathbf{f}_{a}\right)=\cancel{\mathbf{T}^{a}}\wedge\mathbf{f}_{a}-\mathbf{e}^{a}\wedge\mathbf{S}_{a}$.
Setting $\mathbf{d}\chi=\chi_{c}\mathbf{e}^{c}+\chi^{c}\mathbf{f}_{c}$,
expanding the co-torsion into components, and combining like forms
yields three independent equations,

\begin{eqnarray}
\left(1+\chi\right)S_{\left[acd\right]} & = & 0\label{eee dilatation  Bianchi}\\
\left(1+\chi\right)\left(S_{c\;\;\quad d}^{\quad a}-S_{d\;\;\quad c}^{\quad a}\right) & = & \chi_{d}\delta_{c}^{a}-\chi_{c}\delta_{d}^{a}\label{eef dilatation Bianchi}\\
\left(1+\chi\right)S_{a}^{\quad cd} & = & \chi^{d}\delta_{a}^{c}-\chi^{c}\delta_{a}^{d}\label{eff dilatation Bianchi}
\end{eqnarray}
We may now use the co-torsion field equations to gain insight into
$\chi$.

With vanishing torsion, the co-torsion field equations Eqs.(\ref{TFE1})-(\ref{TFE4})
reduce to
\begin{eqnarray}
S_{e}^{\;\;ae} & = & 0\label{Co-torsion FE1}\\
S_{c\quad a}^{\;\;a}-S_{a\quad c}^{\;\;a} & = & 0\label{Co-torsion FE2}\\
\alpha\Delta_{sb}^{ar}\left(S_{c\quad a}^{\;\;\;b}-\delta_{c}^{b}S_{d\quad a}^{\;\;\;d}\right) & = & 0\label{Co-torsion FE3}
\end{eqnarray}
Using the field equation Eq.(\ref{Co-torsion FE2}) to replace the
co-torsion terms in the trace of the Bianchi identity, Eq.(\ref{eef dilatation Bianchi}),
gives
\begin{eqnarray*}
\left(1+\chi\right)\left(S_{c\;\;\quad a}^{\quad a}-S_{a\;\;\quad c}^{\quad a}\right) & = & -\left(n-1\right)\chi_{c}\\
\chi_{c} & = & 0
\end{eqnarray*}
Then, combining Eq.(\ref{Co-torsion FE1}) with the $ad$ trace of
Eq.(\ref{eff dilatation Bianchi}),
\begin{eqnarray*}
\left(1+\chi\right)S_{a}^{\quad ca} & = & -\left(n-1\right)\chi^{c}\\
\chi^{c} & = & 0
\end{eqnarray*}
and therefore,
\begin{equation}
\mathbf{d}\boldsymbol{\chi}=0\label{Constancy of chi}
\end{equation}
The dilatation therefore takes the form
\[
\boldsymbol{\Omega}=\chi\mathbf{e}^{a}\wedge\mathbf{f}_{a}
\]
with $\chi$ \emph{constant}. The remainder of the development of
the solution continues as in the homogeneous case but with a different
constant value,
\begin{eqnarray*}
\chi & = & \frac{1}{\left(n-1\right)\alpha-\beta}\left(\frac{1}{n-1}\Lambda-\frac{\kappa}{n\left(n-1\right)}W_{\;\;\;c}^{c}\right)
\end{eqnarray*}
for the magnitude of the dilatation cross-term.

Importantly, the constancy of $\chi$ implies the constancy of $W_{\;\;\;c}^{c}$,
and via the second source constraint, Eq.(\ref{Second source constraint}),
the (mostly zero) constancy of all of $W_{\;\;\;b}^{a}$.

\subsubsection{The Frobenius theorem and the final reduction \label{subsec:The-Frobenius-theorem}}

With vanishing torsion, \ref{Torsion} shows that the solder form
becomes involute, and we may write
\begin{eqnarray*}
\mathbf{e}^{a} & = & e_{\alpha}^{\;\;\;a}\mathbf{d}x^{\alpha}
\end{eqnarray*}
where $x^{\alpha}$ comprise $n$ of the $2n$ coordinates. Holding
$x^{\alpha}$ constant, $\mathbf{e}^{a}=0$, and the residual field
equations describe a submanifold. Here the discussion exactly parallels
that of \cite{WheelerAugust}:
\begin{enumerate}
\item Solve for the connection on the $\mathbf{e}^{a}=0$ submanifold. Here,
because the curvature and dilatation vanish (see Eqs.(\ref{Vanishing momentum curvature})
and (\ref{Vanishing momentum dilatation})), we may gauge the restricted
components of the spin connection and Weyl vector to zero. A careful
coordinate choice puts the submanifold basis in the form $\mathbf{h}_{a}=e_{a}^{\;\;\;\mu}\mathbf{d}y_{\mu}$.
\item Now let $x^{\alpha}$ vary, extending the solution back to the full
biconformal space. This allows all connection forms to acquire an
additional $\mathbf{d}x^{\alpha}$ or $\mathbf{e}^{a}$ term,
\begin{eqnarray}
\boldsymbol{\omega}_{\;\;\;b}^{a} & = & \boldsymbol{\omega}_{\;\;\;bc}^{a}\mathbf{e}^{c}\nonumber \\
\mathbf{e}^{a} & = & e_{\alpha}^{\;\;\;a}\mathbf{d}x^{\alpha}\nonumber \\
\mathbf{f}_{a} & = & e_{a}^{\;\;\;\mu}\mathbf{d}y_{\mu}+c_{ab}\mathbf{e}^{b}\nonumber \\
\boldsymbol{\omega} & = & W_{a}\mathbf{e}^{a}\label{Reduced connection forms}
\end{eqnarray}
\item Note that Eq.(\ref{Torsion}) is now purely quadratic in $\mathbf{e}^{a}$,
and therefore requires the coefficients to depend only on the $x$-coordinates,
$e_{\alpha}^{\;\;\;a}=e_{\alpha}^{\;\;\;a}\left(x\right)$. Solving
for the connection separates it into a compatible piece and a Weyl
vector piece,
\[
\boldsymbol{\omega}_{\;\;\;b}^{a}=\boldsymbol{\alpha}_{\;\;\;b}^{a}-2\Delta_{db}^{ac}W_{c}\mathbf{e}^{d}
\]
where $\mathbf{d}\mathbf{e}^{a}=\mathbf{e}^{b}\boldsymbol{\alpha}_{\;\;\;b}^{a}$.
\item Substitute these reduced forms of $\mathbf{e}^{a},\mathbf{f}_{b}$
into the dilatation, Eq.(\ref{Dilatation}) and solve for the Weyl
vector. This yields
\[
\boldsymbol{\omega}=-\left(1+\chi\right)y_{a}\mathbf{e}^{a}
\]
where $y_{a}=e_{a}^{\;\;\;\mu}y_{\mu}$.
\end{enumerate}
These steps give the final expressions for the connection forms, except
for the form of $c_{ab}=c_{ba}$ in the expansion of the co-solder
form $\mathbf{f}_{a}$.

We note that the submanifolds found by setting either $\mathbf{e}^{a}=0$
by holding $x^{\mu}$ constant, or $\mathbf{h}_{a}=0$ by holding
$y_{\mu}$ constant are Lagrangian submanifolds.

\subsubsection{The curvature \label{subsec:The-curvature}}

The final steps in the gravitational reduction are to substitute the
partial solution for the connection forms Eq.(\ref{Reduced connection forms})
into Eqs.(\ref{Curvature}) and (\ref{Co-torsion}) to impose the
final field equations.

To express the remaining undetermined component of the curvature,
$\Omega_{\;\;\;bcd}^{a}$, we define the Schouten tensor
\begin{eqnarray*}
\boldsymbol{\mathcal{R}}_{a} & = & \frac{1}{n-2}\left(R_{ab}-\frac{1}{2\left(n-1\right)}\eta_{ab}R\right)\mathbf{e}^{b}
\end{eqnarray*}
where $R_{ab}=\left(n-2\right)\mathcal{R}_{ab}+\eta_{ab}\mathcal{R}$
is the Ricci tensor. The generalization of the Schouten tensor to
an integrable Weyl geometry is then (see \cite{WeylGeom})
\begin{eqnarray*}
\boldsymbol{\mathscr{R}}_{a} & = & \boldsymbol{\mathcal{R}}_{a}+\mathbf{D}_{\left(\alpha,x\right)}W_{a}+W_{a}\boldsymbol{\omega}-\frac{1}{2}\eta_{ab}W^{2}\mathbf{e}^{b}
\end{eqnarray*}
In terms of the Schouten tensor, the decomposition of the Riemann
curvature 2-form into the Weyl curvature 2-form and trace parts is
\[
\mathbf{R}_{\;\;\;b}^{a}=\mathbf{C}_{\;\;\;b}^{a}-2\Delta_{db}^{ac}\mathcal{\boldsymbol{R}}_{c}\wedge\mathbf{e}^{d}
\]
Because of the manifest involution of $\mathbf{h}_{a}=e_{a}^{\;\;\;\mu}\left(x\right)\mathbf{d}y_{\mu}$,
the subspace spanned by the solder form, $\mathbf{e}^{a}$, is a submanifold.
Because $\Omega_{ab}=0$ the submanifold geometry is always an \emph{integrable}
Weyl geometry, so the Weyl vector may be removed from the spacetime
submanifold by a gauge transformation. The spacetime submanifold is
simply a Riemannian geometry with local scale invariance.

Now, introducing the reduced form of the connection into Eq.(\ref{Curvature})
and imposing the corresponding field equation, Eq,(\ref{Ricci eq 1})
shows that
\begin{eqnarray}
\frac{1}{1+\chi}\mathscr{R}_{ab}+c_{ab} & = & -\frac{\kappa}{\alpha}T_{ab}\label{Inhomogeneous solution for cab}
\end{eqnarray}
with the full spacetime component of the biconformal curvature given
by the Weyl curvature, $\Omega_{\;\;\;bcd}^{a}=C_{\;\;\;bcd}^{a}$.

\subsubsection{\label{subsec:The-co-torsion}The co-torsion}

A similar introduction of the reduced connection into Eq.(\ref{Co-torsion})
for the co-torsion shows that the momentum and cross-terms vanish,
while (following the somewhat intricate calculation of \cite{WheelerAugust})
the remaining component is given by
\begin{eqnarray}
\mathbf{S}_{a} & = & \mathbf{d}_{\left(x\right)}\mathbf{c}_{a}+\mathbf{c}_{b}\wedge\boldsymbol{\omega}_{\;\;a}^{b}+\boldsymbol{\omega}\wedge\mathbf{c}_{a}\label{Co-torsion as Dc}
\end{eqnarray}
where $\mathbf{c}_{a}$, in turn, is determined by Eq.(\ref{Inhomogeneous solution for cab}).

Expanding $\mathbf{c}_{a}$ fully to separate the Weyl vector parts,
\begin{eqnarray*}
\mathbf{c}_{a} & = & -\frac{1}{1+\chi}\left(\boldsymbol{\mathcal{R}}_{a}+\mathbf{D}_{\left(\alpha,x\right)}W_{a}+W_{a}\boldsymbol{\omega}-\frac{1}{2}\eta_{ab}W^{2}\mathbf{e}^{b}\right)-\frac{\kappa}{\alpha}\mathbf{T}_{a}\\
 & = & \mathbf{b}_{a}-\frac{1}{1+\chi}\left(\mathbf{D}_{\left(\alpha,x\right)}W_{a}+W_{a}\boldsymbol{\omega}-\frac{1}{2}\eta_{ab}W^{2}\mathbf{e}^{b}\right)
\end{eqnarray*}
where $\mathbf{T}_{a}=T_{ab}\mathbf{e}^{b}$ is the remaining source
field and $\mathbf{b}_{a}\equiv-\frac{1}{1+\chi}\boldsymbol{\mathcal{R}}_{a}-\frac{\kappa}{\alpha}\mathbf{T}_{a}$.
Then substituting into Eq.(\ref{Co-torsion as Dc}), after multiple
cancellations the co-torsion becomes
\begin{eqnarray}
\mathbf{S}_{a} & = & \frac{1}{1+\chi}W_{b}\mathbf{R}{}_{\;\;\;a}^{b}-\mathbf{D}_{\left(\alpha,x\right)}\left(\frac{1}{1+\chi}\boldsymbol{\mathcal{R}}_{a}+\frac{\kappa}{\alpha}\mathbf{T}_{a}\right)\nonumber \\
 &  & +2\Delta_{ca}^{db}W_{d}\left(\frac{1}{1+\chi}\boldsymbol{\mathcal{R}}_{b}+\frac{\kappa}{\alpha}\mathbf{T}_{b}\right)\wedge\mathbf{e}^{c}\label{Final co-torsion}
\end{eqnarray}
with the cross term and momentum term of the co-torsion vanishing.

This result is quite similar to an integrability condition. It is
shown in \cite{WeylGeom} that the condition for the existence of
a conformal gauge in which the Einstein equation, $G_{ab}=\kappa T_{ab}$,
holds is
\begin{equation}
0=\phi_{,b}\mathbf{R}{}_{\;\;\;a}^{b}-\mathbf{D}_{\left(\alpha,x\right)}\left(\boldsymbol{\mathcal{R}}_{a}-\kappa\boldsymbol{\mathcal{T}}_{a}\right)+2\Delta_{ca}^{db}\phi_{,d}\left(\boldsymbol{\mathcal{R}}_{b}-\kappa\boldsymbol{\mathcal{T}}_{b}\right)\wedge\mathbf{e}^{c}\label{Integrability for conformal Einstein}
\end{equation}
where
\[
\boldsymbol{\mathcal{T}}_{a}=\frac{1}{n-2}\left(T_{ab}-\frac{1}{n-1}T\eta_{ab}\right)
\]
When $\boldsymbol{\mathcal{T}}_{a}=0$, Eq.(\ref{Integrability for conformal Einstein})
reduces to the well-known condition, $\mathbf{D}_{\left(\alpha,x\right)}\boldsymbol{\mathcal{R}}_{a}-\varphi_{,b}\mathbf{C}_{\;\;\;a}^{b}=0$,
for the existence of a Ricci flat conformal gauge.

There are two differences between Eq.(\ref{Final co-torsion}) and
Eq.(\ref{Integrability for conformal Einstein}). First, the co-torsion
on the left hand side of Eq.(\ref{Final co-torsion}) obstructs the
integrability condition, Eq.(\ref{Integrability for conformal Einstein}),
and we cannot set $\mathbf{S}_{a}=0$ because the Triviality Theorem
shows that when both torsion and co-torsion vanish, biconformal space
must be trivial. The second difference is that the Weyl vector on
the right is not integrable on the \emph{full} biconformal space.

These issues have a common solution. The part of structure equation
for the co-solder form involving $\mathbf{h}_{a}$ is
\begin{eqnarray}
\mathbf{d}\mathbf{h}_{a} & = & \boldsymbol{\omega}_{\;\;\;a}^{c}\land\mathbf{h}_{c}+\mathbf{h}_{a}\land\boldsymbol{\omega}\label{Structure equation for h}
\end{eqnarray}
Therefore, as briefly noted above, $\mathbf{h}_{a}=e_{a}^{\;\;\;\mu}\mathbf{d}y_{\mu}$
is in involution. Holding $y_{\mu}=y_{\mu}^{0}$ constant shows that
$\mathbf{e}^{a}$ spans a submanifold. On that submanifold, the Weyl
vector becomes exact,
\[
\boldsymbol{\omega}=W_{a}\mathbf{e}^{a}=\mathbf{d}\left(y_{\mu}^{0}x^{\mu}\right)
\]
This means that on the $y_{\mu}=y_{\mu}^{0}$ spacetime submanifold,
the right side takes the form of the integrability condition.

At the same time, we may use the form of $\mathbf{S}_{a}$ as the
covariant derivative of $\mathbf{c}_{a}$, Eq.(\ref{Co-torsion as Dc})
with a suitable choice of $c_{ab}$. In \cite{WheelerAugust} it is
shown that $c_{ab}$ is symmetric and divergence free. While the interpretation
given in \cite{WheelerAugust} of a phenomenological energy tensor
is consistent, it is at odds with the more fundamental interpretation
of sources given here. Instead, we identify $c_{ab}$ as proportional
to the Minkowski metric--the only invariant, symmetric tensor available.
It is also divergence free with respect to the compatible connection,
since $D_{\left(\alpha,x\right)}\eta_{ab}=0$. However, as noted in
the previous Section the fully biconformal-covariant derivative of
$\eta_{ab}$ does not necessarily vanish. Since by Eq.(\ref{Co-torsion as Dc})
the co-torsion is given by the full biconformal derivative of $c_{ab}$,
the identification $c_{ab}=\Lambda_{0}\eta_{ab}$, implies
\[
\mathbf{S}_{a}=2\left(1+\chi\right)\Delta_{ea}^{bc}y_{b}\Lambda_{0}\eta_{cd}\mathbf{e}^{d}\wedge\mathbf{e}^{e}
\]
thereby avoiding the Triviality Theorem. This residual form of the
co-torsion may now be combined into the right hand side of Eq.(\ref{Final co-torsion}).

Combining these observations, on the $\mathbf{h}_{a}=0$ spacetime
submanifold with $c_{ab}=\Lambda_{0}\eta_{ab}$, setting $\phi_{,a}=y_{a}^{0}$,
and using $\mathbf{D}_{\left(\alpha,x\right)}\left(\Lambda_{0}\eta_{ab}\right)=0$,
it follows that
\begin{eqnarray}
0 & = & \frac{1}{1+\chi}\phi_{,b}\mathbf{R}{}_{\;\;\;a}^{b}-\mathbf{D}_{\left(\alpha,x\right)}\left(\frac{1}{1+\chi}\boldsymbol{\mathcal{R}}_{a}+\frac{\kappa}{\alpha}\mathbf{T}_{a}+\Lambda_{0}\eta_{ab}\mathbf{e}^{b}\right)\nonumber \\
 &  & +2\Delta_{ca}^{db}\phi_{,d}\left(\frac{1}{1+\chi}\boldsymbol{\mathcal{R}}_{b}+\frac{\kappa}{\alpha}\mathbf{T}_{b}+\Lambda_{0}\eta_{be}\mathbf{e}^{e}\right)\wedge\mathbf{e}^{c}\label{Integrability}
\end{eqnarray}
This is now the condition for the existence of a conformal transformation
such that
\begin{eqnarray*}
\frac{1}{1+\chi}\boldsymbol{\mathcal{R}}_{b}+\Lambda_{0}\eta_{be}\mathbf{e}^{e} & = & -\frac{\kappa}{\alpha}\mathbf{T}_{b}
\end{eqnarray*}
Expressed in terms of the Einstein tensor,
\begin{eqnarray}
G_{ab}+\Lambda_{C}\eta_{ab} & = & -\left(n-2\right)\frac{\kappa\left(1+\chi\right)}{\alpha}\left(T_{ab}-\eta_{ab}T\right)\label{Einstein equation}
\end{eqnarray}
where the net effect of $c_{ab}$ is a cosmological constant, $\Lambda_{C}=-\left(n-1\right)\left(n-2\right)\left(1+\chi\right)\Lambda_{0}$.

If we make the conformal transformation that produces Eq.(\ref{Einstein equation}),
the co-torsion equation (\ref{Integrability}) reduces to $\phi_{,b}\mathbf{R}{}_{\;\;a}^{b}=0$.
In generic spacetimes this requires $W_{\mu}=y_{\mu}^{0}=0$.

\section{The source for gravity\label{sec:The-source-for}}

\subsection{The reduction of sources forced by coupling to gravity}

The necessary source constraints from the gravitational couplings,
Eqs.(\ref{First source constraint}) and (\ref{Second source constraint}),
may be written as
\begin{eqnarray}
0 & = & \mathcal{H}_{\;\;ac}^{i}\mathcal{H}_{i\:bd}\eta^{cd}+\left(\mathcal{H}_{\;\;ac}^{i}F_{i\:bd}+\mathcal{H}_{\;\;bc}^{i}F_{i\:ad}\right)\eta^{cd}+3\mathcal{G}_{\;\;ac}^{i}\mathcal{G}_{i\:bd}\eta^{cd}\label{S equation}\\
0 & = & \mathcal{G}^{i\;ca}\left(F_{\;\;cb}^{i}+\mathcal{H}_{\;\;cb}^{i}\right)-\frac{1}{n}\delta_{\;\;\;b}^{a}\mathcal{G}^{i\;cd}\left(F_{\;\;cd}^{i}+\mathcal{H}_{\;\;cd}^{i}\right)\label{W equation}
\end{eqnarray}
where the full contraction of the second, $\mathcal{G}^{i\;cd}\left(F_{\;\;cd}^{i}+\mathcal{H}_{\;\;cd}^{i}\right)$,
is constant.

These conditions must continue to hold for small physical variations
of the independent potentials, $A_{\;\;\;a}^{i}$ and $B^{ia}$. We
may imagine two nearby solutions differing only in one or both of
the potentials and look at their difference. The change in $F_{\;\;\;ab}^{i}$
as we change $A_{\;\;\;a}^{i}$ is given by
\begin{eqnarray*}
\delta F_{\;\;\;ab}^{i} & = & \delta A_{\;\;\;b;a}^{i}-\delta A_{\;\;\;a;b}^{i}-c_{\;\;\;jk}^{i}\delta A_{\;\;\;a}^{j}A_{\;\;\;b}^{k}-c_{\;\;\;jk}^{i}A_{\;\;\;a}^{j}\delta A_{\;\;\;b}^{k}\\
 & = & \left(\delta A_{\;\;\;b;a}^{i}-\alpha_{\;\;\;ka}^{i}\delta A_{\;\;\;b}^{k}\right)-\left(\delta A_{\;\;\;a;b}^{i}-\alpha_{\;\;\;jb}^{i}\delta A_{\;\;\;a}^{j}\right)\\
 & = & \mathcal{D}_{a}\left(\delta A_{\;\;\;b}^{i}\right)-\mathcal{D}_{b}\left(\delta A_{\;\;\;a}^{i}\right)
\end{eqnarray*}
where $\mathcal{D}_{a}$ is covariant with respect to local Lorentz,
dilatational and $SU\left(N\right)$ transformations. Similarly we
find for $G_{\;\;\quad b}^{i\;a}$ and $H^{i\;\;ab}$,
\begin{eqnarray*}
\delta_{A}G_{\;\;\quad b}^{i\;a} & = & \mathcal{D}^{a}\left(\delta A_{\;\;\;b}^{i}\right)\\
\delta_{A}H^{i\;\;ab} & = & 0
\end{eqnarray*}
Of course, under changes of gauge, these fields are invariant.

The conditions (\ref{S equation}) and (\ref{W equation}) must continue
to hold throughout such small changes. Substituting these variations
into the first constraint,
\begin{equation}
0=\delta F_{\;\:ed}^{i}\left(\mathcal{H}_{i\;ac}\delta_{b}^{e}+\mathcal{H}_{i\;bc}\delta_{a}^{e}\right)\eta^{cd}+3\left(\eta_{ae}\delta G_{\;\;\;\;c}^{i\;e}\mathcal{G}_{i\:bd}+\mathcal{G}_{i\;ac}\eta_{be}\delta G_{\;\;\;\;d}^{i\;e}\right)\eta^{cd}\label{Variation of A}
\end{equation}
The first term of Eq.(\ref{S equation}) has dropped out because $H^{i\;\;ab}$
is independent of $A_{\;\;\;a}^{i}$.

Now we expand Eq.(\ref{Variation of A}) in terms of the variation
$\delta A_{\;\;\;a}^{i}$ and its derivatives,
\begin{eqnarray*}
0 & = & \delta A_{\;\;\;d,e}^{i}\left(\mathcal{H}_{i\;ac}\delta_{b}^{e}\eta^{cd}+\mathcal{H}_{i\;bc}\delta_{a}^{e}\eta^{cd}-\mathcal{H}_{i\;ac}\delta_{b}^{d}\eta^{ce}-\mathcal{H}_{i\;bc}\delta_{a}^{d}\eta^{ce}\right)\\
 &  & +\delta A_{\;\;\;c}^{i\quad\;\;,e}3\left(\eta_{ae}\mathcal{G}_{i\:bd}+\mathcal{G}_{i\;ad}\eta_{be}\right)\eta^{cd}\\
 &  & \delta A_{\;\;\;f}^{k}\left(-\omega_{\;\;\;de}^{f}+W_{e}\delta_{d}^{f}+\alpha_{\;\;\;kd}^{i}\delta_{e}^{f}\right)\left(\mathcal{H}_{k\;ac}\delta_{b}^{e}\eta^{cd}+\mathcal{H}_{k\;bc}\delta_{a}^{e}\eta^{cd}-\mathcal{H}_{k\;ac}\delta_{b}^{d}\eta^{ce}-\mathcal{H}_{k\;bc}\delta_{a}^{d}\eta^{ce}\right)\\
 &  & +\delta A_{\;\;\;f}^{k}\left(3c_{\;\;\;kj}^{i}B^{j\;e}\right)\left(\eta_{ae}\mathcal{G}_{i\:bd}\eta^{fd}+\mathcal{G}_{i\;ad}\eta_{be}\eta^{fd}\right)
\end{eqnarray*}
where we collect terms proportional to $\delta A_{\;\;\;f}^{k},\delta A_{\;\;\;f,e}^{k}$
and $\delta A_{\;\;\;f}^{k\;\;\quad,e}$ separately, noting that the
gravitational solution reduces the $y$-covariant derivative to a
partial, $\delta A_{\;\;\;b}^{i\quad\;\;;a}=\delta A_{\;\;\;b}^{i\quad\;\;,a}$.

While the field equations determine the second derivatives of the
potentials, the potential itself and its first derivative are arbitrary
initial conditions on any Cauchy surface. Therefore, the three variations
$\delta A_{\;\;\;f}^{k},\delta A_{\;\;\;f,e}^{k}$ and $\delta A_{\;\;\;f}^{k\;\;\quad,e}$
are independent, and the coefficient of each must vanish separately:
\begin{eqnarray}
0 & = & \mathcal{H}_{i\;ac}\delta_{b}^{e}\eta^{cd}+\mathcal{H}_{i\;bc}\delta_{a}^{e}\eta^{cd}-\mathcal{H}_{i\;ac}\delta_{b}^{d}\eta^{ce}-\mathcal{H}_{i\;bc}\delta_{a}^{d}\eta^{ce}\label{x derivative constraint}\\
0 & = & 3\left(\eta_{ae}\mathcal{G}_{i\:bd}+\mathcal{G}_{i\;ad}\eta_{be}\right)\eta^{cd}\label{y derivative constraint}\\
0 & = & \left(-\omega_{\;\;\;de}^{f}+W_{e}\delta_{d}^{f}+\alpha_{\;\;\;kd}^{i}\delta_{e}^{f}\right)\left(\mathcal{H}_{k\;ac}\delta_{b}^{e}\eta^{cd}+\mathcal{H}_{k\;bc}\delta_{a}^{e}\eta^{cd}-\mathcal{H}_{k\;ac}\delta_{b}^{d}\eta^{ce}-\mathcal{H}_{k\;bc}\delta_{a}^{d}\eta^{ce}\right)\nonumber \\
 &  & +\left(3c_{\;\;\;kj}^{i}B^{j\;e}\right)\left(\eta_{ae}\mathcal{G}_{i\:bd}\eta^{fd}+\mathcal{G}_{i\;ad}\eta_{be}\eta^{fd}\right)\label{Value of potential constraint}
\end{eqnarray}
For the $x$-derivative part of the constraint, Eq.(\ref{x derivative constraint}),
we contract $eb$ and lower the $d$ index to show that $\mathcal{H}_{i\;ac}$
must vanish,
\[
0=n\mathcal{H}_{i\;ac}
\]
Similarly, contracting $ac$ in Eq.(\ref{y derivative constraint})
expressing the independence of the $y$-derivative, shows that $\mathcal{G}_{i\:be}$
must also vanish.
\begin{eqnarray*}
0 & = & 3\mathcal{G}_{i\:be}
\end{eqnarray*}
With these two conditions, the final equation Eq.(\ref{Value of potential constraint})
is identically satisfied.

These conditions satisfy both gravitational conditions on the sources,
Eqs.(\ref{First source constraint}) and (\ref{Second source constraint}).

\subsection{The source for gravity}

We have shown that
\begin{eqnarray*}
\frac{1}{1+\chi}\boldsymbol{\mathcal{R}}_{b}+\Lambda_{0}\eta_{be}\mathbf{e}^{e} & = & -\frac{\kappa}{\alpha}\mathbf{T}_{b}
\end{eqnarray*}
Expressed in terms of the Einstein tensor this is
\begin{eqnarray*}
G_{ab}+\Lambda_{C}\eta_{ab} & = & -\left(n-2\right)\left(1+\chi\right)\frac{\kappa}{\alpha}\left(T_{ab}-\eta_{ab}T\right)
\end{eqnarray*}
where the vanishing of $\mathcal{H}_{i\;ab}$ and $\mathcal{G}_{i\;ab}$
leave us with
\begin{eqnarray*}
T_{ab} & = & F_{\;\;ca}^{i}F_{i\:db}\eta^{cd}
\end{eqnarray*}
The trace of this is well-known to be gauge dependent, and the conformal
symmetry requires the energy tensor to be trace free. Therefore, we
are justified in adjusting the $SU\left(N\right)$ gauge to give
\begin{eqnarray*}
G_{ab}+\Lambda_{C}\eta_{ab} & = & -\lambda\left(F_{\;\;ca}^{i}F_{i\:db}\eta^{cd}-\frac{1}{4}\eta_{ab}\left(\eta^{ce}\eta^{df}F_{\;\;cd}^{i}F_{i\:ef}\right)\right)
\end{eqnarray*}
where
\[
\lambda=\left(n-2\right)\left(1+\chi\right)\frac{\kappa}{\alpha}
\]

\subsection{The Yang-Mills equation}

With $\mathcal{H}_{\;\;\;ab}^{k}=G_{\;\;\quad b}^{i\;a}=0$, Eqs.(\ref{Nonabelian YM field equation 1})
and (\ref{Nonabelian YM field equation 2}) reduce to
\begin{eqnarray*}
0 & = & \eta^{bc}F_{k\;ac}^{\;\;\quad;a}+\eta^{bc}F_{i\;ac}\beta_{\;\;\;k}^{i\;\;\quad a}\\
0 & = & \eta^{bc}\left(F_{j\;ac;b}+F_{i\;ac}\alpha_{\;\;\;jb}^{i}\right)
\end{eqnarray*}
where
\begin{eqnarray*}
\boldsymbol{\mathcal{A}}_{\;\;\;k}^{i} & = & c_{\;\;\;jk}^{i}A_{\;\;\;a}^{j}\mathbf{e}^{a}+c_{\;\;\;jk}^{i}B^{j\;a}\mathbf{f}_{a}\\
 & = & \boldsymbol{\alpha}_{\;\;\;k}^{i}+\boldsymbol{\beta}_{\;\;\;k}^{i}
\end{eqnarray*}

We may use these results and the form of the co-torsion,
\[
\mathbf{S}_{a}=2\left(1+\chi\right)\Delta_{ea}^{bc}y_{b}\Lambda\eta_{cd}\mathbf{e}^{d}\wedge\mathbf{e}^{e}
\]
to solve for the potentials,
\begin{eqnarray}
F_{\;\;\;ab}^{i} & = & A_{\;\;\;b;a}^{i}-A_{\;\;\;a;b}^{i}-c_{\;\;\;jk}^{i}A_{\;\;\;a}^{j}A_{\;\;\;b}^{k}+\left(1+\chi\right)\Lambda B^{i\;c}\left(\Delta_{bc}^{gf}y_{g}\eta_{fa}-\Delta_{ac}^{gf}y_{g}\eta_{fb}\right)\nonumber \\
0 & = & A_{\;\;\;b}^{i\quad\;\;,a}-B_{\;\;\;;b}^{i\;a}-c_{\;\;\;jk}^{i}B^{j\;a}A_{\;\;\;b}^{k}\nonumber \\
0 & = & B^{i\;\;b,a}-B^{i\;\;a,b}-c_{\;\;\;jk}^{i}B^{j\;a}B^{k\;b}\label{Fields in term of potentials}
\end{eqnarray}
The third equation is the vanishing of the Yang-Mills field strength
on the $y$-submanifold,
\[
\mathbf{d}_{\left(y\right)}\mathbf{B}^{i}=-\frac{1}{2}c_{\;\;\;jk}^{i}\mathbf{B}^{j}\wedge\mathbf{B}^{k}
\]
so that $\mathbf{B}^{k}$ is a pure-gauge connection for any fixed
$x^{\alpha}$. Therefore, for each $x_{0}^{\alpha}$ we may choose
an $SU\left(N\right)$ gauge $\Lambda\left(x_{0}^{\alpha},y_{\beta}\right)$
such that $\mathbf{B}^{k}=0$. But this makes the value of $\mathbf{B}^{k}$
independent of $x^{\alpha}$ as well, so $\mathbf{B}^{k}=0$ everywhere.
As a result, the fields in terms of the potentials reduce to
\begin{eqnarray*}
F_{\;\;\;ab}^{i} & = & A_{\;\;\;b;a}^{i}-A_{\;\;\;a;b}^{i}-c_{\;\;\;jk}^{i}A_{\;\;\;a}^{j}A_{\;\;\;b}^{k}\\
A_{\;\;\;b}^{i\quad\;\;,a} & = & 0
\end{eqnarray*}

Now, when we write the field equations in terms of the potentials
and set $B^{k\,a}=0$, we have
\begin{eqnarray*}
0 & = & F_{k\;ab}^{\;\;\quad,a}\\
0 & = & \eta^{bc}\left(F_{j\;ac;b}+F_{i\;ac}c_{\;\;\;jk}^{i}A_{\;\;\;a}^{j}\right)
\end{eqnarray*}
The first shows that $F_{k\;ab}$ is independent of $y_{\alpha}$
and the second shows it to be covariantly divergence free.

\section{Conclusions}

Gravitational field theories in doubled dimensions include biconformal
gravity \cite{Ivanov,NCG,WW,WheelerAugust}, double field theory \cite{Siegel1,Siegel2,Siegel3,Duff,Brandenburger,Pedagogical double field theory},
and gravity on a Kähler manifold \cite{Hazboun Wheeler,WheelerAugust}.
Each of these cases starts as a fully $2n$-dimensional theory but
ultimately is intended to describe gravity on an $n$-dimensional
submanifold. We have found a satisfactory $2n$-dimensional form of
Yang-Mills matter sources and shown that they also reduce to the expected
$n$-dimensional sources \emph{as a consequence of the field equations}.
Our gravitational reduction and the consequent reduction of the Yang-Mills
fields and field equations \emph{does }\textbf{not}\emph{ require
a section condition.}

While we discussed the issue in biconformal space, our results hold
in the related forms of double field theory and Kähler manifolds \cite{Hazboun Wheeler,WheelerAugust}.

For matter fields we restrict our attention to gauged $SU\left(N\right)$
sources (Yang-Mills type). While we find that the usual form of $2n$-dimensional
Yang-Mills action gives incorrect coupling to gravity, including a
``twist'' matrix in the action corrects the problem.

For the gravitational fields we use the most general action linear
in the biconformal curvatures. The variation is taken with respect
to all $\frac{\left(n+1\right)\left(n+2\right)}{2}$ conformal gauge
fields. In the absence of sources, the use of the gravitational field
equations to reduce fully $2n$-dimensional gravity theory to dependence
only on the fields of $n$-dimensional gravity is well established.
The field equations of torsion-free biconformal space restrict the
$\frac{1}{2}\left(n+1\right)\left(n+2\right)$ curvature components,
each initially dependent on $2n$ independent coordinates, to the
usual locally scale covariant Riemannian curvature tensor in $n$
dimensions. Ultimately, the $n$-dim solder form determines all fields,
up to coordinate and gauge transformations. Generic, torsion-free,
vacuum solutions describe $n$-dimensional scale-covariant general
relativity on the co-tangent bundle.

Here we have shown that the same reduction occurs when gauged $SU\left(N\right)$
field strengths are included as matter sources. The result goes well
beyond any previous work. With two exceptions \cite{BCSMatterWW,AW}
studies of biconformal spaces \cite{NCG,WW,AW,AWQM,Spencer Wheeler,Hazboun Wheeler,Hazboun dissertation,WheelerAugust,Lovelady,Hazboun}
have considered the pure gravity biconformal spaces, leading to vacuum
general relativity. With $SU\left(N\right)$ Yang-Mills fields as
gravitational sources, the central issue was to show that a completely
general $SU\left(N\right)$ gauge theory over a $2n$-dimensional
biconformal space does not disrupt the gravitational reduction to
general relativity, but rather itself reduces to a suitable $n$-dim
gravitational source and Yang-Mills field equation.

As with the Riemann-Cartan construction of general relativity above,
the development of biconformal spaces from group symmetry made it
straightforward to include the additional symmetry of sources. By
extending the quotient to
\[
\mathcal{M}^{2n}=\left[SO\left(p+1,q+1\right)\times SU\left(N\right)\right]/\left[SO\left(p,q\right)\times SO\left(1,1\right)\times SU\left(N\right)\right]
\]
the local symmetry is enlarged by $SU\left(N\right)$. We considered
the effects of adding an $SU\left(N\right)$ action to the gravitational
action Eq.(\ref{Action}). As central results we successfully showed:
\begin{enumerate}
\item The number of field components in $2n$ dimensions reduces by a factor
of $\frac{n-2}{2\left(2n-1\right)}$ to the expected number $\frac{n\left(n-1\right)}{2}\left(N^{2}-1\right)$
on $n$-dimensional spacetime.
\item The functional dependence of the fields reduces from $2n$ to $n$
independent variables.
\item The usual form of Yang-Mills stress-energy tensor provides the source
for the scale-covariant Einstein equation on $n$-dimensional spacetime.
\item The usual Yang-Mills field equation holds on the spacetime submanifold.
\end{enumerate}
To accomplish these goals we required two interdependent intermediate
steps:
\begin{enumerate}
\item We considered alternate forms of $2n$-dimensional Yang-Mills action,
showing that the usual action, $S_{YM}^{0}=\int tr\left(\mathbf{F}\land^{*}\mathbf{F}\right)$
gives nonstandard coupling to gravity. Instead, including a ``twist''
matrix in the action $S_{YM}=\int tr\left(\bar{\mathbf{F}}\land^{*}\mathbf{F}\right)$
with twisted form
\[
\bar{\mathcal{F}}_{AB}=\frac{1}{2}\left(K_{A}^{\;\;\;C}\mathcal{F}_{CB}+\mathcal{F}_{AC}K_{\;\;\;B}^{C}\right)
\]
leads to both the usual $n$-dimensional Yang-Mills source to the
Einstein tensor and the usual Yang-Mills equation for the $SU\left(N\right)$
fields. A similar twist has been found in other double field theory
studies \cite{SUSY Twist,Twisted DFT,Gauged double field theory}
in order to enforce supersymmetry. Here, the twist is required for
the bosonic fields alone. Interestingly, the twist matrix $K_{\;\;\;B}^{A}=\bar{K}^{AC}g_{CB}$
makes use of both the Kähler and Killing forms, $g_{AB}$ and $K_{AB}$,
respectively.
\item We considered two naturally occurring inner products for the orthonormal
frame fields: the restriction to the base manifold of the Killing
form, and the Kähler metric. We showed the Kähler form cannot lead
to the usual field equations while the variation of the Killing form
in the twisted action gives usual Yang-Mills equations and usual coupling
to gravity. Previous results in biconformal gravity did not require
the inner product.
\end{enumerate}
We also gave a brief discussion of sources in Weyl geometry and in
Weyl gravity. For Weyl gravity we indicated the possibility of alternative,
fourth order source fields. We give two examples--scalar and electromagnetic
fields--of fourth order generalizations for sources. The fourth order
fields satisfy field equations which differ from the field equations
for the corresponding second order fields only by terms linear in
the product of the fields and the curvature. The difference between
solutions to Weyl gravity with fourth order sources and general relativity
with the corresponding second order sources need not introduce a mass
scale.

\section{Appendices}

\section*{Appendix A: Failure of the Kähler inner product}

In this Appendix, we show that using the Kähler metric to define orthonormality
of the solder and co-solder forms cannot lead to the usual Yang-Mills
gravitational coupling to gravity, for either the usual Yang-Mills
action or the twisted Yang-Mills action.

We first find the Hodge dual of the Yang-Mills field here using the
inverse Kähler metric $\bar{g}^{AB}$. In general terms the Hodge
dual of a 2-form is given by Eq.(\ref{Hodge dual of a 2 form}), where
we now substitute the Kähler metric,
\begin{eqnarray}
^{*}\boldsymbol{\mathcal{F}} & = & \frac{1}{n!\left(n-2\right)!}\left(\frac{1}{2}F_{ab}\bar{g}^{am}\bar{g}^{bn}+G_{\;\;\;b}^{a}\bar{g}_{a}^{\;\;\;m}\bar{g}^{bn}+\frac{1}{2}H^{ab}\bar{g}_{a}^{\;\;\;m}\bar{g}_{b}^{\;\;\;n}\right)\varepsilon_{\;\;\;\qquad mne\cdots f}^{c\cdots d}\mathbf{f}_{c\cdots d}\mathbf{e}^{e\cdots f}\nonumber \\
 &  & +\frac{\left(-1\right)^{n-1}}{\left(n-1\right)!\left(n-1\right)!}\left(\frac{1}{2}F_{ab}\bar{g}_{\;\;\;m}^{a}\bar{g}^{bn}+G_{\;\;\;b}^{a}\bar{g}_{am}\bar{g}^{bn}+\frac{1}{2}H^{ab}\bar{g}_{am}\bar{g}_{b}^{\;\;\;n}\right)\varepsilon_{\;\;\;\qquad ne\cdots f}^{mc\cdots d}\mathbf{f}_{c\cdots d}\mathbf{e}^{e\cdots f}\nonumber \\
 &  & +\frac{\left(-1\right)^{n}}{\left(n-1\right)!\left(n-1\right)!}\left(\frac{1}{2}F_{ab}\bar{g}^{am}\bar{g}_{\;\;\;n}^{b}+G_{\;\;\;b}^{a}\bar{g}_{a}^{\;\;\;m}\bar{g}_{\;\;\;n}^{b}+\frac{1}{2}H^{ab}\bar{g}_{a}^{\;\;\;m}\bar{g}_{bn}\right)\varepsilon_{\;\;\;\qquad me\cdots f}^{nc\cdots d}\mathbf{f}_{c\cdots d}\mathbf{e}^{e\cdots f}\nonumber \\
 &  & +\frac{1}{n!\left(n-2\right)!}\left(\frac{1}{2}F_{ab}\bar{g}_{\;\;\;m}^{a}\bar{g}_{\;\;\;n}^{b}+G_{\;\;\;b}^{a}\bar{g}_{am}\bar{g}_{\;\;\;n}^{b}+\frac{1}{2}H^{ab}\bar{g}_{am}\bar{g}_{bn}\right)\varepsilon_{\;\;\;\qquad e\cdots f}^{mnc\cdots d}\mathbf{f}_{c\cdots d}\mathbf{e}^{e\cdots f}\label{Hodge dual of a 2 form-1}
\end{eqnarray}
Forming the usual Yang-Mills Lagrangian density as the wedge product,
$\boldsymbol{\mathcal{F}}\wedge{}^{*}\boldsymbol{\mathcal{F}}$, and
eliminating the basis forms in favor of the volume form $\boldsymbol{\Phi}$
yields

\begin{eqnarray}
\boldsymbol{\mathcal{F}}\wedge{}^{*}\boldsymbol{\mathcal{F}} & = & \left(\frac{1}{2}F_{mn}\bar{g}^{am}\bar{g}^{bn}+G_{\;\;\;n}^{m}\bar{g}_{\;\;\;m}^{a}\bar{g}^{bn}+\frac{1}{2}H^{mn}\bar{g}_{\;\;\;m}^{a}\bar{g}_{\;\;\;n}^{b}\right)F_{ab}\boldsymbol{\Phi}\nonumber \\
 &  & +\left(F_{mn}\bar{g}_{a}^{\;\;\;m}\bar{g}^{bn}+G_{\;\;\;n}^{m}\left(\bar{g}_{am}\bar{g}^{bn}-\bar{g}_{a}^{\;\;\;n}\bar{g}_{\;\;\;m}^{b}\right)+H^{mn}\bar{g}_{am}\bar{g}_{\;\;\;n}^{b}\right)G_{\;\;\;b}^{a}\boldsymbol{\Phi}\nonumber \\
 &  & +\left(\frac{1}{2}F_{mn}\bar{g}_{a}^{\;\;\;m}\bar{g}_{b}^{\;\;\;n}+G_{\;\;\;n}^{m}\bar{g}_{am}\bar{g}_{b}^{\;\;\;n}+\frac{1}{2}H^{mn}\bar{g}_{am}\bar{g}_{bn}\right)H^{ab}\boldsymbol{\Phi}\label{YM Lagrange density with Kahler}
\end{eqnarray}
For the diagonal form of the Kähler metric, Eq.(\ref{K=0000E4hler metric}),
this Lagrange density reduces to
\begin{eqnarray}
\boldsymbol{\mathcal{F}}\wedge{}^{*}\boldsymbol{\mathcal{F}} & = & \left(\frac{1}{2}\eta^{am}\eta^{bn}F_{ab}F_{mn}+\eta_{am}\eta^{bn}G_{\;\;\;n}^{m}G_{\;\;\;b}^{a}+\frac{1}{2}\eta_{am}\eta_{bn}H^{mn}H^{ab}\right)\boldsymbol{\Phi}\label{F F dual Lagrangian density Kahler}
\end{eqnarray}

Varying the potentials in Eq.(\ref{F F dual Lagrangian density Kahler})
yields the usual Yang-Mills field equation. However, metric variation
of Eq.(\ref{YM Lagrange density with Kahler}) gives a nonstandard
coupling to gravity. Varying and reducing the integral of Eq.(\ref{YM Lagrange density with Kahler}),
the resulting gravitational field equations are:
\begin{eqnarray}
\alpha\left(\Omega_{\;\;\;b\quad m}^{n\quad b}-\Omega_{\;\;\;b\quad a}^{a\quad b}\delta_{m}^{n}\right)+\beta\left(\Omega_{\;\;\;m}^{n}-\Omega_{\;\;\;a}^{a}\delta_{m}^{n}\right)+\Lambda\delta_{m}^{n} & = & -\left(2\eta^{be}F_{me}F_{ab}+2G_{\;\;\;a}^{d}G_{\;\;\;m}^{e}\eta_{de}\right)\eta^{an}\label{FE1 Kahler-1}\\
\left[\alpha\left(\Omega_{\;\;\;n\quad a}^{a\quad m}-\Omega_{\;\;\;b\quad a}^{a\quad b}\delta_{n}^{m}\right)+\beta\left(\Omega_{\;\;\;n}^{m}-\Omega_{\;\;\;a}^{a}\delta_{n}^{m}\right)+\Lambda\delta_{n}^{m}\right] & = & -\left(2G_{\;\;\;c}^{d}G_{\;\;\;b}^{n}\eta_{dm}\eta^{bc}+2\eta_{bc}\eta_{dm}H^{dc}H^{nb}\right)\label{FE2 Kahler-1}\\
\alpha\Omega_{\;\;\;nam}^{a}+\beta\Omega_{nm} & = & \left(2\eta^{ca}G_{\;\;\;a}^{b}F_{mc}+2\eta_{ac}G_{\;\;\;m}^{c}H^{ab}\right)\eta_{bn}\label{FE3 Kahler-1}\\
\alpha\Omega_{\;\;\;b}^{n\quad bm}+\beta\Omega^{nm} & = & -\left(2\eta^{ca}G_{\;\;\;a}^{m}F_{bc}+2\eta_{ac}G_{\;\;\;b}^{c}H^{am}\right)\eta^{nb}\label{FE4 Kahler-1}
\end{eqnarray}
The remaining four field equations involving the torsion and co-torsion
are unchanged.

Notice that the sources on the right sides of Eqs.(\ref{FE3 Kahler-1})
and (\ref{FE4 Kahler-1}) differ only in the overall sign. This means
that both the spacetime curvature, $\Omega_{\;\;\;nam}^{a}$, and
the momentum space curvature, $\Omega_{\;\;\;b}^{n\quad bm}$, are
driven with equal strength. Thus, if spacetime curvature is nonzero,
the momentum space must also be correspondingly curved. In torsion-free
solutions the left hand side of Eq.(\ref{FE4 Kahler-1}) vanishes
independently of the sources, implying a constraint on the source
fields. Thus, for the $\int\boldsymbol{\mathcal{F}}\wedge^{*}\boldsymbol{\mathcal{F}}$
source and the Kähler case,
\begin{eqnarray*}
\eta^{ca}G_{\;\;\;a}^{m}F_{bc}+\eta_{ac}G_{\;\;\;b}^{c}H^{am} & = & 0
\end{eqnarray*}
and this immediately shows that the source for the Einstein equation,
Eq.(\ref{FE3 Kahler-1}), vanishes.

The necessity for equal spacetime and momentum curvatures suggests
the possibility implementing Born reciprocity. This idea will be explored
elsewhere. However, momentum curvature also requires some part of
the torsion to be nonvanishing, and this in turn requires a different
gravitational solution than that known to reproduce general relativity.
Thus, we cannot maintain vanishing torsion without forcing the spacetime
source to vanish.

In addition to the inescapability of torsion and momentum space curvature,
the independence of the sources to Eqs.(\ref{FE1 Kahler-1}) and (\ref{FE2 Kahler-1})
also raises issues with the Kähler variation, because the method of
solution employed in \cite{WheelerAugust} makes use of the near identity
of these two equations. At the very least, an entirely different form
of reduction of the equations would be required, with no guarantee
that the Einstein equation would emerge.

A similar calculation shows that the same difficulties arise from
the twisted form of the action using the Kähler metric.

These issues do not arise with the Killing variation, Eqs.(\ref{Variation of the Killing form})--the
source for the spacetime curvature and momentum space curvature are
independent while remaining two variations are identical. The use
of the Killing form as metric also makes good geometric sense, since
it arises directly as a symmetric form in the Lie algebra and thus
as metric of the co-tangent space. As such, it respects the conformal
invariance of the full model. The Kähler structure, by contrast, reflects
symmetries and dynamical properties within the conformal group and
depend for their existence on the solution on the biconformal space.
It is not conformally invariant.

\section*{Appendix B: Symmetry of the twist matrix}

Writing the twist matrix while keeping factors of $\eta^{ab}$ explicit,
we have
\begin{eqnarray*}
K_{\;\;\;B}^{A} & = & \left(\begin{array}{cc}
\bar{K}^{ac} & \bar{K}_{\;\;\;e}^{a}\eta^{ec}\\
\eta^{ae}\bar{K}_{e}^{\;\;\;c} & \eta^{ae}\eta^{cf}\bar{K}_{ef}
\end{array}\right)\left(\begin{array}{cc}
\eta_{cb} & 0\\
0 & \eta_{cm}\eta_{bn}\eta^{mn}
\end{array}\right)\\
 & = & \left(\begin{array}{cc}
\bar{K}^{ac}\eta_{cb} & \bar{K}_{\;\;\;b}^{a}\\
\eta^{ae}\bar{K}_{e}^{\;\;\;c}\eta_{cb} & \eta^{ae}\bar{K}_{eb}
\end{array}\right)\\
K_{A}^{\;\;\;B} & = & \left(\begin{array}{cc}
\eta_{ad}\bar{K}^{db} & \bar{K}_{a}^{\;\;\;b}\\
\eta_{ad}\bar{K}_{\;\;\;e}^{d}\eta^{eb} & \bar{K}_{ae}\eta^{eb}
\end{array}\right)
\end{eqnarray*}
Symmetry of $K_{\;\;\;B}^{A}$ follows from the symmetry of the Killing
and Kähler forms, $K_{AB}=K_{BA},g_{AB}=g_{BA}$:

\begin{eqnarray*}
K_{\;\;\;B}^{A}\equiv\bar{K}^{AC}g_{CB} & = & g_{BC}\bar{K}^{CA}=K_{B}^{\;\;\;A}
\end{eqnarray*}
Comparing the expressions,
\begin{eqnarray*}
K_{\;\;\;B}^{A} & = & K_{B}^{\;\;\;A}\\
\left(\begin{array}{cc}
\bar{K}^{ac}\eta_{cb} & \bar{K}_{\;\;\;b}^{a}\\
\eta^{ae}\bar{K}_{e}^{\;\;\;c}\eta_{cb} & \eta^{ae}\bar{K}_{eb}
\end{array}\right) & = & \left(\begin{array}{cc}
\eta_{bc}\bar{K}^{ca} & \eta_{bc}\bar{K}_{\;\;\;e}^{c}\eta^{ea}\\
\bar{K}_{b}^{\;\;\;a} & \bar{K}_{bc}\eta^{ca}
\end{array}\right)
\end{eqnarray*}
Therefore, from the upper right quadrant we must have $\bar{K}_{\;\;\;b}^{a}=\eta_{bc}\bar{K}_{\;\;\;e}^{c}\eta^{ea}$
and similarly $\eta^{ae}\bar{K}_{e}^{\;\;\;c}\eta_{cb}=\bar{K}_{b}^{\;\;\;a}$
from the lower left. However we also have symmetry $\bar{K}^{AB}=\bar{K}^{BA}$
of $\bar{K}^{AB}$ itself:
\begin{eqnarray*}
\bar{K}^{AB} & = & \left(\begin{array}{cc}
\bar{K}^{ab} & \bar{K}_{\;\;\;e}^{a}\eta^{eb}\\
\eta^{ae}\bar{K}_{e}^{\;\;\;b} & \eta^{ae}\eta^{bf}\bar{K}_{ef}
\end{array}\right)\\
\left[\bar{K}^{t}\right]^{BA} & = & \left(\begin{array}{cc}
\bar{K}^{ba} & \eta^{be}\bar{K}_{e}^{\;\;\;a}\\
\bar{K}_{\;\;\;e}^{b}\eta^{ea} & \eta^{ae}\eta^{bf}\bar{K}_{ef}
\end{array}\right)
\end{eqnarray*}
This shows that $\bar{K}_{\;\;\;e}^{a}\eta^{eb}=\eta^{be}\bar{K}_{e}^{\;\;\;a}$
from which it follows that
\begin{eqnarray*}
\bar{K}_{\;\;\;b}^{a} & = & \bar{K}_{b}^{\;\;\;a}
\end{eqnarray*}
Combine this with $\eta^{ae}\bar{K}_{e}^{\;\;\;c}\eta_{cb}=\bar{K}_{b}^{\;\;\;a}$
and we see that all forms are equivalent,
\[
\eta^{ae}\bar{K}_{e}^{\;\;\;c}\eta_{cb}=\bar{K}_{b}^{\;\;\;a}=\bar{K}_{\;\;\;b}^{a}=\eta_{bc}\bar{K}_{\;\;\;e}^{c}\eta^{ea}
\]
With this, we may write
\begin{eqnarray*}
K_{\;\;\;B}^{A} & = & \left(\begin{array}{cc}
\bar{K}^{ac}\eta_{cb} & \bar{K}_{\;\;\;b}^{a}\\
\bar{K}_{\;\;\;b}^{a} & \eta^{ae}\bar{K}_{eb}
\end{array}\right)\\
K_{B}^{\;\;\;A} & = & \left(\begin{array}{cc}
\eta_{bd}\bar{K}^{da} & \bar{K}_{b}^{\;\;\;a}\\
\bar{K}_{b}^{\;\;\;a} & \bar{K}_{be}\eta^{ea}
\end{array}\right)=K_{\;\;\;B}^{A}
\end{eqnarray*}

\section*{Appendix C: Details of the metric variation of the action}

The variation of the twisted action is lengthy and includes some subtleties,
so we include details here.

We have the dual field,
\begin{eqnarray*}
^{*}\boldsymbol{\mathcal{F}} & = & \frac{1}{n!\left(n-2\right)!}\left(\frac{1}{2}F_{ab}\bar{K}^{am}\bar{K}^{bn}+\mathcal{G}_{gb}\eta^{ga}\bar{K}_{a}^{\;\;\;m}\bar{K}^{bn}+\frac{1}{2}\mathcal{H}_{gh}\eta^{ga}\bar{K}_{a}^{\;\;\;m}\eta^{hb}\bar{K}_{b}^{\;\;\;n}\right)\varepsilon_{\;\;\;\qquad mne\cdots f}^{c\cdots d}\mathbf{f}_{c\cdots d}\mathbf{e}^{e\cdots f}\\
 &  & +\frac{\left(-1\right)^{n-1}}{\left(n-1\right)!\left(n-1\right)!}\left(\frac{1}{2}F_{ab}\bar{K}_{\;\;\;m}^{a}\bar{K}^{bn}+\mathcal{G}_{gb}\eta^{ga}\bar{K}_{am}\bar{K}^{bn}+\frac{1}{2}\mathcal{H}_{gh}\eta^{ga}\eta^{hb}\bar{K}_{am}\bar{K}_{b}^{\;\;\;n}\right)\varepsilon_{\;\;\;\qquad ne\cdots f}^{mc\cdots d}\mathbf{f}_{c\cdots d}\mathbf{e}^{e\cdots f}\\
 &  & +\frac{\left(-1\right)^{n}}{\left(n-1\right)!\left(n-1\right)!}\left(\frac{1}{2}F_{ab}\bar{K}^{am}\bar{K}_{\;\;\;n}^{b}+\mathcal{G}_{gb}\eta^{ga}\bar{K}_{a}^{\;\;\;m}\bar{K}_{\;\;\;n}^{b}+\frac{1}{2}\mathcal{H}_{gh}\eta^{ga}\eta^{hb}\bar{K}_{a}^{\;\;\;m}\bar{K}_{bn}\right)\varepsilon_{\;\;\;\qquad me\cdots f}^{nc\cdots d}\mathbf{f}_{c\cdots d}\mathbf{e}^{e\cdots f}\\
 &  & +\frac{1}{n!\left(n-2\right)!}\left(\frac{1}{2}F_{ab}\bar{K}_{\;\;\;m}^{a}\bar{K}_{\;\;\;n}^{b}+\mathcal{G}_{gb}\eta^{ga}\bar{K}_{am}\bar{K}_{\;\;\;n}^{b}+\frac{1}{2}\mathcal{H}_{gh}\eta^{ga}\eta^{hb}\bar{K}_{am}\bar{K}_{bn}\right)\varepsilon_{\;\;\;\qquad e\cdots f}^{mnc\cdots d}\mathbf{f}_{c\cdots d}\mathbf{e}^{e\cdots f}
\end{eqnarray*}
and, changing indices to avoid duplication, the barred field,
\begin{eqnarray*}
\bar{\boldsymbol{\mathcal{F}}} & = & \frac{1}{2}\left(F_{rq}\bar{K}^{qt}\eta_{ts}+\bar{K}_{r}^{\;\;\;q}\mathcal{G}_{qs}\right)\mathbf{e}^{r}\wedge\mathbf{e}^{s}\\
 &  & +\frac{1}{2}\left(F_{rc}\bar{K}_{\;\;\;s}^{c}+\bar{K}_{r}^{\;\;\;c}\mathcal{H}_{cs}\right)\eta^{sw}\mathbf{e}^{r}\wedge\mathbf{f}_{w}\\
 &  & +\frac{1}{2}\left(-\mathcal{G}_{sq}\bar{K}^{qt}\eta_{tr}-\mathcal{G}_{qr}\eta^{qt}\bar{K}_{ts}\right)\eta^{sw}\mathbf{e}^{r}\wedge\mathbf{f}_{w}\\
 &  & +\frac{1}{2}\left(\mathcal{G}_{rq}\bar{K}_{\;\;\;s}^{q}+\mathcal{H}_{rq}\eta^{qt}\bar{K}_{ts}\right)\eta^{rw}\eta^{sx}\mathbf{f}_{w}\wedge\mathbf{f}_{x}
\end{eqnarray*}
We must wedge these together, then vary the metric.

Wedging, 
\begin{eqnarray*}
\bar{\boldsymbol{\mathcal{F}}}\land{}^{*}\boldsymbol{\mathcal{F}} & = & \frac{1}{2}\frac{1}{n!\left(n-2\right)!}\left(F_{rq}\bar{K}^{qt}\eta_{ts}+\bar{K}_{r}^{\;\;\;q}\mathcal{G}_{qs}\right)\\
 &  & \times\left(\mathcal{G}_{gb}\eta^{ga}\bar{K}_{a}^{\;\;\;m}\bar{K}^{bn}+\frac{1}{2}\mathcal{H}_{gh}\eta^{ga}\bar{K}_{a}^{\;\;\;m}\eta^{hb}\bar{K}_{b}^{\;\;\;n}\right)\varepsilon_{\;\;\;\qquad mne\cdots f}^{c\cdots d}\mathbf{f}_{c\cdots d}\mathbf{e}^{rse\cdots f}\\
 &  & -\frac{1}{2}\frac{1}{\left(n-1\right)!\left(n-1\right)!}\left(F_{rc}\bar{K}_{\;\;\;s}^{c}+\bar{K}_{r}^{\;\;\;c}\mathcal{H}_{cs}\right)\eta^{sw}\\
 &  & \times\left(\frac{1}{2}F_{ab}\bar{K}_{\;\;\;m}^{a}\bar{K}^{bn}+\frac{1}{2}\mathcal{H}_{gh}\eta^{ga}\eta^{hb}\bar{K}_{am}\bar{K}_{b}^{\;\;\;n}\right)\varepsilon_{\;\;\;\qquad ne\cdots f}^{mc\cdots d}\mathbf{f}_{wc\cdots d}\mathbf{e}^{re\cdots f}\\
 &  & +\frac{1}{2}\frac{1}{\left(n-1\right)!\left(n-1\right)!}\left(F_{rc}\bar{K}_{\;\;\;s}^{c}+\bar{K}_{r}^{\;\;\;c}\mathcal{H}_{cs}\right)\eta^{sw}\\
 &  & \times\left(\frac{1}{2}F_{ab}\bar{K}^{am}\bar{K}_{\;\;\;n}^{b}+\mathcal{G}_{gb}\eta^{ga}\bar{K}_{a}^{\;\;\;m}\bar{K}_{\;\;\;n}^{b}+\frac{1}{2}\mathcal{H}_{gh}\eta^{ga}\eta^{hb}\bar{K}_{a}^{\;\;\;m}\bar{K}_{bn}\right)\varepsilon_{\;\;\;\qquad me\cdots f}^{nc\cdots d}\mathbf{f}_{wc\cdots d}\mathbf{e}^{re\cdots f}\\
 &  & -\frac{1}{2}\frac{1}{\left(n-1\right)!\left(n-1\right)!}\left(-\mathcal{G}_{sq}\bar{K}^{qt}\eta_{tr}-\mathcal{G}_{qr}\eta^{qt}\bar{K}_{ts}\right)\eta^{sw}\\
 &  & \times\left(\frac{1}{2}F_{ab}\bar{K}_{\;\;\;m}^{a}\bar{K}^{bn}+\frac{1}{2}\mathcal{H}_{gh}\eta^{ga}\eta^{hb}\bar{K}_{am}\bar{K}_{b}^{\;\;\;n}\right)\varepsilon_{\;\;\;\qquad ne\cdots f}^{mc\cdots d}\mathbf{f}_{wc\cdots d}\mathbf{e}^{re\cdots f}\\
 &  & +\frac{1}{2}\frac{1}{\left(n-1\right)!\left(n-1\right)!}\left(-\mathcal{G}_{sq}\bar{K}^{qt}\eta_{tr}-\mathcal{G}_{qr}\eta^{qt}\bar{K}_{ts}\right)\eta^{sw}\\
 &  & \times\left(\frac{1}{2}F_{ab}\bar{K}^{am}\bar{K}_{\;\;\;n}^{b}+\mathcal{G}_{gb}\eta^{ga}\bar{K}_{a}^{\;\;\;m}\bar{K}_{\;\;\;n}^{b}+\frac{1}{2}\mathcal{H}_{gh}\eta^{ga}\eta^{hb}\bar{K}_{a}^{\;\;\;m}\bar{K}_{bn}\right)\varepsilon_{\;\;\;\qquad me\cdots f}^{nc\cdots d}\mathbf{f}_{wc\cdots d}\mathbf{e}^{re\cdots f}\\
 &  & +\frac{1}{2}\frac{1}{n!\left(n-2\right)!}\left(\mathcal{G}_{rq}\bar{K}_{\;\;\;s}^{q}+\mathcal{H}_{rq}\eta^{qt}\bar{K}_{ts}\right)\eta^{rw}\eta^{sx}\\
 &  & \times\left(\frac{1}{2}F_{ab}\bar{K}_{\;\;\;m}^{a}\bar{K}_{\;\;\;n}^{b}+\mathcal{G}_{gb}\eta^{ga}\bar{K}_{am}\bar{K}_{\;\;\;n}^{b}\right)\varepsilon_{\;\;\;\qquad e\cdots f}^{mnc\cdots d}\mathbf{f}_{wxc\cdots d}\mathbf{e}^{e\cdots f}
\end{eqnarray*}
Now use the relation between the basis forms and the volume element,
and the Kronecker reduction of pairs of Levi-Civita tensors,
\begin{eqnarray*}
\mathbf{f}_{c\cdots d}\wedge\mathbf{e}^{e\cdots f} & = & \bar{e}_{c\cdots d}^{\qquad e\cdots f}\boldsymbol{\Phi}\\
\varepsilon_{\;\;\;\qquad e\cdots f}^{mnc\cdots d}\bar{e}_{pqc\cdots d}^{\qquad e\cdots f} & = & n!\left(n-2\right)!\left(\delta_{p}^{m}\delta_{q}^{n}-\delta_{p}^{n}\delta_{q}^{m}\right)\\
\varepsilon_{\;\;\;\qquad ne\cdots f}^{mc\cdots d}\bar{e}_{pc\cdots d}^{\qquad qe\cdots f} & = & \left(n-1\right)!\left(n-1\right)!\delta_{p}^{m}\delta_{n}^{q}
\end{eqnarray*}
to reduce the Lagrange density to
\begin{eqnarray*}
\bar{\boldsymbol{\mathcal{F}}}\land{}^{*}\boldsymbol{\mathcal{F}} & = & \frac{1}{2}\left(F_{rq}\bar{K}^{qt}\eta_{ts}+\bar{K}_{r}^{\;\;\;q}\mathcal{G}_{qs}\right)\left(\mathcal{G}_{gb}\eta^{ga}\bar{K}_{a}^{\;\;\;m}\bar{K}^{bn}+\frac{1}{2}\mathcal{H}_{gh}\eta^{ga}\bar{K}_{a}^{\;\;\;m}\eta^{hb}\bar{K}_{b}^{\;\;\;n}\right)\left(\delta_{m}^{r}\delta_{n}^{s}-\delta_{m}^{s}\delta_{n}^{r}\right)\boldsymbol{\Phi}\\
 &  & -\frac{1}{2}\left(F_{rc}\bar{K}_{\;\;\;s}^{c}+\bar{K}_{r}^{\;\;\;c}\mathcal{H}_{cs}\right)\eta^{sw}\left(\frac{1}{2}F_{ab}\bar{K}_{\;\;\;m}^{a}\bar{K}^{bn}+\frac{1}{2}\mathcal{H}_{gh}\eta^{ga}\eta^{hb}\bar{K}_{am}\bar{K}_{b}^{\;\;\;n}\right)\delta_{w}^{m}\delta_{n}^{r}\boldsymbol{\Phi}\\
 &  & +\frac{1}{2}\left(F_{rc}\bar{K}_{\;\;\;s}^{c}+\bar{K}_{r}^{\;\;\;c}\mathcal{H}_{cs}\right)\eta^{sw}\left(\frac{1}{2}F_{ab}\bar{K}^{am}\bar{K}_{\;\;\;n}^{b}+\mathcal{G}_{gb}\eta^{ga}\bar{K}_{a}^{\;\;\;m}\bar{K}_{\;\;\;n}^{b}+\frac{1}{2}\mathcal{H}_{gh}\eta^{ga}\eta^{hb}\bar{K}_{a}^{\;\;\;m}\bar{K}_{bn}\right)\delta_{w}^{n}\delta_{m}^{r}\boldsymbol{\Phi}\\
 &  & -\frac{1}{2}\left(-\mathcal{G}_{sq}\bar{K}^{qt}\eta_{tr}-\mathcal{G}_{qr}\eta^{qt}\bar{K}_{ts}\right)\eta^{sw}\left(\frac{1}{2}F_{ab}\bar{K}_{\;\;\;m}^{a}\bar{K}^{bn}+\frac{1}{2}\mathcal{H}_{gh}\eta^{ga}\eta^{hb}\bar{K}_{am}\bar{K}_{b}^{\;\;\;n}\right)\delta_{w}^{m}\delta_{n}^{r}\boldsymbol{\Phi}\\
 &  & +\frac{1}{2}\left(-\mathcal{G}_{sq}\bar{K}^{qt}\eta_{tr}-\mathcal{G}_{qr}\eta^{qt}\bar{K}_{ts}\right)\eta^{sw}\left(\frac{1}{2}F_{ab}\bar{K}^{am}\bar{K}_{\;\;\;n}^{b}+\mathcal{G}_{gb}\eta^{ga}\bar{K}_{a}^{\;\;\;m}\bar{K}_{\;\;\;n}^{b}+\frac{1}{2}\mathcal{H}_{gh}\eta^{ga}\eta^{hb}\bar{K}_{a}^{\;\;\;m}\bar{K}_{bn}\right)\delta_{w}^{n}\delta_{m}^{r}\boldsymbol{\Phi}\\
 &  & +\frac{1}{2}\left(\mathcal{G}_{rq}\bar{K}_{\;\;\;s}^{q}+\mathcal{H}_{rq}\eta^{qt}\bar{K}_{ts}\right)\eta^{rw}\eta^{sx}\left(\frac{1}{2}F_{ab}\bar{K}_{\;\;\;m}^{a}\bar{K}_{\;\;\;n}^{b}+\mathcal{G}_{gb}\eta^{ga}\bar{K}_{am}\bar{K}_{\;\;\;n}^{b}\right)\left(\delta_{w}^{m}\delta_{x}^{n}-\delta_{w}^{n}\delta_{x}^{m}\right)\boldsymbol{\Phi}
\end{eqnarray*}
Then, absorbing the $\eta_{ab}$ and $\delta_{b}^{a}$ factors,
\begin{eqnarray*}
\bar{\boldsymbol{\mathcal{F}}}\land{}^{*}\boldsymbol{\mathcal{F}} & = & \frac{1}{2}\left(F_{mq}\bar{K}^{qt}\eta_{tn}-F_{nq}\bar{K}^{qt}\eta_{tm}+\bar{K}_{m}^{\;\;\;q}\mathcal{G}_{qn}-\bar{K}_{n}^{\;\;\;q}\mathcal{G}_{qm}\right)\left(\mathcal{G}_{gb}\eta^{ga}\bar{K}_{a}^{\;\;\;m}\bar{K}^{bn}+\frac{1}{2}\mathcal{H}_{gh}\eta^{ga}\bar{K}_{a}^{\;\;\;m}\eta^{hb}\bar{K}_{b}^{\;\;\;n}\right)\boldsymbol{\Phi}\\
 &  & -\frac{1}{2}\left(F_{nc}\bar{K}_{\;\;\;s}^{c}+\bar{K}_{n}^{\;\;\;c}\mathcal{H}_{cs}\right)\eta^{sm}\left(\frac{1}{2}F_{ab}\bar{K}_{\;\;\;m}^{a}\bar{K}^{bn}+\frac{1}{2}\mathcal{H}_{gh}\eta^{ga}\eta^{hb}\bar{K}_{am}\bar{K}_{b}^{\;\;\;n}\right)\boldsymbol{\Phi}\\
 &  & +\frac{1}{2}\left(F_{mc}\bar{K}_{\;\;\;s}^{c}+\bar{K}_{m}^{\;\;\;c}\mathcal{H}_{cs}\right)\eta^{sn}\left(\frac{1}{2}F_{ab}\bar{K}^{am}\bar{K}_{\;\;\;n}^{b}+\mathcal{G}_{gb}\eta^{ga}\bar{K}_{a}^{\;\;\;m}\bar{K}_{\;\;\;n}^{b}+\frac{1}{2}\mathcal{H}_{gh}\eta^{ga}\eta^{hb}\bar{K}_{a}^{\;\;\;m}\bar{K}_{bn}\right)\boldsymbol{\Phi}\\
 &  & -\frac{1}{2}\left(-\mathcal{G}_{sq}\bar{K}^{qt}\eta_{tn}-\mathcal{G}_{qn}\eta^{qt}\bar{K}_{ts}\right)\eta^{sm}\left(\frac{1}{2}F_{ab}\bar{K}_{\;\;\;m}^{a}\bar{K}^{bn}+\frac{1}{2}\mathcal{H}_{gh}\eta^{ga}\eta^{hb}\bar{K}_{am}\bar{K}_{b}^{\;\;\;n}\right)\boldsymbol{\Phi}\\
 &  & +\frac{1}{2}\left(-\mathcal{G}_{sq}\bar{K}^{qt}\eta_{tm}-\mathcal{G}_{qm}\eta^{qt}\bar{K}_{ts}\right)\eta^{sn}\left(\frac{1}{2}F_{ab}\bar{K}^{am}\bar{K}_{\;\;\;n}^{b}+\mathcal{G}_{gb}\eta^{ga}\bar{K}_{a}^{\;\;\;m}\bar{K}_{\;\;\;n}^{b}+\frac{1}{2}\mathcal{H}_{gh}\eta^{ga}\eta^{hb}\bar{K}_{a}^{\;\;\;m}\bar{K}_{bn}\right)\boldsymbol{\Phi}\\
 &  & +\frac{1}{2}\left(\mathcal{G}_{rq}\bar{K}_{\;\;\;s}^{q}\eta^{rm}\eta^{sn}-\mathcal{G}_{rq}\bar{K}_{\;\;\;s}^{q}\eta^{rn}\eta^{sm}+\mathcal{H}_{rq}\eta^{qt}\bar{K}_{ts}\eta^{rm}\eta^{sn}-\mathcal{H}_{rq}\eta^{qt}\bar{K}_{ts}\eta^{rn}\eta^{sm}\right)\\
 &  & \times\left(\frac{1}{2}F_{ab}\bar{K}_{\;\;\;m}^{a}\bar{K}_{\;\;\;n}^{b}+\mathcal{G}_{gb}\eta^{ga}\bar{K}_{am}\bar{K}_{\;\;\;n}^{b}\right)\boldsymbol{\Phi}
\end{eqnarray*}
After varying the metric, the null-orthonormal form of the metric
is restored, so we may anticipate this and drop terms in the product
such as $\mathcal{G}_{gb}\eta^{ga}\bar{K}_{a}^{\;\;\;m}\bar{K}^{bn}\mathcal{G}_{sq}\bar{K}^{qt}\eta_{tr}$
which will ultimately vanish. Terms with two or more factors of $\bar{K}^{ab}$
and/or $\bar{K}_{ab}$ will vanish, so we have only terms with one
of these and two off diagonal factors such as $\bar{K}_{a}^{\;\;\;b}$,
or terms with three off diagonal factors. In all cases where there
is one factor of $\bar{K}^{ab}$ or $\bar{K}_{ab}$, it is this factor
that must be varied so the off diagonal factors may be replaced. For
example, once the null orthonormal basis is restored, the only surviving
term of the variation of
\[
F_{mq}\eta_{tn}\frac{1}{2}\mathcal{H}_{gh}\eta^{ga}\eta^{hb}\bar{K}^{qt}\bar{K}_{a}^{\;\;\;m}\bar{K}_{b}^{\;\;\;n}
\]
will be
\[
F_{mq}\eta_{tn}\frac{1}{2}\mathcal{H}_{gh}\eta^{ga}\eta^{hb}\delta\bar{K}^{qt}\bar{K}_{a}^{\;\;\;m}\bar{K}_{b}^{\;\;\;n}
\]
which we may immediately write as
\[
F_{mq}\eta_{tn}\frac{1}{2}\mathcal{H}_{gh}\eta^{ga}\eta^{hb}\delta\bar{K}^{qt}\delta_{a}^{\;\;\;m}\delta_{b}^{\;\;\;n}=\frac{1}{2}\left(F_{aq}\eta_{tb}\mathcal{H}_{gh}\eta^{ga}\eta^{hb}\right)\delta\bar{K}^{qt}
\]
Terms with three off-diagonal components of the metric must be retained
until after variation.

Distributing fully, then making these reductions where possible, we
collect terms

\begin{eqnarray*}
\bar{\boldsymbol{\mathcal{F}}}\land{}^{*}\boldsymbol{\mathcal{F}} & = & \frac{1}{2}\left(\frac{1}{2}F_{ec}\eta^{cb}F_{db}-\frac{1}{2}F_{ec}\eta^{ca}F_{ad}+\frac{1}{2}F_{ad}\eta_{eb}H^{ab}+\frac{1}{2}\mathcal{H}_{ec}\eta^{cb}F_{db}-\frac{1}{2}F_{bd}\eta_{ea}H^{ab}-\frac{1}{2}\mathcal{H}_{ec}\eta^{ca}F_{ad}\right)\bar{K}^{de}\boldsymbol{\Phi}\\
 &  & +\frac{1}{2}\left(-\mathcal{G}_{ea}\mathcal{G}_{cd}\eta^{ca}+\mathcal{G}_{ae}\mathcal{G}_{cd}\eta^{ca}-\mathcal{G}_{cd}\eta_{ea}\eta^{cb}\mathcal{G}_{gb}\eta^{ga}\right)\bar{K}^{de}\boldsymbol{\Phi}\\
 &  & +\frac{1}{2}\left(-\frac{1}{2}F_{bc}\eta^{ce}H^{db}-\mathcal{H}_{bc}\eta^{ce}\frac{1}{2}H^{db}+\frac{1}{2}F_{ac}\eta^{ce}H^{ad}+\frac{1}{2}\mathcal{H}_{ac}\eta^{ce}H^{ad}+\frac{1}{2}H^{ad}\eta^{eb}F_{ab}-\frac{1}{2}H^{bd}\eta^{ea}F_{ab}\right)\bar{K}_{de}\boldsymbol{\Phi}\\
 &  & +\frac{1}{2}\left(\mathcal{G}_{rs}\eta^{re}\eta^{sb}\mathcal{G}_{gb}\eta^{gd}-\mathcal{G}_{rs}\eta^{rb}\eta^{se}\mathcal{G}_{gb}\eta^{gd}-\mathcal{G}_{qa}\eta^{qd}\eta^{eb}\mathcal{G}_{gb}\eta^{ga}\right)\bar{K}_{de}\boldsymbol{\Phi}\\
 &  & +\frac{1}{2}\left(\frac{1}{2}\mathcal{G}_{qn}H^{ab}\delta_{m}^{c}-\frac{1}{2}\mathcal{G}_{qm}H^{ab}\delta_{n}^{c}+F_{mq}\eta^{cb}\mathcal{G}_{gn}\eta^{ga}\right)\bar{K}_{\;\;\;c}^{q}\bar{K}_{a}^{\;\;\;m}\bar{K}_{b}^{\;\;\;n}\boldsymbol{\Phi}\\
 &  & +\frac{1}{2}\left(\mathcal{H}_{qs}\eta^{sb}\mathcal{G}_{gn}\eta^{ga}\delta_{m}^{c}+\frac{1}{2}\mathcal{G}_{rq}\eta^{ra}\eta^{cb}F_{mn}-\frac{1}{2}\mathcal{G}_{rq}\eta^{rb}\eta^{ca}F_{mn}\right)\bar{K}_{\;\;\;a}^{m}\bar{K}_{\;\;\;b}^{n}\bar{K}_{\;\;\;c}^{q}\boldsymbol{\Phi}
\end{eqnarray*}
and consolidate using symmetries,
\begin{eqnarray*}
\bar{\boldsymbol{\mathcal{F}}}\land{}^{*}\boldsymbol{\mathcal{F}} & = & \frac{1}{2}\left(F_{dc}\eta^{cb}F_{eb}+2F_{ad}H^{ab}\eta_{be}\right)\bar{K}^{de}\boldsymbol{\Phi}\\
 &  & +\frac{1}{2}\left(\mathcal{G}_{ae}\mathcal{G}_{cd}\eta^{ca}-2\mathcal{G}_{ea}\mathcal{G}_{cd}\eta^{ca}\right)\bar{K}^{de}\boldsymbol{\Phi}\\
 &  & +\frac{1}{2}\left(\mathcal{H}_{ab}H^{ad}\eta^{be}+2F_{ac}H^{ad}\eta^{ce}\right)\bar{K}_{de}\boldsymbol{\Phi}\\
 &  & +\frac{1}{2}\left(\mathcal{G}_{ca}\mathcal{G}_{db}\eta^{ab}-2\mathcal{G}_{da}\mathcal{G}_{bc}\eta^{ab}\right)\eta^{de}\eta^{cf}\bar{K}_{ef}\boldsymbol{\Phi}\\
 &  & +\frac{1}{2}\left(\frac{1}{2}\mathcal{G}_{dn}H^{ab}\delta_{m}^{c}-\frac{1}{2}\mathcal{G}_{dm}H^{ab}\delta_{n}^{c}+F_{md}\eta^{cb}\mathcal{G}_{gn}\eta^{ga}\right)\bar{K}_{a}^{\;\;\;m}\bar{K}_{b}^{\;\;\;n}\bar{K}_{\;\;\;c}^{d}\boldsymbol{\Phi}\\
 &  & +\frac{1}{2}\left(\mathcal{H}_{ds}\eta^{sb}\mathcal{G}_{gn}\eta^{ga}\delta_{m}^{c}+\frac{1}{2}\mathcal{G}_{rd}\eta^{ra}\eta^{cb}F_{mn}-\frac{1}{2}\mathcal{G}_{rd}\eta^{rb}\eta^{ca}F_{mn}\right)\bar{K}_{\;\;\;a}^{m}\bar{K}_{\;\;\;b}^{n}\bar{K}_{\;\;\;c}^{d}\boldsymbol{\Phi}
\end{eqnarray*}
Checking the limit in the orthonormal basis, we find the correct form,
$\bar{\boldsymbol{\mathcal{F}}}\land{}^{*}\boldsymbol{\mathcal{F}}=\left(\mathcal{G}_{ab}H^{ab}+F_{ab}\eta^{ac}\eta^{bd}\mathcal{G}_{cd}\right)\boldsymbol{\Phi}$.
Proceeding, we vary the metric

\begin{eqnarray*}
\delta\left(\bar{\boldsymbol{\mathcal{F}}}\land{}^{*}\boldsymbol{\mathcal{F}}\right) & = & \frac{1}{2}\left(F_{dc}\eta^{cb}F_{eb}+2F_{ad}H^{ab}\eta_{be}+\left(\mathcal{G}_{ae}-2\mathcal{G}_{ea}\right)\mathcal{G}_{cd}\eta^{ca}\right)\delta\bar{K}^{de}\boldsymbol{\Phi}\\
 &  & +\frac{1}{2}\bar{K}_{b}^{\;\;\;n}\bar{K}_{\;\;\;c}^{d}\left(\frac{1}{2}\mathcal{G}_{db}H^{ab}\delta_{m}^{c}-\frac{1}{2}\mathcal{G}_{dm}H^{ab}\delta_{b}^{c}+F_{md}\eta^{cb}\mathcal{G}_{gb}\eta^{ga}\right)\delta\bar{K}_{a}^{\;\;\;m}\boldsymbol{\Phi}\\
 &  & +\frac{1}{2}\left(\frac{1}{2}\mathcal{G}_{dn}H^{ab}\delta_{m}^{c}-\frac{1}{2}\mathcal{G}_{dm}H^{ab}\delta_{n}^{c}+F_{md}\eta^{cb}\mathcal{G}_{gn}\eta^{ga}\right)\bar{K}_{a}^{\;\;\;m}\delta\bar{K}_{b}^{\;\;\;n}\bar{K}_{\;\;\;c}^{d}\boldsymbol{\Phi}\\
 &  & +\frac{1}{2}\left(\frac{1}{2}\mathcal{G}_{dn}H^{ab}\delta_{m}^{c}-\frac{1}{2}\mathcal{G}_{dm}H^{ab}\delta_{n}^{c}+F_{md}\eta^{cb}\mathcal{G}_{gn}\eta^{ga}\right)\bar{K}_{a}^{\;\;\;m}\bar{K}_{b}^{\;\;\;n}\delta\bar{K}_{\;\;\;c}^{d}\boldsymbol{\Phi}\\
 &  & +\frac{1}{2}\left(\mathcal{H}_{ds}\eta^{sb}\mathcal{G}_{gn}\eta^{ga}\delta_{m}^{c}+\frac{1}{2}\mathcal{G}_{rd}\eta^{ra}\eta^{cb}F_{mn}-\frac{1}{2}\mathcal{G}_{rd}\eta^{rb}\eta^{ca}F_{mn}\right)\delta\bar{K}_{\;\;\;a}^{m}\bar{K}_{\;\;\;b}^{n}\bar{K}_{\;\;\;c}^{d}\boldsymbol{\Phi}\\
 &  & +\frac{1}{2}\left(\mathcal{H}_{ds}\eta^{sb}\mathcal{G}_{gn}\eta^{ga}\delta_{m}^{c}+\frac{1}{2}\mathcal{G}_{rd}\eta^{ra}\eta^{cb}F_{mn}-\frac{1}{2}\mathcal{G}_{rd}\eta^{rb}\eta^{ca}F_{mn}\right)\bar{K}_{\;\;\;a}^{m}\delta\bar{K}_{\;\;\;b}^{n}\bar{K}_{\;\;\;c}^{d}\boldsymbol{\Phi}\\
 &  & +\frac{1}{2}\left(\mathcal{H}_{ds}\eta^{sb}\mathcal{G}_{gn}\eta^{ga}\delta_{m}^{c}+\frac{1}{2}\mathcal{G}_{rd}\eta^{ra}\eta^{cb}F_{mn}-\frac{1}{2}\mathcal{G}_{rd}\eta^{rb}\eta^{ca}F_{mn}\right)\bar{K}_{\;\;\;a}^{m}\bar{K}_{\;\;\;b}^{n}\delta\bar{K}_{\;\;\;c}^{d}\boldsymbol{\Phi}\\
 &  & +\frac{1}{2}\left(\mathcal{H}_{ab}H^{ad}\eta^{be}+2F_{ac}H^{ad}\eta^{ce}\right)\bar{K}_{de}\boldsymbol{\Phi}+\frac{1}{2}\mathcal{G}_{ca}\eta^{ab}\left(\mathcal{G}_{db}-2\mathcal{G}_{bd}\right)\eta^{de}\eta^{cf}\bar{K}_{ef}\boldsymbol{\Phi}\\
 &  & +\left(\mathcal{G}_{ab}H^{ab}+F_{ab}\eta^{ac}\eta^{bd}\mathcal{G}_{cd}\right)\delta\boldsymbol{\Phi}
\end{eqnarray*}
At this point we may replace remaining unvaried metric components,
leaving
\begin{eqnarray*}
\delta\left(\bar{\boldsymbol{\mathcal{F}}}\land{}^{*}\boldsymbol{\mathcal{F}}\right) & = & \frac{1}{2}\left(F_{dc}\eta^{cb}F_{eb}+2F_{ad}H^{ab}\eta_{be}+\left(\mathcal{G}_{ae}-2\mathcal{G}_{ea}\right)\mathcal{G}_{cd}\eta^{ca}\right)\delta\bar{K}^{de}\boldsymbol{\Phi}\\
 &  & +\frac{1}{2}\left(\frac{1}{2}\mathcal{G}_{cb}H^{ab}\delta_{m}^{c}-\frac{1}{2}\mathcal{G}_{cm}H^{ab}\delta_{b}^{c}+F_{mc}\eta^{cb}\mathcal{G}_{gb}\eta^{ga}\right)\delta\bar{K}_{a}^{\;\;\;m}\boldsymbol{\Phi}\\
 &  & +\frac{1}{2}\left(\frac{1}{2}\mathcal{G}_{an}H^{ab}-\frac{1}{2}\mathcal{G}_{na}H^{ab}+F_{ac}\eta^{cn}\mathcal{G}_{gb}\eta^{ga}\right)\delta\bar{K}_{b}^{\;\;\;n}\boldsymbol{\Phi}\\
 &  & +\frac{1}{2}\left(\frac{1}{2}\mathcal{G}_{db}H^{cb}-\frac{1}{2}\mathcal{G}_{da}H^{ac}+F_{ad}\eta^{cb}\mathcal{G}_{gb}\eta^{ga}\right)\bar{K}_{b}^{\;\;\;n}\delta\bar{K}_{\;\;\;c}^{d}\boldsymbol{\Phi}\\
 &  & +\frac{1}{2}\left(\mathcal{H}_{cs}\eta^{sb}\mathcal{G}_{gb}\eta^{ga}\delta_{m}^{c}+\frac{1}{2}\mathcal{G}_{rc}\eta^{ra}\eta^{cb}F_{mb}-\frac{1}{2}\mathcal{G}_{rc}\eta^{rb}\eta^{ca}F_{mb}\right)\delta\bar{K}_{\;\;\;a}^{m}\boldsymbol{\Phi}\\
 &  & +\frac{1}{2}\left(\mathcal{H}_{cs}\eta^{sb}\mathcal{G}_{gn}\eta^{gc}+\frac{1}{2}\mathcal{G}_{rc}\eta^{ra}\eta^{cb}F_{an}-\frac{1}{2}\mathcal{G}_{rc}\eta^{rb}\eta^{ca}F_{an}\right)\delta\bar{K}_{\;\;\;b}^{n}\boldsymbol{\Phi}\\
 &  & +\frac{1}{2}\left(\mathcal{H}_{ds}\eta^{sb}\mathcal{G}_{gb}\eta^{gc}+\frac{1}{2}\mathcal{G}_{rd}\eta^{ra}\eta^{cb}F_{ab}-\frac{1}{2}\mathcal{G}_{rd}\eta^{rb}\eta^{ca}F_{ab}\right)\delta\bar{K}_{\;\;\;c}^{d}\boldsymbol{\Phi}\\
 &  & +\frac{1}{2}\left(\left(\mathcal{H}_{ab}H^{ae}\eta^{bf}+2F_{ac}H^{ae}\eta^{cf}\right)+\mathcal{G}_{ca}\eta^{ab}\left(\mathcal{G}_{db}-2\mathcal{G}_{bd}\right)\eta^{de}\eta^{cf}\right)\bar{K}_{ef}\boldsymbol{\Phi}\\
 &  & +\left(\mathcal{G}_{ab}H^{ab}+F_{ab}\eta^{ac}\eta^{bd}\mathcal{G}_{cd}\right)\delta\boldsymbol{\Phi}
\end{eqnarray*}
Finally, we collect all terms by type of variation,
\begin{eqnarray*}
\delta\left(\bar{\boldsymbol{\mathcal{F}}}\land{}^{*}\boldsymbol{\mathcal{F}}\right) & = & \frac{1}{2}\left(F_{dc}\eta^{cb}F_{eb}+2F_{ad}H^{ab}\eta_{be}+\left(\mathcal{G}_{ae}-2\mathcal{G}_{ea}\right)\mathcal{G}_{cd}\eta^{ca}\right)\delta\bar{K}^{de}\boldsymbol{\Phi}\\
 &  & +\frac{1}{2}\left(\frac{1}{2}\mathcal{G}_{cb}H^{nb}\delta_{m}^{c}-\frac{1}{2}\mathcal{G}_{cm}H^{nb}\delta_{b}^{c}+F_{mc}\eta^{cb}\mathcal{G}_{gb}\eta^{gn}\right)\delta\bar{K}_{n}^{\;\;\;m}\boldsymbol{\Phi}\\
 &  & +\frac{1}{2}\left(\frac{1}{2}\mathcal{G}_{am}H^{an}-\frac{1}{2}\mathcal{G}_{ma}H^{an}+F_{ac}\eta^{cn}\mathcal{G}_{gm}\eta^{ga}\right)\delta\bar{K}_{n}^{\;\;\;m}\boldsymbol{\Phi}\\
 &  & +\frac{1}{2}\left(\frac{1}{2}\mathcal{G}_{mb}H^{nb}-\frac{1}{2}\mathcal{G}_{ma}H^{an}+F_{am}\eta^{nb}\mathcal{G}_{gb}\eta^{ga}\right)\delta\bar{K}_{\;\;\;n}^{m}\boldsymbol{\Phi}\\
 &  & +\frac{1}{2}\left(\mathcal{H}_{cs}\eta^{sb}\mathcal{G}_{gb}\eta^{gn}\delta_{m}^{c}+\frac{1}{2}\mathcal{G}_{rc}\eta^{rn}\eta^{cb}F_{mb}-\frac{1}{2}\mathcal{G}_{rc}\eta^{rb}\eta^{cn}F_{mb}\right)\delta\bar{K}_{\;\;\;n}^{m}\boldsymbol{\Phi}\\
 &  & +\frac{1}{2}\left(\mathcal{H}_{cs}\eta^{sn}\mathcal{G}_{gm}\eta^{gc}+\frac{1}{2}\mathcal{G}_{rc}\eta^{ra}\eta^{cn}F_{am}-\frac{1}{2}\mathcal{G}_{rc}\eta^{rn}\eta^{ca}F_{am}\right)\delta\bar{K}_{\;\;\;n}^{m}\boldsymbol{\Phi}\\
 &  & +\frac{1}{2}\left(\mathcal{H}_{ms}\eta^{sb}\mathcal{G}_{gb}\eta^{gn}+\frac{1}{2}\mathcal{G}_{rm}\eta^{ra}\eta^{nb}F_{ab}-\frac{1}{2}\mathcal{G}_{rm}\eta^{rb}\eta^{na}F_{ab}\right)\delta\bar{K}_{\;\;\;n}^{m}\boldsymbol{\Phi}\\
 &  & +\frac{1}{2}\left(\left(\mathcal{H}_{ab}H^{ae}\eta^{bf}+2F_{ac}H^{ae}\eta^{cf}\right)+\mathcal{G}_{ca}\eta^{ab}\left(\mathcal{G}_{db}-2\mathcal{G}_{bd}\right)\eta^{de}\eta^{cf}\right)\bar{K}_{ef}\boldsymbol{\Phi}\\
 &  & +\left(\mathcal{G}_{ab}H^{ab}+F_{ab}\eta^{ac}\eta^{bd}\mathcal{G}_{cd}\right)\delta\boldsymbol{\Phi}
\end{eqnarray*}
This allows further simplifications, then including the variation
of the volume form we have the final result,
\begin{eqnarray*}
\delta\left(\bar{\boldsymbol{\mathcal{F}}}\land{}^{*}\boldsymbol{\mathcal{F}}\right) & = & \frac{1}{2}\left(F_{dc}\eta^{cb}F_{eb}+2F_{ad}H^{ab}\eta_{be}+\left(\mathcal{G}_{ae}-2\mathcal{G}_{ea}\right)\mathcal{G}_{cd}\eta^{ca}\right)\delta\bar{K}^{de}\boldsymbol{\Phi}\\
 &  & +H^{na}\left(2\mathcal{G}_{ma}-\mathcal{G}_{am}\right)\delta\bar{K}_{n}^{\;\;\;m}\boldsymbol{\Phi}\\
 &  & +F_{ma}\left(\mathcal{G}^{na}-2\mathcal{G}^{an}\right)\delta\bar{K}_{\;\;\;n}^{m}\boldsymbol{\Phi}\\
 &  & +\frac{1}{2}\left(\left(\mathcal{H}_{ab}H^{ae}\eta^{bf}+2F_{ac}H^{ae}\eta^{cf}\right)+\mathcal{G}_{ca}\eta^{ab}\left(\mathcal{G}_{db}-2\mathcal{G}_{bd}\right)\eta^{de}\eta^{cf}\right)\bar{K}_{ef}\boldsymbol{\Phi}\\
 &  & +\left(\mathcal{G}_{ab}H^{ab}+F_{ab}\eta^{ac}\eta^{bd}\mathcal{G}_{cd}\right)\delta\boldsymbol{\Phi}
\end{eqnarray*}

\pagebreak{}

\end{document}